\documentclass[
fontsize =11pt , % 10pt , 12 pt ( Schriftgroesse im Dokument )
paper =a4, % DIN -A4- Format
draft =false , % Probeversion , falls draft = true
titlepage=false , % gesonderte Titelseite
twoside =false , % einseitiges Layout ( sonst doppelseitig )
open =right , % neue Kapitel beginnen rechts (left , any )
parskip=half , % Absatzformat full <*+->, half <*+->, no
DIV =12]{scrartcl}
\usepackage[utf8]{inputenc}
\usepackage[T1]{fontenc}
\usepackage[english]{babel}
\usepackage{amsmath}
\usepackage{amsfonts}
\usepackage{amssymb}
\usepackage{graphicx}
\usepackage{bm}
\usepackage{csquotes}
\usepackage{tikz}
\usepackage{subcaption}
\usepackage{booktabs}
\usepackage{array}
\usepackage{enumitem}
\usepackage{adjustbox}
\usepackage{multirow}
\usepackage{tabularx}
\usepackage{authblk}
\usepackage{setspace}
\usepackage{anysize}
\marginsize{2cm}{2.5cm}{2.5cm}{2cm}

\usetikzlibrary{arrows.meta}

\title{Spatio-temporal count autoregression}
\author[1]{Steffen Maletz}
\author[2]{Konstantinos Fokianos}
\author[1]{Roland Fried}
\affil[1]{Department of Statistics, TU Dortmund University, Germany}
\affil[2]{Department of Mathematics and Statistics, University of Cyprus, Cyprus}

\usepackage[backend=biber, 
			style=authoryear-comp, 
			citestyle = authoryear-comp, 
			maxcitenames=2, 
			maxbibnames=20, 
			dashed = false, 
			giveninits=true, 
			url = false, 
			natbib=true, 
			doi = false, 
			eprint =false, 
			bibstyle =authoryear, 
			isbn=false, clearlang=true, 
            uniquename=false,
			uniquelist=false]{biblatex}

\renewbibmacro*{volume+number+eid}{%
	\printfield{volume}%
	\setunit*{\addnbthinspace}% NEW (optional); there's also \addnbthinspace
	\printfield{number}%
	\setunit{\addcomma\space}%
	\printfield{eid}}
\DeclareFieldFormat[article]{number}{\mkbibparens{#1}}

\AtEveryBibitem{\clearfield{month}}
\AtEveryCitekey{\clearfield{month}}
\AtEveryBibitem{\clearfield{day}}
\AtEveryCitekey{\clearfield{day}}
\AtEveryBibitem{\clearfield{note}}
\AtEveryCitekey{\clearfield{note}}

\DefineBibliographyStrings{german}{andothers={\mbox{et~al.}}}

\DeclareMathOperator{\vect}{vec}

\DeclareMathOperator{\var}{Var}
\providecommand{\keywords}[1]
{
  \small	
  \textbf{\textit{Keywords---}} #1
}

\NewBibliographyString{toappear}
\DefineBibliographyStrings{english}{%
  toappear = {to appear},
  version = {R package version},
}

\renewbibmacro*{in:}{%
  \iffieldundef{pubstate}
    {}
    {\printfield{pubstate}%
     \setunit{\addspace}%
     \clearfield{pubstate}}%
  \printtext{%
    \bibstring{in}\intitlepunct}}

\usepackage[labelformat=simple]{subcaption}

\newcolumntype{C}[1]{>{\centering\let\newline\\\arraybackslash\hspace{0pt}}m{#1}}

\begin{document}
	\maketitle

	%{\centering \section*{Abstract}}
\begin{center}
    \sffamily\Large\bfseries Abstract
\end{center}

We study the problem of modeling and inference for spatio-temporal count processes. Our approach uses parsimonious parameterisations of multivariate autoregressive count time series models, including possible regression on covariates. We control the number of parameters by specifying spatial neighbourhood structures for possibly huge matrices that take into account spatio-temporal dependencies. This work is motivated by real data applications which call for suitable models. Extensive simulation studies show that our approach yields reliable estimators.

\keywords{covariates, generalized linear models, INGARCH, log-linear models, space time data, spatio-temporal autocovariance}

	\section{Introduction}
The analysis of spatio-temporal count data is a challenging problem in many areas, such as epidemiology, traffic, ecology and financial markets. Applications vary from modeling diseases, such as dengue fever in Thailand \citep{chen_markov_2019} or the COVID-19 pandemic in Italy \citep{farcomeni_ensemble_2021}, crime data \citep{liesenfeld_likelihood-based_2017, clark_class_2021}, traffic data (\cite{ermagun_spatiotemporal_2018}) including road accidents \citep{liu_exploring_2017} or the number of emergency calls for identifying major cost factors in health care \citep{simoes_spatio-temporal_2019}.

Spatio-temporal data are often collected at a fix set of locations at equidistant time points. Then there can be time-lagged dependencies between the measurements at different locations. Univariate time series models, applied to each location separately, do not make use of all the information available and neglect such relevant information, which may be crucial for statistical inference and prediction. The multivariate count time series models proposed by \textcite{fokianos_multivariate_2020} offer an alternative approach. These models are multivariate extensions of popular univariate linear and log-linear time series models for counts derived in the framework of generalized linear models \citep{ferland_integer-valued_2006, fokianos_log-linear_2011}. 
However, application of these time series models poses computational issues as the number of unknown parameters increases quadratically with the number of component processes. In this work, the components of a multivariate count process correspond to the, possibly many, locations at which we observe the data. 

We approach this problem in the same spirit as \textcite{pfeifer_three-stage_1980}, who propose  space-time autoregressive moving-average (STARMA) processes as modified versions of vector autoregressive moving-average (VARMA) models (see also \cite{martins_space-time_2023}), i.e., a model class different from those considered here. Replacing the parameter matrices with simple linear combinations of prespecified neighborhood matrices makes it possible to reduce the number of unknown parameters considerably. The choice of neighborhood matrices forming parametric combinations is based on the application at hand; for instance, we can employ adjacency matrices for networks, distance matrices between different locations, or spatial lagging on lattice structures. We give some examples later on.

Our work extends the unpublished master thesis by \textcite{maletz_spatio-temporal_2021} to higher order temporal model orders. Simple special cases of the models considered in this paper have also been investigated in \textcite{armillotta_count_2023, armillotta_nonlinear_2023}. \textcite{jahn_approximately_2023} present an approximately linear model following this methodology, but the authors use a different link function.  \textcite{martins_space-time_2023} study a similar approach but in the framework of multivariate integer-valued autoregressive moving-average (INARMA) processes. Such models are based on the notion of the thinning operator and their properties are different from those of the models studied here. In addition, likelihood inference is challenging in the INARMA framework. We consider both linear and log-linear models and the modelling of covariate effects.
We additionally derive an asymptotic Wald test to check the significance of individual parameters based on likelihood inference. Further, we analyse the weekly number of reported Rota virus infections in Germany and identify spatial dependencies, population effects as well as seasonal and spatial effects.  Another data example is monthly burglary counts in Chicago, where we analyse spatial dependencies and demographic effects as well as temperature effects and a trend.

The rest of this paper is structured as follows. In Section~\ref{sec:methods} we describe the model class we put forward in this contribution and discuss its theoretical properties. Theoretical properties of the quasi-maximum likelihood estimation, model selection criteria, and an asymptotic Wald test are provided in Section~\ref{sec:estimation}. In  Section~\ref{sec:simulation} we present a simulation study in which we check these theoretical properties  and consider the effects of model misspecification. Section~\ref{sec:real} provides  applications to real data representing rota virus cases in Germany and burglary counts in Chicago. We compare the models with models from the literature to illustrate the usefulness of our approach. The paper concludes with a summary and some comments on possible future research.
	\subsection*{Notations}
Let $\bm{1}_p = (1, \ldots, 1)' \in \mathbb{R}^p$ be the $p$-dimensional vector of ones and $\bm{I}_p$ the $(p \times p)$  identity matrix. For two vectors $\bm{x}, \bm{y} \in \mathbb{R}^p$ we write $\bm{x} \preceq \bm{y}$ iff $x_i \leq y_i \,\forall i = 1, \ldots, p$. For a vector $\bm{x} \in \mathbb{R}^p$ we denote by $\Vert\bm{x}\Vert_r = \left( \sum_{i = 1}^r |x_i|^r \right)^{1/r}$ the $\ell^r$-norm of the vector, and for a matrix $\bm{A} \in \mathbb{R}^{p \times p}$, $\Vert \bm{A} \Vert_r$ is the matrix norm induced by the $\ell^r$ vector norm, i.e. $||\bm{A}||_r = \max_{||\bm{x}||_r = 1} ||\bm{Ax}||_r$. For $r = 1$ this results in the column sum norm $||\bm{A}||_1 = \max_{j = 1, \ldots, p} |a_{ij}|$, and for $r = \infty$ in the row sum norm $||\bm{A}||_{\infty} = \max_{i = 1, \ldots, p} |a_{ij}|$.
The spectral norm, for $r = 2$, is determined via $||\bm{A}||_2 = \sqrt{\lambda_{\max}(\bm{A}^{\mathrm{T}}\bm{A})}$, where $\lambda_{\max}(\bm{A}^{\mathrm{T}}\bm{A})$ is the largest eigenvalue of $\bm{A}^{\mathrm{T}}\bm{A}$. With $\text{trace}(\cdot)$ we denote the trace of a square matrix, i.e., the sum of the diagonal elements.

\section{Methodology}\label{sec:methods}
In this section we modify models for multivariate count time series, as introduced by  \textcite{fokianos_multivariate_2020}, to suit high-dimensional spatio-temporal processes. For this we apply ideas similar to the literature on STARMA-models, as introduced by \textcite{pfeifer_three-stage_1980}, which provide flexible and yet parsimonious parameterisations of VARMA models.

\subsection{Assumptions}
\label{sec:assumptions}
Assume that a $p$-dimensional count data time series $\lbrace \bm{Y}_t = (Y_{1, t}, \ldots, Y_{p, t})' \, | \, t = 0, 1, \ldots \rbrace=\lbrace \bm{Y}_t \rbrace$, and $m$ \textcolor{black}{$p$-dimensional} covariate processes $\lbrace \bm{X}_{k, t}=(X_{1, k, t},\ldots,X_{p, k, t})' \, | \, t = 0, 1, \ldots \rbrace$, $k = 1, \ldots, m$, are observed simultaneously. Let $\lbrace \bm{Y}_t \rbrace$ have an associated intensity process $\lbrace \bm{\lambda}_t \, | \, t = 0, 1, \ldots \rbrace$, where $\bm{\lambda}_t  = \mathbb{E}\left( \bm{Y}_t \, | \, \mathcal{F}_{t - 1}^{\bm{Y}, \bm{\lambda}, \bm{X}} \right)$. Here we denote with $\mathcal{F}_{t}^{\bm{Y}, \bm{\lambda}, \bm{X}} = \sigma\left( \bm{Y}_s,  \bm{X}_{k, s}, k = 1, \ldots, m, s \leq t, \bm{\lambda}_{0} \right)$ the $\sigma$-algebra comprising all information on the observed and the covariate processes up to time $t$, including an initial value $\bm{\lambda}_{0}$. We assume the same general form for the data-generating process as in  \textcite{fokianos_multivariate_2020}. \textcolor{black}{Conditional} on $\mathcal{F}_{t}^{\bm{Y}, \bm{\lambda}, \bm{X}}$, the components $Y_{i, t}$, marginally, are assumed to follow a Poisson distribution with parameter $\lambda_{i, t}$. Each component $\lbrace Y_{i, t}, t=0,1,\ldots \rbrace$, $\lbrace \lambda_{i, t}, t=0,1,\ldots \rbrace$, and $\lbrace X_{i, k, t}, t=0,1,\ldots \rbrace$ represents the trajectory of these variables at one of $p$ fixed locations or regions. A copula is used to \textcolor{black}{take into account} contemporaneous dependencies between the components of $\bm{Y}_t$ given the effects of the past. \textcolor{black}{Hence, }the joint distribution of $\bm{Y}_t \, | \, \mathcal{F}_{t - 1}^{\bm{Y}, \bm{\lambda}, \bm{X}}$ \textcolor{black}{implies that} the components $Y_{i,t}$ are \textcolor{black}{correlated}. Note that the data generating process is based on simple properties of Poisson processes, but it can be extended to accommodate other desired marginal distributions.

\sloppy Following \textcite{pfeifer_three-stage_1980}, we make use of \textcolor{black}{a given set of}  spatial weight matrices $\bm{W}^{(\ell)} \textcolor{black}{= \left(w_{ij}^{(\ell)}\right)_{i, j = 1, \ldots, p}} \in \mathbb{R}^{p \times p}, \ell = 0, 1, \ldots, \textcolor{black}{\ell_{\text{max}}}$, which characterise the spatial dependencies between the locations. \textcolor{black}{The index $\ell$ denotes the spatial order of the neighbourhood defined by $\bm{W}^{(\ell)}$.} We assume that the matrices $\bm{W}^{(\ell)}$ contain only non-negative entries and are row normalised, i.e., $\bm{W}^{(\ell)} \bm{1}_p = \bm{1}_p$. This normalisation is not necessary, in general, but it simplifies the conditions for stochastic stability of the processes\textcolor{black}{, see \ref{sec:stationarity} in the Supplement}. We set $\bm{W}^{(0)} = \bm{I}_p$, while the diagonal elements of $\bm{W}^{(\ell)}$ are set to $0$ for $\ell > 0$ to achieve identifiability. Locations that belong to a neighbourhood of low order are supposed to be closer to the originating locations than the locations in neighbourhoods of higher order. In the neighbourhood of order $0$, all locations are therefore only neighbouring themselves.

In applications, there are various ways to define the matrices $\bm{W}^{(\ell)}$, see \textcite{getis_weights_2004}. For example, on grid data, the spatial weights may be defined via 
\begin{align}
\label{eq:weights}
    w_{ij}^{(\ell)} = \begin{cases}
        1 / \# \mathcal{N}_i^{(\ell)}, & \text{if } j \in \mathcal{N}_i^{(\ell)},\\
        0, & \text{otherwise,}
    \end{cases}
\end{align}
with $\mathcal{N}_i^{(\ell)}$ the set of all neighbours of order $\ell$ of location $i$. A common way of defining neighbourhoods is to consider spatially contiguous locations. Then, for example, the neighbourhood $\mathcal{N}_i^{(1)}$ may include all locations that share a common boundary with location $i$. Simple examples of these weights on a regular grid are discussed in the next subsection and in Section~\ref{sec:simulation}.
If locations can be represented by coordinates, the weight matrices may also be defined based on distances. This can be done, for example, by using inverse distances or a neighborhood based on $k$-nearest neighbours. The first approach implies that more distant locations have less influence, while with $k$-nearest neighbours, all locations in the neighbourhood are assigned the same weight.
It is important to choose a reasonable neighbourhood structure that is suitable for answering the research question and to interpret it correctly. Fine-tuning the neighborhood matrices using several weights does not necessarily lead to substantially better conclusions, as they often share many elements, causing correlated parameter estimations \citep{lesage_myth_2014}. 
Moreover, including a large number of neighbors may include places that actually have no influence.

\subsection{Linear Poisson-STARMA Models}

We investigate linear Poisson-STARMA processes, assuming the latent intensity process $\bm{\lambda}_t$ is of the form
\begin{align}
	\label{method:linear_extended}
	\bm{\lambda}_t = \bm{\delta} + \sum_{i = 1}^{q} \sum_{\ell = 0}^{a_i} \alpha_{i\ell}\bm{W}^{(\ell)} \bm{\lambda}_{t - i} + \sum_{j = 1}^{r} \sum_{\ell = 0}^{b_j} \beta_{j\ell} \bm{W}^{(\ell)}\bm{Y}_{t - j}.
\end{align}
In the above, $\bm{\delta} \succeq \bm{0}$ is the vector of intercept terms, $r \in \mathbb{N}$ denotes the maximum time lag for regression on past values $\bm{Y}_{t - j}$ of the observed process, $q \in \mathbb{N}$ indicates the maximum time lag for regression on past values $\bm{\lambda}_{t - i}$ of the intensity process, which corresponds to a feedback mechanism, and $a_i$ and $b_j$ represent the maximum spatial orders at time lag $i=1,\ldots,q$ and $j=1,\ldots,r$, respectively. 
The parameters $\alpha_{i\ell} \geq 0$ and $\beta_{j\ell} \geq 0$ measure the impact of past conditional moments $\bm{\lambda}_{t - i}$ and past observations $\bm{Y}_{t - j}$ in the neighborhood defined by $\bm{W}^{(\ell)}$. All parameters are assumed to be non-negative so that $\lambda_{i, t}$ is guaranteed to be positive.

\textcite{fokianos_multivariate_2020} consider a multivariate time series setting and studied the first order model, i.e. $q=r=1$. They allow for arbitrary linear effects of the previous observation $\bm{Y}_{t - 1}$ and the previous intensity $\bm{\lambda}_{t - 1}$ by using arbitrary matrices $\bm{A}$ and $\bm{B}$ in \eqref{method:linear_extended}. The number of parameters in their model for the intensity process is $2p^2+p$, while the first order model
\begin{align}
	\label{method:linear_without_covariates}
	\bm{\lambda}_t = \bm{\delta} + \sum\limits_{\ell = 0}^{a_1} \alpha_{1\ell} \bm{W}^{(\ell)} \bm{\lambda}_{t - 1} + \sum\limits_{\ell = 0}^{b_1} \beta_{1\ell} \bm{W}^{(\ell)} \bm{Y}_{t - 1}
\end{align}
\textcolor{black}{has} $a_1+b_1+p+2$ parameters, 
which will usually be much smaller than the numbers of parameters of the full model if $p$ is large. 
As $\bm{Y}_{t}$ corresponds to observations of the same variable at different locations in the applications we \textcolor{black}{considered}, we further reduce the number of parameters when assuming a common intercept term, by setting $\bm{\delta} = \delta_0\bm{1}_p$. Then we call the model \textit{homogeneous}, while the model is called \textit{inhomogeneous} without this restriction. 
The linear Poisson network autoregression (PNAR) model of order 1 of \textcite{armillotta_count_2023} corresponds to the homogeneous case with \textcolor{black}{$b_1 = 1$} and $\alpha_{1\ell} = 0$ for all~$\ell$.

In view of the notation for STARMA processes, we refer to model~\eqref{method:linear_extended} as a linear PSTARMA$(q_{a_1, \ldots, a_q},\allowbreak r_{b_1, \ldots, b_r})$ model. Thus, model~\eqref{method:linear_without_covariates} is a linear PSTARMA$(1_{a_1}, 1_{b_1})$ model.  The restriction that all parameters $\bm{\delta}, \alpha_{i\ell}, \beta_{j\ell}\:$ $\forall i,j, \ell$ need to be non-negative guarantees that the intensity process $\bm{\lambda}_t$ is non-negative as well.

By assuming $\bm{W}^{(0)} = \bm{I}_p$, the parameters $\alpha_{i0}$ and $\beta_{i0}$ describe the effect of past expectations or observations on the expectation at the same location. The interpretation for higher spatial orders $\ell$ depends on the matrices $\bm{W}^{(\ell)}$. Sparse models, such as the PNAR processes of \textcite{armillotta_count_2023}, only use a single row-normalised adjacency matrix to include spatial dependencies. In this case, the $i$-th component of $\bm{W}^{(1)}\bm{Y}_{t - 1}$ corresponds to a weighted mean of the observations at time $t - 1$ from the locations that are adjacent to location $i$. The weights $w_{ij}^{(\ell)}$ indicate the influence of the location $j$ on the location $i$ for the spatial order $\ell$.

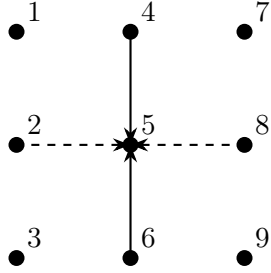
\begin{figure}[bt]
    \centering
    \begin{tikzpicture}[>=Stealth, thick, scale=1.5]
		% Gitter zeichnen und Punkte beschriften
		\foreach \i in {0,1,2}
		\foreach \j in {0,1,2}
		\fill (\i,\j) circle (2pt) node[above right]{\the\numexpr\i*3+3-\j};
		
		% Pfeile zeichnen
		\draw[<-] (1,1) -- (1,2);
		\draw[<-] (1,1) -- (1,0);
		\draw[<-, dashed] (1,1) -- (2,1);
		\draw[<-, dashed] (1,1) -- (0,1);
\end{tikzpicture}
    
    \caption{4-Nearest-Neighbors of central location (5) of a $3 \times 3$ grid}
    \label{fig:neighbor}
\end{figure}

Figure~\ref{fig:neighbor} illustrates a subset of a two-dimensional rectangular grid. Assuming a 4-nearest-neighbour structure, only locations 2, 4, 6 and 8 have an impact on location 5. If we set the weights $w_{5j}^{(1)}$ to 0.25 for $j = 2, 4, 6, 8$, there is a dependence of location 5 on the neighbouring locations which is independent of direction. This is generally referred to as isotropy, see \textcite{cressie_statistics_2011}. Directed spatial dependencies, i.e. anisotropy, can be realised by using more than one neighbourhood matrix. For example, define $\bm{W}^{(1)}$ such that it captures columnwise dependencies of direct neighbours from the north and south. For location 5 in Figure~\ref{fig:neighbor}, these are the two locations 4 and 6 with the associated weights $w_{5j}^{(1)} = 0.5$ for $j = 4, 6$ and $w_{5j}^{(1)} = 0$ otherwise. Similarly, the weights in the matrix $\bm{W}^{(2)}$ for rowwise dependencies on locations in west and east direction are given by $w_{5j}^{(2)} = 0.5$ for $j = 2, 8$. The parameter $\beta_{1i}$ then corresponds to the influence of the averaged observations of the locations from the north and south at time lag $i$, while the parameter $\beta_{2i}$ corresponds to the dependence in the west and east directions. In the case of $\beta_{1i} = \beta_{2i}$, the anisotropic model reduces to an isotropic one.

Linear PSTARMA(X) models provide several advantages: (a) they are simple to study and provide key insights for the properties of such processes, (b) their interpretation is straightforward, as all variables are included linearly in the (conditional) expectation and (c) they are useful for modeling integer-valued data because usually they fit well positively correlated data. However, due to their limitations, we introduce and study log-linear models below. The log-linear models provide a natural non-linear model in the context of dependent count data and such approaches are scarce in the STARMA literature.

\subsection{Log-Linear Poisson-STARMA Models}
\sloppy In this subsection, we define log-linear models which are based on the logarithmic link function. Applying the logarithm componentwise, we set $\bm{\nu}_t = \log(\bm{\lambda}_t)$. Define the log-linear PSTARMA$(q_{a_1, \ldots, a_q}, p_{b_1, \ldots, b_p})$ model by
\begin{align}
	\label{method:log_without_covariates}
	\bm{\nu}_t = \bm{\delta} + \sum_{i = 1}^{q} \sum_{\ell = 0}^{a_i} \alpha_{i\ell}\bm{W}^{(\ell)} \bm{\nu}_{t - i} + \sum_{j = 1}^{r} \sum_{\ell = 0}^{b_j} \beta_{j\ell} \bm{W}^{(\ell)}\log(\bm{Y}_{t - j} + \bm{1}_p).
\end{align}

This is analogous to model~\eqref{method:linear_extended}, but sign restrictions on the parameters are not necessary anymore. The log-linear model is defined on a logarithmic scale, meaning that the model parameters influence the conditional expectation not linearly, but multiplicatively. An advantage of the log-linear model is that it can capture both positive and negative dependencies, while the linear model is limited to positive correlations. In addition, covariates can easily be included without requiring positivity restrictions on their regression coefficients.

\subsection{Covariates}
In practice, in addition to autoregressive effects of the past there can be other spatial or temporal effects which can be explained by external factors. These factors can include demographic data, like population size or unemployment rates, as well as environmental variables such as temperature or air pollution. 
These effects can be included by adding covariates in models~\eqref{method:linear_extended} and~\eqref{method:log_without_covariates}, i.e.,
\begin{align}
	\label{method:linear_with_covariates}
	\bm{\lambda}_t = \bm{\delta} + \sum_{i = 1}^{q} \sum_{\ell = 0}^{a_i} \alpha_{i\ell}\bm{W}^{(\ell)} \bm{\lambda}_{t - i} + \sum_{j = 1}^{r} \sum_{\ell = 0}^{b_j} \beta_{j\ell} \bm{W}^{(\ell)}\bm{Y}_{t - j}  + \sum\limits_{k = 1}^m \sum\limits_{\ell = 0}^{s_k} \gamma_{k,\ell} \bm{W}^{(\ell)} \bm{X}_{k, t},
\end{align}
and similarly
{\footnotesize 
\begin{align}
	\label{method:log_with_covariates}
	\bm{\nu}_t = \bm{\delta} + \sum_{i = 1}^{q} \sum_{\ell = 0}^{a_i} \alpha_{i\ell}\bm{W}^{(\ell)} \bm{\nu}_{t - i} + \sum_{j = 1}^{r} \sum_{\ell = 0}^{b_j} \beta_{j\ell} \bm{W}^{(\ell)}\log(\bm{Y}_{t - j} + \bm{1}_p)  + \sum\limits_{k = 1}^m \sum\limits_{\ell = 0}^{s_k} \gamma_{k,\ell} \bm{W}^{(\ell)} \bm{X}_{k, t}.
\end{align}
}%
For each covariate process indexed by $k = 1, \ldots, m$, $s_k \in \mathbb{N}_0$ denotes the maximum spatial order for regression on the $k$-th covariate $\lbrace \bm{X}_{k, t} \rbrace$. Analogous to autoregressive moving-average processes with exogenous input (ARMAX), we refer to a model of the form \eqref{method:linear_with_covariates} resp. \eqref{method:log_with_covariates}, featuring regression on $m$ covariates, as a linear/log-linear PSTARMAX$(q_{a_1, \ldots, a_q}, r_{b_1, \ldots, b_r}, m_{s_1, \ldots, s_m})$ model. In model~\eqref{method:linear_with_covariates} it is required that $\gamma_{k, \ell} \geq 0 \, \forall k, \ell$, and that the covariates only take non-negative values, while for \eqref{method:linear_with_covariates} this is not necessary.

\subsection{Properties}
In this subsection we study some properties of linear and log-linear PSTARMA models.

\subsubsection{A Note on the Identifiability}
\label{sec:identifiability}
The identifiability of the model parameters is not always guaranteed and should be ensured by the user.

Similar conditions apply to the neighbourhood matrices as in the case of STARMA processes, see \textcite{pfeifer_three-stage_1980}. In particular, we assume that the main diagonal elements are equal to 0 for all neighbourhood matrices $\bm{W}^{(\ell)}$ with $\ell > 0$. Furthermore, the neighbourhood matrices must be linearly independent of each other \textcolor{black}{, i.e., $\bm{W}^{(\ell)} \neq \sum_{\tilde{\ell} \neq \ell}c_{\tilde{\ell}}\bm{W}^{(\tilde{\ell})}$ for all $\ell$ and all $c_{\tilde{\ell}} \in \mathbb{R}$.} This can be achieved by spatial lagging, for example, which implies for the weights $w_{i,j}^{(\ell)} = 0$ if $\exists \,\tilde{\ell} \neq \ell: w_{i,j}^{(\tilde{\ell})} \neq 0$.

Furthermore, the same conditions apply as for the models from \textcite{fokianos_multivariate_2020}. This means, for the identifiability of the parameters in models~\eqref{method:linear_extended} and~\eqref{method:log_without_covariates}, i.e., for PSTARMA processes that do not include covariates, it is necessary that there exist $j, \ell$ such that $\beta_{j\ell} \neq 0$. If $\beta_{j\ell} = 0 \: \forall j, \ell$ then the model is not identifiable. In other words, in general, it is not possible to fit models that regress solely on the feedback process $\lbrace \bm{\lambda}_{t} \rbrace$ or $\lbrace \bm{\nu}_{t} \rbrace$ since they are not identifiable.

In the case of PSTARMAX process~\eqref{method:linear_with_covariates} (respectively \eqref{method:log_with_covariates}) the covariates and model orders must fulfil basic conditions to ensure the identifiability of the model parameters. As in other regression models, the design matrix of the observed explanatory variables must have full rank. For each $t$, \textcolor{black}{the design matrix} is formed columnwise with the components of past observations, i.e., $\bm{W}^{(0)}\bm{Y}_{t - 1}, \ldots, \bm{W}^{(a_1)}\bm{Y}_{t - 1}, \ldots, \bm{W}^{(a_r)}\bm{Y}_{t - r}$ and of the covariate processes, i.e., $\bm{W}^{(0)}\bm{X}_{1, t}, \ldots, \bm{W}^{(s_1)}\bm{X}_{1, t}, \ldots, \bm{W}^{(s_m)}\bm{X}_{m, t}$. If a homogenous intercept is present, then $\bm{1}_p$ is added as a colum, otherwise $\bm{I}_p$ is added. For the full rank condition we discuss two special cases.

In addition to covariates that vary across space and time, such as precipitation, there are two other important types of covariate processes. One such type includes purely temporal information, such as weekday or data useful for modeling spatially constant temporal trends. For such covariates all locations share the same temporal information, i.e., $\forall\, t: \bm{X}_{k, t} = \tilde{X}_{k, t}\cdot \bm{1}_p$ for a univariate process $\lbrace \tilde{X}_{k, t} \rbrace$. The choice of spatial orders $s_k > 0$ for such a spatially constant covariate process $\bm{X}_{k, t}$ yields non-identifiable covariate parameters, we have $\sum_{\ell = 0}^{s_k} \gamma_{k, \ell}\bm{W}^{(\ell)}\bm{X}_{k, t} = \tilde{\gamma}_{k, 0} \bm{W}^{(0)}\bm{X}_{k, t}$ with $\tilde{\gamma}_{k, 0} = \sum_{\ell = 0}^{s_k} \gamma_{k, \ell}$. For such covariates we need to select spatial orders $s_k = 0$, \textcolor{black}{so} that $\gamma_{0, k}$ becomes the global effect of the covariate.

Moreover, there can also be covariates that do not change during the observation period. Such covariates may describe spatial attributes of the locations, aiming to capture effects like longitude or latitude, or nearly constant covariates such as the population size. Formally, we define such covariates as $\bm{X}_{k, t} = \tilde{\bm{X}}_k$ for a $\tilde{\bm{X}}_k \in \mathbb{R}^p$. Such time-constant covariates interact with the intercept $\bm{\delta}$. In the case of an inhomogeneous model with a different intercept term for each location, these parameters cannot be distinguished from those for a time-constant covariate. Effects of a variable that is constant in both time and space are already captured in the intercept.

\subsubsection{Relationship to STARMA processes}\label{subsec:starma}
Processes defined by \eqref{method:linear_extended} and \eqref{method:log_without_covariates}  are based on similar ideas as STARMA processes introduced by \textcite{pfeifer_three-stage_1980}. A linear PSTARMA$(q_{a_1, \ldots, a_q}, r_{b_1, \ldots, b_r})$ process according to equation~\eqref{method:linear_extended} in fact satisfies the model equation of a STARMA$(v_{u_1, \ldots, u_v}, q_{a_1, \ldots, a_q})$ process with $v = \max \lbrace q, r \rbrace$ and $u_i = \max \lbrace a_i, b_i \rbrace$ for $i = 1, \ldots, v$ with $a_i = -1$ for $i > q$ and $b_i = -1$ for $i > r$; see~\ref{definition:starma} in the supplementary material for a definition. Therefore, second order properties of linear PSTARMA processes can be derived directly from those of STARMA processes, which we provide in the supplementary material.

For the log-linear PSTARMA processes such relations do not hold exactly, so that obtaining exact properties is difficult. However, we have $\log(\bm{x}) \approx \log(\bm{x} + \bm{1}_p)$ if the components of $\bm{x}$ are large, so that a log-linear PSTARMA process approximately fulfills a STARMA model in terms of the process $\lbrace \log(\bm{Y}_t + \bm{1}_p) \rbrace$.

%%%%%%%%%%%%%%%%%%%%%%%%%%%%%%%%%%%%%%%%%%%%%%%
\section{Model Estimation and Inference}\label{sec:estimation}
The data-generating process proposed by \textcite{fokianos_multivariate_2020} and assumed in subsection \ref{sec:assumptions} allows correlated contemporary components with Poisson distributions as marginal distributions. This is achieved by incorporating a copula in the data-generating process which is based on a Poisson process. Because this is a simulation based method it is hard to determine the likelihood function and implement exact maximum likelihood estimation. We use quasi-maximum likelihood estimation and derive its asymptotic properties. The methodology still takes into account the temporal and spatial dependencies through the equations \eqref{method:linear_extended} and \eqref{method:linear_with_covariates} (respectively \eqref{method:log_without_covariates}  and \eqref{method:log_with_covariates}), although we use the working likelihood, which assumes the contemporaneous components to be conditionally independent given the past.

\subsection{Quasi Maximum Likelihood Estimation}
Consider $\lbrace \bm{Y}_t, t = 0, \ldots, T \rbrace$ as a realization of a PSTARMA or PSTARMAX process. We estimate the model parameters $\bm{\delta}, \alpha_{i\ell}, \beta_{j\ell}$ and $\gamma_{k\ell}$ of the process \eqref{method:linear_with_covariates} or \eqref{method:log_with_covariates}. Let $\bm{\theta} = (\bm{\delta}, \alpha_{i\ell}, (i = 1, \ldots, q, \ell = 0, \ldots, a_i), \beta_{j\ell}, (j = 1, \ldots, r, \ell = 0, \ldots, b_j) \gamma_{k, \ell}, (k = 1, \ldots, m, \ell = 0, \ldots, s_k))'$ be the vector of all unknown parameters. 
The resulting quasi-log likelihood is given by
\begin{align}
	\label{eq:qll}
	\ell (\bm{\theta}) = \sum_{t = 1}^{T} \sum_{i = 1}^{p} \left( Y_{i, t} \log (\lambda_{i, t} (\bm{\theta})) - \lambda_{i, t}(\bm{\theta}) \right).
\end{align}

For linear PSTARMAX processes the quasi score is given by
\begin{align}
	S_T(\bm{\theta}) = \sum_{t = 1}^{T} \frac{\partial \bm{\lambda}_t'}{\partial \bm{\theta}} \bm{D}_t^{-1}(\bm{\theta}) \left(\bm{Y}_t - \bm{\lambda}_t(\bm{\theta})\right),
\end{align}
where $\bm{D}_t(\bm{\theta}) = \text{diag}(\lambda_{i, t}(\bm{\theta}), i = 1, \ldots, p)$ and $\partial \bm{\lambda}_t' / \partial \bm{\theta}$ is the derivative of \eqref{method:linear_with_covariates} with respect to $\theta$, which is calculated componentwise by the following recursions: 
\begin{equation}
	\label{eq:recursion}
	\begin{aligned}
		\frac{\partial \bm{\lambda}_t}{\partial \bm{\delta}} &= \bm{I}_p + \sum_{j = 1}^q \bm{A}_j \frac{\partial \bm{\lambda}_{t - j}}{\partial \bm{\delta}}, \qquad
		&&\frac{\partial \bm{\lambda}_t}{\partial \alpha_{i\ell}} = \bm{W}^{(\ell)} \bm{\lambda}_{t - i} + \sum_{j = 1}^q \bm{A}_j \frac{\partial \bm{\lambda}_{t - j}}{\partial \alpha_{j\ell}},\\
		\frac{\partial \bm{\lambda}_t}{\partial \beta_{i\ell}} &= \bm{W}^{(\ell)} \bm{Y}_{t - i} + \sum_{j = 1}^q \bm{A}_j \frac{\partial \bm{\lambda}_{t - j}}{\partial \beta_{i\ell}}, \qquad
		&&\frac{\partial \bm{\lambda}_t}{\partial \gamma_{k,\ell}} = \bm{W}^{(\ell)} \bm{X}_{k, t} + \sum_{j = 1}^q \bm{A}_j \frac{\partial \bm{\lambda}_{t - j}}{\partial \gamma_{k, \ell}}.
	\end{aligned}
\end{equation}
In case of a homogenous model, the derivative $\partial \bm{\lambda}_t / \partial \bm{\delta}$ is replaced by $\partial \bm{\lambda}_t / \partial \delta_0 = \bm{1}_p + \sum_{j = 1}^q \bm{A}_j \partial \bm{\lambda}_{t - j} / \partial \delta_0$.

For a log-linear PSTARMAX process the quasi score is equal to
\begin{align}
	S_T(\bm{\theta}) = \sum_{t = 1}^{T} \frac{\partial \bm{\nu}_t'}{\partial \bm{\theta}} \left(\bm{Y}_t - \exp(\bm{\nu}_t(\bm{\theta}))\right).
\end{align}
The componentwise derivatives in $\partial \bm{\nu}_t' / \partial \bm{\theta} $ are equal to those in \eqref{eq:recursion}, but with $\bm{\lambda}$ replaced by $\bm{\nu}$ and $\log(\bm{Y}_{t - i} + \bm{1}_p)$ instead of $\bm{Y}_{t - i}$.

The quasi information matrix of a linear PSTARMAX process is given by the formula
\begin{align}
	\label{eq:information_linear}
	\bm{G}_T(\theta) = \sum_{t = 1}^{T} \frac{\partial \bm{\lambda}_t'}{\partial \bm{\theta}} \bm{D}_t^{-1}(\bm{\theta}) \bm{\Sigma}_t(\bm{\theta}) \bm{D}_t^{-1}(\bm{\theta}) \frac{\partial \bm{\lambda}_t}{\partial \bm{\theta}}.
\end{align}
The matrix $\bm{\Sigma}_t$ is the conditional variance of $\bm{Y}_t$ given $\mathcal{F}_{t - 1}^{\bm{Y}, \bm{\lambda}, \bm{X}}$, which is specified by the copula function.
The quasi information \textcolor{black}{matrix} of a log-linear PSTARMAX process is equal to
\begin{align}
	\label{eq:information_log}
	\bm{G}_T(\bm{\theta}) = \sum_{t = 1}^T \sum_{i = 1}^{p} \exp(\nu_{i, t}(\bm{\theta})) \frac{\partial\nu_{i, t}}{\partial \bm{\theta}} \frac{\partial\nu_{i, t}'}{\partial \bm{\theta}}.
\end{align}

Under mild assumptions, see \ref{subsec:asymptotic} in the Supplementary Material, the quasi-maximum likelihood estimator, denoted by $\bm{\hat{\theta}} = {\arg\max}_{\bm{\theta}}\ell(\bm{\theta})$, is strongly consistent and asymptotically normally distributed for $T \to \infty$, with 
\begin{align}
\label{eq:asymptotics}
    \sqrt{T} \left(\bm{\hat{\theta}} - \bm{\theta}_{\mathrm{true}} \right) \; \overset{D}{\to}\; \mathcal{N}(\bm{0}, \bm{H}^{-1} \bm{G} \bm{H}^{-1}).
\end{align}

The matrices $\bm{H}$ and $\bm{G}$ in~\eqref{eq:asymptotics} for the linear PSTARMA-process~\eqref{method:linear_extended} are given by 
\begin{align}
	\label{eq:var_linear}
	\bm{G}(\bm{\theta}) = \mathbb{E} \left[ \frac{\partial \bm{\lambda}_t'}{\partial \bm{\theta}} \bm{D}_t^{-1}(\bm{\theta}) \bm{\Sigma}_t(\bm{\theta}) \bm{D}_t^{-1}(\bm{\theta}) \frac{\partial \bm{\lambda}_t}{\partial \bm{\theta}} \right] \text{ and } \bm{H}(\bm{\theta}) = \mathbb{E} \left[ \frac{\partial \bm{\lambda}_t'}{\partial \bm{\theta}} \bm{D}_t^{-1}(\bm{\theta}) \frac{\partial \bm{\lambda}_t}{\partial \bm{\theta}} \right],
\end{align}
and for the log-linear PSTARMA-process~\eqref{method:log_without_covariates} by
\begin{align}
	\label{eq:var_log}
	\bm{G}(\bm{\theta}) = \mathbb{E} \left[ \frac{\partial \bm{\nu}_t'}{\partial \bm{\theta}}  \bm{\Sigma}_t(\bm{\theta})  \frac{\partial \bm{\nu}_t}{\partial \bm{\theta}} \right] \text{ and } \bm{H}(\bm{\theta}) = \mathbb{E} \left[ \frac{\partial \bm{\nu}_t'}{\partial \bm{\theta}} \bm{D}_t(\bm{\theta}) \frac{\partial \bm{\nu}_t}{\partial \bm{\theta}} \right].
\end{align}

For the calculation of standard errors, the matrices $\bm{H}$ and $\bm{G}$ can be estimated by their empirical counterparts by substituting $\bm{\hat{\theta}}$. The assumptions required for the asymptotic normality imply, that $\bm{\theta}_{\mathrm{true}}$ is in the interior of the parameter space, i.e. that the linear PSTARMA-processes are only allowed to have positive parameter values $\bm{\theta}_{\mathrm{true}} \succ \bm{0}$.
A detailed discussion of the assumptions required can be found in the Supplementary Material.

\subsection{Model choice}

In regression analysis, the Akaike information criterion (AIC) \citep{akaike_new_1974} or the Bayesian information criterion (BIC) \citep{schwarz_estimating_1978} are commonly used for the purpose of model selection. Nevertheless, their effectiveness relies on the presumption of a correctly specified likelihood function. Hence, their application can introduce bias in a model selection procedure when quasi-maximum likelihood estimation is employed as is done here. To address this problem, \textcite{pan_akaikes_2001} introduced the quasi information criterion (QIC), which adjusts for the additional uncertainty resulting from a potentially misspecified likelihood. It is computed via 
\begin{align}
\label{method:qic}
    \text{QIC} = -2\ell(\bm{\theta}) + 2 \cdot \text{trace}(\bm{G}\bm{H}^{-1}) 
\end{align}
by recalling \eqref{eq:qll}, \eqref{eq:var_linear} and \eqref{eq:var_log}.

\subsection{Testing of linear hypotheses}
The asymptotic distribution \eqref{eq:asymptotics} of the QMLE is used to implement an asymptotic Wald test for testing hypotheses of the type
\begin{align*}
	H_0: \bm{C}\bm{\theta} = \bm{c}_0 \qquad \text{ vs. }\qquad H_1: \bm{C}\bm{\theta} \neq \bm{c}_0,
\end{align*}
where  $\bm{C} \in \mathbb{R}^{u \times v}$ with $u \leq v$,  $\mathrm{rank}(\bm{C}) = f$ and $\bm{\theta} \in \mathbb{R}^v$. Under the null hypothesis it holds that, as $T \to \infty$,
\begin{align}
\label{eq:wald}
    T\cdot(\bm{C}\bm{\theta} - \bm{c}_0)' (\bm{C} \bm{\hat{\Sigma}}_T \bm{C}')^{-1}(\bm{C}\bm{\theta} - \bm{c}_0) \; \overset{D}{\to}\; \chi^2(f),
\end{align}
where $\chi^2(f)$ is the chi-square distribution with $f$ degrees of freedom. The variance $\bm{\hat{\Sigma}_T} = \bm{\hat{H}}^{-1} \bm{\hat{G}} \bm{\hat{H}}^{-1}$ is estimated by using the empirical counterparts of \eqref{eq:var_linear} and \eqref{eq:var_log}. 

\paragraph{Remark on linear PSTARMAX-Processes}
In the linear PSTARMAX process, it is assumed that all parameters are non-negative, i.e. $\bm{\theta} \succeq \bm{0}$. If the true parameter vector $\bm{\theta}_{\mathrm{true}}$ lies on the boundary of the admissible parameter space, i.e., if one or several components are equal to 0, the QMLE is not asymptotically normally distributed and the above asymptotic distribution of the test statistic does not hold. 
This problem also occurs e.g. for GARCH processes \citep[Sec. 8]{francq_garch_2019}. The asymptotic distribution of the test statistic is then a chi-bar squared distribution \citep{self_asymptotic_1987}, i.e., a weighted sum of chi-squared distributions. The hypothesis to be tested determines the exact distribution, which can have a simple form \citep{shapiro_unified_1988}. A more detailed discussion can be found in \textcite{mohlenberghs_likelihood_2007}, as well as procedures for determining the critical values of this distribution. \citep[Sec. 8.2]{francq_garch_2019}

In practice, it is often of interest whether a single parameter $\theta_i$ of $\bm{\theta}$ is significantly different from 0. This can be tested using the Wald test, with $\bm{C} = (0, \ldots, 0, 1, 0, \ldots, 0) \in \mathbb{R}^{1 \times v}$ the transpose of the $i$-th unit vector of $\mathbb{R}^{v}$ and $c_0 = 0$. In the case of the linear model, the  hypothesis reduces to   
\begin{align*}
	H_0: \theta_i = 0 \qquad \text{ vs. }\qquad H_1: \theta_i > 0.
\end{align*}
For this special case, the asymptotic distribution of the test statistic is $0.5\chi^2_0 + 0.5\chi^2_1$, where $\chi^2_0$ is a degenerate distribution at 0. The critical values of this distribution can be determined from the $\chi^2_1$ distribution, since for $\alpha < 0.5$, the $1 - \alpha$ quantile of the mixed distribution equals the $1 - 2\alpha$ quantile of the $\chi^2_1$ distribution.
Thus, it is sufficient to double the local significance level, i.e. set $\Tilde{\alpha} = 2\alpha$.

Such adjustments are not necessary for the log-linear model.
The hypothesis being tested there is 
\begin{align*}
	H_0: \theta_i = 0 \qquad \text{ vs. }\qquad H_1: \theta_i \neq 0,
\end{align*}
so the asymptotic distribution is a $\chi^2_1$, since in this case we do not have a boundary issue.

	\section{Simulation Study}\label{sec:simulation}
In this section, we describe an extensive simulation study by examining thoroughly the usefulness of the asymptotical  theory in finite samples. For the reason of simplicity we restrict ourselves to rectangular spatial grids with the same number of locations in both directions (squares).

We use two different types of neighbourhood matrices in the simulations, which have already been described as examples in Figure \ref{fig:neighbor}. The first implements a 4-nearest-neighbour structure, which results in a spatial isotropic model, i.e., direction-independent dependency on the neighbours. We denote the neighbourhood matrix in this case by $\bm{W}_{\text{4NN}} = (w_{ij}^{\text{4NN}})$. The second type of neighbourhood matrices allows directional dependencies on the neighbours. This is achieved by using two matrices $\bm{W}_{\text{NS}} = (w_{ij}^{\text{NS}})$ and $\bm{W}_{\text{WE}} = (w_{ij}^{\text{WE}})$, where $\bm{W}_{\text{NS}}$ captures the column-wise dependencies on the neighbours in the north and south direction and $\bm{W}_{\text{WE}}$ captures the row-wise dependencies neighbours in the west and east. Assume there are $p = k^2$ locations distributed on a $k \times k$ grid and the locations are column-wise numbered from 1 (top left) to $p$ (bottom right). The weights $w_{ij}^{\text{WE}}$ are then given as follows, where the weight 1 results for locations on the left or right boundary of the grid, i.e.,
\begin{align}
    w_{ij}^{\text{WE}} = \begin{cases}
        0, & \text{if } |i - j| \neq k, \\
        1, & \text{if } i \in \lbrace 1, \ldots, k, k(k - 1) + 1, \ldots, k^2 \rbrace \text{ and } |i - j| = k, \\
        0.5, & \text{otherwise}. \\
    \end{cases}
\end{align}
Similarly, the weights $w_{ij}^{\text{NS}}$ for the north-south case are given with weights of 1 for locations at the top and bottom of the grid
\begin{align}
    w_{ij}^{\text{NS}} = \begin{cases}
        0, & \text{if } |i - j| \neq \textcolor{black}{1}, \\
        1, & \text{if } i \in \lbrace \ell(k - 1) + 1, \: \ell k ;\: \ell = 1, \ldots, k \rbrace \text{ and } |i - j| = \textcolor{black}{1}, \\
        0.5, & \text{otherwise.} \\
    \end{cases}
\end{align}
In the 4-nearest-neighbour case the weights are given by
\begin{align}
    w_{ij}^{\text{4NN}} = \begin{cases}
        0, & \text{if } w_{ij}^{\text{WE}} = w_{ij}^{\text{NS}} = 0, \\
        0.5, & \text{if } w_{ij}^{\text{WE}} = 1 \wedge w_{ij}^{\text{NS}} = 1, \\
        1/3, & \text{if } (w_{ij}^{\text{WE}} = 1 \wedge w_{ij}^{\text{NS}} = 0.5) \vee (w_{ij}^{\text{WE}} = 0.5 \wedge w_{ij}^{\text{NS}} = 1), \\
        0.25, & \text{otherwise}, \\
    \end{cases}
\end{align}
where \enquote{$\wedge$} denotes the logical \enquote{and}, and \enquote{$\vee$} denotes the logical \enquote{or}.

The settings of the simulations performed here are summarized in Table~\ref{tab:settings}. We use 1000 repetitions in all simulations.
\begin{table}[tbp]
		\centering
		\resizebox{0.845\textwidth}{!}{%
		\begin{tabular}{|C{0.09\textwidth}|p{0.3\textwidth}|p{0.35\textwidth}|@{}p{0.45\textwidth}@{}|}
			\hline
			\multicolumn{1}{|c|}{\textbf{Study}} & \multicolumn{1}{|c|}{\textbf{Description}} & \multicolumn{1}{|c|}{\textbf{Model Orders \& Data}} &\multicolumn{1}{|c|}{\textbf{True Parameters}} \\ \hline
			\parbox[t]{2mm}{\multirow{2}{*}{\rotatebox[origin=c]{90}{Feedback Initialization}}}
			% \parbox[t]{2mm}{\rotatebox[origin=c]{90}{\multirow{3}{*}{Feedback Initialization}}}
			 & 
			 \multirow[t]{2}{*}{\parbox{\linewidth}{%
			 	\par\vspace{6\baselineskip}
			 	Impact of different initialisations of the feedback term on parameter estimation. 
			 	\begin{itemize}[left=0em]
			 		\item Linear: 
			 		\begin{itemize}[left=0.2em, topsep=0pt]
			 			\item $\bm{\lambda}_i = \bm{Y}_i$
			 			\item $\bm{\lambda}_i = \frac{1}{T}\sum_{t = 1}^T\bm{Y}_t$
			 			\item $\bm{\lambda}_i = \bm{0}_p$
			 		\end{itemize}
			 		\item Log-Linear: 
			 		\begin{itemize}[left=0em]
			 			\item $\bm{\nu}_i = \log(\bm{Y}_i + \bm{1}_p)$
			 			\item $\bm{\nu}_i = \log(\bar{\bm{Y}} + \bm{1}_p)$
			 			%\item $\bm{\nu}_i = \frac{1}{T}\sum_{t = 1}^T\log(\bm{Y}_t + \bm{1}_p)$
			 			\item $\bm{\nu}_i = \bm{0}_p$
			 		\end{itemize}
			 	\end{itemize}}}
			 & 
			 	 \multirow[t]{2}{*}{\parbox{\linewidth}{%
			 	 	\par\vspace{1.2\baselineskip}
			 		 \begin{itemize}[left=0.2em, topsep=0pt]
			 		 	\itemsep0.5pt
			 			\item $T = 250$
			 			\item Grid size: $9 \times 9$  $(p = 81)$
			 			\item Maximum time lags: \newline $(q, r) \in \lbrace (1, 1), (2, 2) \rbrace$
			 			\item Number of covariates: \newline $m \in \lbrace 0, 1 \rbrace$
			 			\item Neighborhood matrices: \newline$\bm{W}^{(0)} = \bm{I}_p$, \newline$\bm{W}^{(1)} = \bm{W}_{\text{4NN}}$
			 		\end{itemize}}}
			 &
			 \begin{tabular}{@{}C{0.2\linewidth}@{}|@{}C{0.2\linewidth}@{}C{0.2\linewidth}@{}C{0.2\linewidth}@{}C{0.2\linewidth}@{}}
			 	\multicolumn{5}{c}{linear}  \\
			 	\hline
			 	$\delta_0$ & 5 & 5 & 5 & 5 \\
			 	$\alpha_{0,1}$ & 0.2 & 0.2 & 0.2 & 0.2 \\
			 	$\alpha_{1,1}$ & 0.1 & 0.1 & 0.1 & 0.1 \\
			 	$\alpha_{0,2}$ & - & 0.05 & - & 0.05 \\
			 	$\alpha_{1,2}$ & - & 0.05 & - & 0.05 \\
			 	$\beta_{0,1}$ & 0.2 & 0.2 & 0.2 & 0.2 \\
			 	$\beta_{1,1}$ & 0.1 & 0.1 & 0.1 & 0.1 \\
			 	$\beta_{0,2}$ & - & 0.05 & - & 0.05 \\
			 	$\beta_{1,2}$ & - & 0.05 & - & 0.05 \\
			 	$\gamma_{0,1}$ & - & - & 2 & 2 \\
			 	%\hline
			 \end{tabular}  \\
			 & & &
			 \begin{tabular}{@{}C{0.2\linewidth}@{}|@{}C{0.2\linewidth}@{}C{0.2\linewidth}@{}C{0.2\linewidth}@{}C{0.2\linewidth}@{}}
			 	\hline
			 	\multicolumn{5}{c}{log-linear} \\
			 	\hline
			 	$\delta_0$ & 0.6 & 0.6 & 0.6 & 0.6 \\
			 	$\alpha_{0,1}$ & 0.2 & 0.2 & 0.2 & 0.2 \\
			 	$\alpha_{1,1}$ & 0.1 & 0.1 & 0.1 & 0.1 \\
			 	$\alpha_{0,2}$ & - & 0.05 & - & 0.05 \\
			 	$\alpha_{1,2}$ & - & 0.05 & - & 0.05 \\
			 	$\beta_{0,1}$ & 0.2 & 0.2 & 0.2 & 0.2 \\
			 	$\beta_{1,1}$ & 0.1 & 0.1 & 0.1 & 0.1 \\
			 	$\beta_{0,2}$ & - & 0.05 & - & 0.05 \\
			 	$\beta_{1,2}$ & - & 0.05 & - & 0.05 \\
			 	$\gamma_{0,1}$ & - & - & 0.9 & 0.9 \\
			 	%\hline
			 \end{tabular}  \\
			 \hline
		% \parbox[c]{5mm}{\rotatebox[origin=c]{90}{\multirow{3}{*}{\shortstack[l]{Misspecification of \\ Spatial Dependencies}}}}
		 \parbox{7mm}{\multirow{2}{*}{\rotatebox[origin=c]{90}{\shortstack[c]{Misspecification of \\ Spatial Dependencies}}}}
		 & 
		 \multirow[t]{2}{*}{\parbox{\linewidth}{%
		 \par\vspace{-3\baselineskip}
		Different types of spatial dependence and detection of missspecification	 
	 }}
		 &
		 \parbox[t]{\linewidth}{%
		 	\par\vspace{-3.5\baselineskip}
		 \begin{itemize}[left=0.2em, topsep=0pt]
		 	\itemsep0.5pt
		 	\item $T \in \lbrace 50, 100, 250, 500 \rbrace$
		 	\item Grid sizes: $9\times9$ $(p = 81)$
		 	\item Maximum time lags: \newline$(q, r) = (1, 1)$
		 	\item Number of covariates: \newline$m  = 0$
		 	\item Neighborhood matrices: \newline $\bm{W}^{(0)} = \bm{I}_p$, \newline$\bm{W}^{(1)} = \bm{W}_{\text{NS}}$, \newline$\bm{W}^{(2)} = \bm{W}_{\text{WE}}$
		 \end{itemize}}
		 & 
		 \begin{tabular}{@{}C{0.2\linewidth}@{}|@{}C{0.4\linewidth}@{}C{0.4\linewidth}@{}}
		 	& linear & log-linear \\
		 	\hline
		 	$\delta_0$ & 5 & 0.6 \\
		 	$\alpha_{0,1}$ & 0.2 & 0.2 \\
		 	$\alpha_{1,1}$ & 0.1 & 0.1 \\
		 	$\alpha_{2,1}$ & 0.05 & 0.05\\
		 	$\beta_{0,1}$ & 0.2 & 0.2 \\
		 	$\beta_{1,1}$ & 0.1 & 0.1 \\
		 	$\beta_{2,1}$ & 0.05 & 0.05 \\ \hline
		 \end{tabular}
		  \\ \hline
		\end{tabular}}
	\caption{Summary of simulations performed. Covariate values for each location are generated independently of the other locations as a realisation of a (stationary) ARMA(1, 1) process with $\mathcal{U}([0, 1])$-distributed innovations. This is done in advance of all further simulations so that the covariate values remain identical in all repetitions. Contemporary correlations between the counts are generated by the data-generating process proposed by the work of \textcite{fokianos_multivariate_2020} using the Clayton copula with parameter 2.}
	\label{tab:settings}
\end{table}

\subsection{Details on the implementation}
\label{sim:implementation}
The simulations and the evaluation of the results were carried out with \texttt{R 4.3.2}. The functions for simulating and estimating the PSTARMA(X) processes have been implemented in \texttt{C++} and integrated in \texttt{R} using the package \texttt{Rcpp} \citep{eddelbuettel_rcpp_2011}. For maximization of the quasi-log likelihood~\eqref{eq:qll}, the \texttt{SLSQP} algorithm \citep{kraft_algorithm_1994} from the \texttt{nloptr} package \citep{ypma_nloptr_2022} is used. Parameters are estimated under the constraint of a stationary solution. The functions for parameter estimation are collected in an R package, whose preliminary version is available on GitHub, with the code to reproduce the simulation results, see \url{https://github.com/stmaletz/PoissonSTARMA}. 
For data generation, a burn-in phase of 100 observation times is used, which are then removed to allow the process to stabilise. Covariates are ignored in this case.

{
Table~\ref{tab:fitting_time} shows average computation times in seconds for linear and log-linear first-order PSTARMAX processes in case of different grid sizes and number of observation times, aggregated over multiple simulation settings. The time was measured with the R package \texttt{tictoc} \citep{tictoc_2024}. The computer cluster we used is equipped with two AMD Epyc 7453 processors. A maximum of 48 GB of RAM was allocated for each setup. 

The runtime of the models is linear in the number of temporal observations, while the number of locations has a quadratic impact. Overall the log-linear model has slightly higher computing times when compared to the linear model. The inclusion of the feedback term leads to an approximate double increase of the computing time, depending on the number of observations. Further computing times for models with larger $T$ and smaller grid sizes are given in the Supplementary Material.

}

\begin{table}[tbp]
		\centering
  
		\begin{tabular}{|c|ccccccc|}
			\hline
\multirow{2}{*}{Feedback order} & \multirow{2}{*}{Model}& \multirow{2}{*}{Gridsize} & \multicolumn{5}{c|}{$T$} \\ \cline{4-8}
%	           Parameter             &            Model             & Gridsize &  
& & & 5   &  10  &  20  &  50  &  100  \\ \hline
			     %	           Parameter       &            Model            &         Gridsize          &       &       &       &   5   &  10   & 20 & 50 & 100 \\ \hline
			\multirow{10}{*}{$q = 1$} & \multirow{5}{*}{log-linear} &      $10 \times 10$       & 0.006 & 0.014 & 0.027 & 0.065 & 0.126 \\
			                                  &                             &      $20 \times 20$       & 0.030 & 0.057 & 0.110 & 0.278 & 0.548 \\
			                                  &                             &      $30 \times 30$       & 0.087 & 0.153 & 0.271 & 0.623 & 1.173 \\
			                                  &                             &      $40 \times 40$       & 0.234 & 0.357 & 0.575 & 1.226 & 2.465 \\
			                                  &                             &      $50 \times 50$       & 0.561 & 0.810 & 1.134 & 2.317 & 4.084 \\
 \cline{2-8}
			                                  &   \multirow{5}{*}{linear}   &      $10 \times 10$       & 0.004 & 0.007 & 0.015 & 0.039 & 0.083 \\
			                                  &                             &      $20 \times 20$       & 0.018 & 0.033 & 0.066 & 0.177 & 0.375 \\
			                                  &                             &      $30 \times 30$       & 0.062 & 0.102 & 0.180 & 0.436 & 0.895 \\
			                                  &                             &      $40 \times 40$       & 0.181 & 0.259 & 0.406 & 0.930 & 1.800 \\
			                                  &                             &      $50 \times 50$       & 0.502 & 0.650 & 0.860 & 1.802 & 3.143 \\ \hline
			\multirow{10}{*}{$q = 0$}  & \multirow{5}{*}{log-linear} &      $10 \times 10$       & 0.003 & 0.005 & 0.010 & 0.025 & 0.051 \\
			                                  &                             &      $20 \times 20$       & 0.016 & 0.026 & 0.046 & 0.112 & 0.231 \\
			                                  &                             &      $30 \times 30$       & 0.058 & 0.082 & 0.129 & 0.286 & 0.563 \\
			                                  &                             &      $40 \times 40$       & 0.164 & 0.210 & 0.304 & 0.596 & 1.106 \\
			                                  &                             &      $50 \times 50$       & 0.492 & 0.550 & 0.680 & 1.129 & 1.863 \\
 \cline{2-8}
			                                  &   \multirow{5}{*}{linear}   &      $10 \times 10$       & 0.002 & 0.004 & 0.008 & 0.019 & 0.040 \\
			                                  &                             &      $20 \times 20$       & 0.013 & 0.020 & 0.035 & 0.086 & 0.178 \\
			                                  &                             &      $30 \times 30$       & 0.052 & 0.067 & 0.105 & 0.228 & 0.446 \\
			                                  &                             &      $40 \times 40$       & 0.162 & 0.189 & 0.254 & 0.478 & 0.876 \\
			                                  &                             &      $50 \times 50$       & 0.406 & 0.438 & 0.579 & 1.009 & 1.657 \\ \hline
		\end{tabular}
  \caption{Mean computation time in seconds to estimate a PSTARMAX process for different grid sizes and observation times.}
  \label{tab:fitting_time}
\end{table}

\subsection{Initialization of the feedback process}
\label{sim:init}
Processes with feedback term, i.e., with regression on $\lbrace \bm{\lambda}_t \rbrace$ or $\lbrace \bm{\nu}_t \rbrace$, require specification of values for initializing $\bm{\lambda}_0, \ldots, \bm{\lambda}_{q - 1}$. The choice of initial values, in general, has an impact to final parameter estimates. Obvious choices for starting values are $\bm{\lambda}_{i} = \bm{\bar{y}}$ or $\bm{\lambda}_{i} = \bm{y}_i$, and $\bm{\nu_i} = \log(\bm{\bar{y}} + \bm{1}_p)$  or $\bm{\nu_i} = \log(\bm{y}_i + \bm{1}_p)$ for $i = 0, \ldots, q - 1$. An alternative strategy is to set $\bm{\lambda}_{i} = \bm{\nu_i} = \bm{0}_p, \:\forall\, i = 0, \ldots, q - 1$. We compare these three initialization options and also include their true values for comparison. We evaluate them using the mean square error (MSE) for the parameter estimation, $\text{MSE} = (\bm{\hat{\theta}} - \bm{\theta}_{\mathrm{true}})'(\bm{\hat{\theta}} - \bm{\theta}_{\mathrm{true}}) / K$. Here, $K = 1 + \sum_{i = 1}^q (a_i + 1) + \sum_{i = 1}^r (b_i + 1)$ is the number of unknown model parameters. 

\begin{figure}[tb]
    \centering
    \begin{subfigure}{0.49\textwidth}
         \includegraphics[keepaspectratio, width = 1\textwidth]{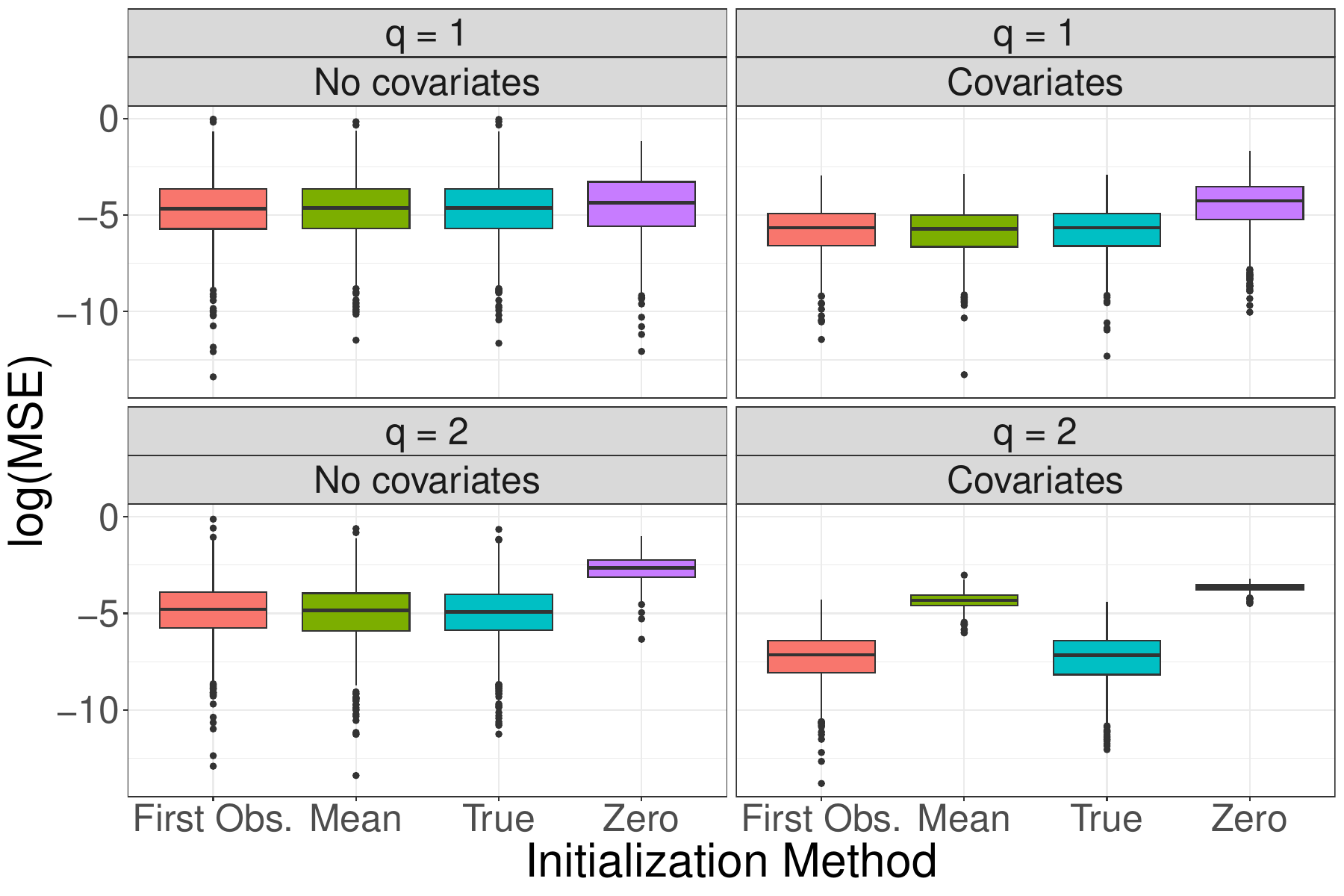}
        \caption{log-linear}
        \label{fig:init_log}
    \end{subfigure}
    \hfill
    \begin{subfigure}{0.49\textwidth}
        \includegraphics[keepaspectratio, width = 1\textwidth]{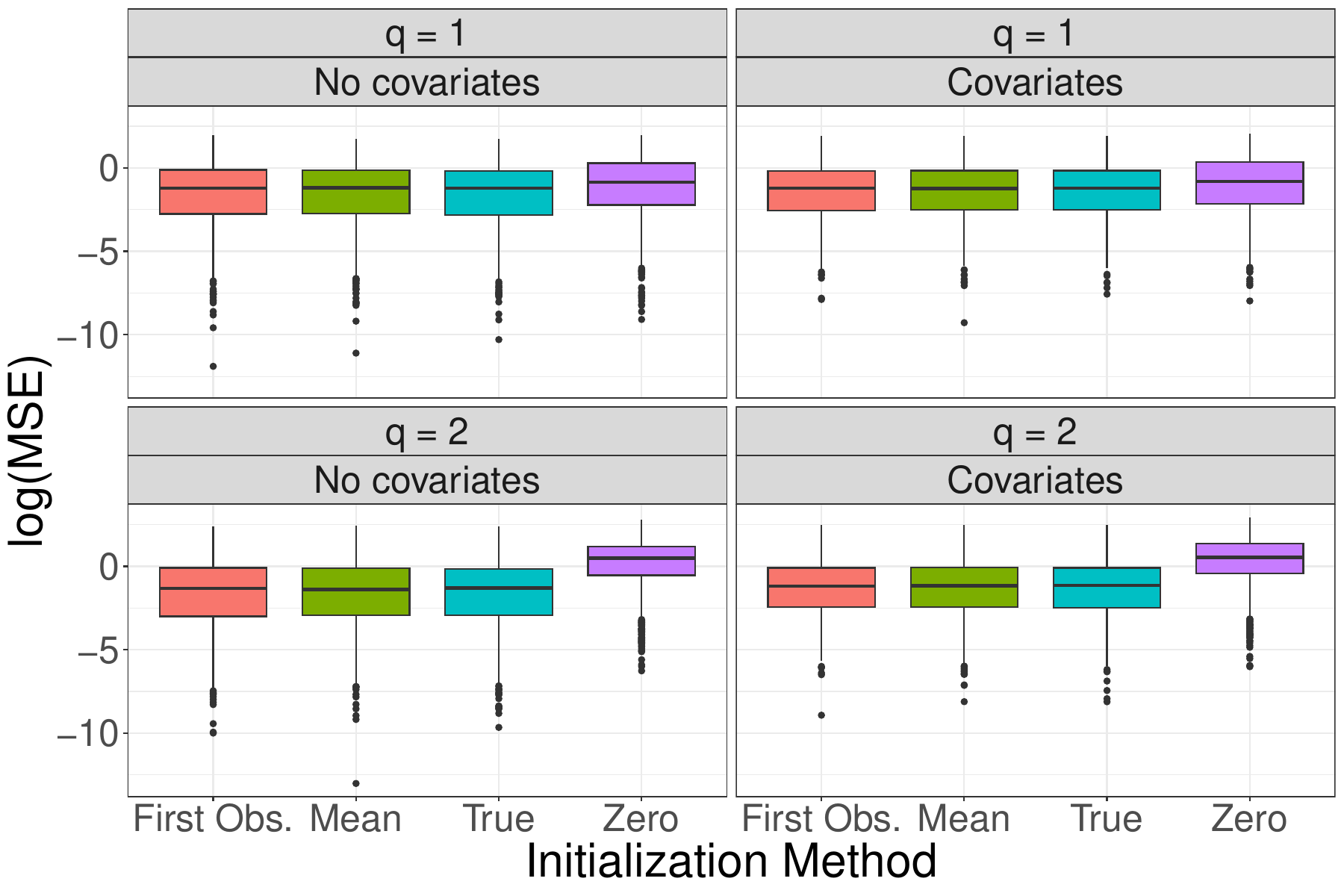}
        \caption{linear}
        \label{fig:init_linear}
    \end{subfigure}
    \caption{MSE (in logarithmic scale) of the QMLE in linear and log-linear model settings for different initialization methods for the feedback process.}
    \label{fig:init}
\end{figure}

Figure~\ref{fig:init_log} shows the logarithmic MSE for log-linear PSTARMA and PSTARMAX processes with temporal model orders 1 and 2 and $T = 250$ observation times for different initializations of the feedback process. In the case of first-order models without covariates, the MSE is the same regardless of the chosen initialization method. For models with temporal order 2, the initialization via $\bm{\nu}_0 = \bm{\nu}_1 = \bm{0}_p$ yields higher values of the MSE than the other methods. Noteworthy differences in the MSEs for the initialization via the global mean and initialization using the first observations occur for models with temporal order 2 with covariates, with higher MSE for the initialization obtained when using the mean. Initialization via the first observations yields similar MSEs as when using the true values of the feedback process, which are unknown in practice. We conclude that setting the  initial values of the hidden process to the first observations is suggested by this simulation study.

These differences are less apparent for linear PSTARMA and PSTARMAX processes; see Figure~\ref{fig:init_linear}. Note again that zero initialization yields higher MSE for the parameter estimation than the other initializations.

Overall, based on the results of these simulations, we use initialization via the first observations in the rest of this work.

%%%%%%%%%%%%%%%%%%%%%%%%%
\subsection{Misspecification of spatial dependence}
\label{sim:miss_dependence}
The choice of the neighborhood matrices determines the spatial dependencies captured by the  models. This study explores the consequences of a slight misspecification, focusing on a PSTARMA process with an anisotropic neighborhood structure (see Figure~\ref{fig:neighbor}) as the true data-generating process. We choose the true parameters for both observations and the feedback term in the east-west direction and the north-south direction as 0.1 and 0.05, respectively, and fit an isotropic model with a single neighborhood matrix assuming the same effect in both directions. Furthermore, we assess the efficacy of identifying the anisotropy and spatial dependence in general.

We first examine whether spatial dependency is detected at all when fitting the isotropic model to data generated by the anisotropic model. For this we employ Wald tests to assess whether the estimated spatial dependency parameters, $\alpha_{1,1}$ and $\beta_{1,1}$, are significantly different from zero in the isotropic model. The empirical powers of the Wald test at the 5\% significance level shown in Table~\ref{tab:isotropy_test} illustrate that spatial dependence is more easily detected on the observations than on the feedback process. For instance, in the log-linear process with $T = 250$ observation times, spatio-temporal dependency is identified in approximately 88\% of cases in the observations, but only around 19\% in the feedback process.

\begin{table}[tb]
\centering
\begin{tabular}{cccccc}
  \hline
  \multirow{2}{*}{Model} & \multirow{2}{*}{Coefficient}& \multicolumn{4}{c}{$T$} \\
 &  & 50 & 100 & 250 & 500 \\ 
  \hline
\multirow{2}{*}{log-linear}&$\beta_{1, 1}$  & 0.264 & 0.488 & 0.876 & 0.990 \\
&$\alpha_{1, 1}$  & 0.173 & 0.153 & 0.186 & 0.304 \\ \hline
 \multirow{2}{*}{linear} &$\beta_{1, 1}$  & 0.345 & 0.627 & 0.944 & 0.999 \\ 
  &$\alpha_{1, 1}$  & 0.147 & 0.180 & 0.311 & 0.430 \\ 
   \hline
\end{tabular}
\caption{Relative frequencies of significant coefficients ($\beta_{1, 1}$ and $\alpha_{1, 1}$) ($p$-value $< 0.05$) in an isotropic model fitted to data from an anisotropic model.}
\label{tab:isotropy_test}
\end{table}

 Table~\ref{tab:anisotropy_qic} reports comparisons in terms of QIC, see \eqref{method:qic}, for selecting between the isotropic and anisotropic models when the data are in fact generated employing anisotropy. For $T = 50$ the isotropic model is favored, approximately 44\% of times in the linear model case. This number decreases to about 25\% for $T = 100$. In total for all models, with a higher number of observation points, the true model (anisotropic) is favored more frequently. This observation is consistent with the results obtained by applying asymptotic Wald tests for anisotropy conducted when fitting the anisotropic model. These tests evaluate whether the disparities $\alpha_{2,1} - \alpha_{1,1}$ and $\beta_{2,1} - \beta_{1,1}$ between the estimated spatial parameters in the anisotropic model deviate significantly from zero. The empirical significance levels ($p < 0.05$) of this test are presented in Table~\ref{tab:anisotropy_test}. For $T = 500$ observation periods, the parameters associated with the observation process are correctly identified as distinct in the vast majority of cases. However, for the feedback process, this occurs in less than 20\% of the cases in both the linear and the log-linear model.

\begin{table}[tb]
    \centering
    \begin{tabular}{lrrrr}
  \hline
  \multirow{2}{*}{Model} & \multicolumn{4}{c}{$T$} \\
  & 50 & 100 & 250 & 500 \\ 
  \hline
log-linear & 0.361 & 0.257 & 0.100 & 0.027 \\ 
linear & 0.444 & 0.247 & 0.075 & 0.017 \\ 
   \hline
\end{tabular}
    \caption{Relative frequencies of cases in which the QIC favors isotropic models over anisotropic models, which are the true ones.}
    \label{tab:anisotropy_qic}
\end{table}

\begin{table}[tb]
    \centering
    \begin{tabular}{cccccc}
  \hline
  \multirow{2}{*}{Model} & \multirow{2}{*}{Anisotropy-Test}& \multicolumn{4}{c}{$T$} \\
 &  & 50 & 100 & 250 & 500 \\ 
  \hline
\multirow{2}{*}{log-linear}&$\beta$  & 0.295 & 0.465 & 0.815 & 0.980 \\
&$\alpha$  & 0.122 & 0.124 & 0.126 & 0.173 \\ \hline
 \multirow{2}{*}{linear} &$\beta$  & 0.216 & 0.510 & 0.876 & 0.992 \\ 
  &$\alpha$  & 0.040 & 0.051 & 0.092 & 0.146 \\
   \hline
\end{tabular}
    \caption{Empirical power of Wald tests at 5\% significance level for testing differences between the coefficients of an anisotropic model using separate neighborhood matrices for different directions;  $\alpha$ refers to the test for the coefficients of the feedback process, while $\beta$ refers to the observation process.}
\label{tab:anisotropy_test}
\end{table}

This study indicates that an isotropic model can capture spatial dependencies to some extent, even in the presence of a moderate degree of anisotropy. When only a moderately large number of observation times is available, an isotropic model might be preferable over an anisotropic model in terms of QIC, as the latter requires more observations to get precise estimates of the dependencies. However, if it is known that directional spatial dependencies exist and there is enough data available, an anisotropic model should be fitted.

\subsection{Further Remarks}

We provide further simulation results e.g. on the asymptotics of the QMLE in the supplementary material. We also consider high-dimensional grids with few observation times. An important result of this study is that linear PSTARMAX processes provide faster convergence of the test statistics than the log-linear models. For models with feedback terms, the asymptotics does not provide satisfactory approximations if there are only a few observation times, $T \leq 50$. As opposed to this, for a linear PSTARMAX process without a feedback term, i.e., $q = 0$, the simulation revealed a good approximation of the  test statistics for autoregressive effects already for $T = 5$.

Further simulations in the supplementary materials are based on \textcite{maletz_spatio-temporal_2021}.
These investigate the impact of contemporaneous dependencies on the QMLE, and further types of misspecification.

The QMLE is based on the assumption of contemporaneous conditional independence between the different locations. Hence the impact of different contemporary dependence structures generated by different copulas with several parameter values on the QMLE has been studied. A plausible finding is that increased contemporary dependencies between the component processes increase the MSE of the QMLE, although the impact on the bias is small. Furthermore, log-linear PSTARMA processes were found to capture the dependencies inherent in linear PSTARMA processes to some degree, while a linear PSTARMA process cannot describe negative dependencies. Therefore, the consideration of both model types is recommended to find a model which fits the data well. The performance of homogeneous models with a global intercept on data from a model with an inhomogeneous intercept has also been investigated. The homogeneous models provided competitive results in case of small sample sizes. A suggestion thus is to fit a homogeneous model with some covariates describing differences between the locations.

	\section{Data analysis}
\label{sec:real}

We apply our models to two real-world datasets to demonstrate the usefulness of the proposed methodology. Our first example is the weekly number of Rota virus infections reported to the Robert Koch Institute (RKI) in Germany \citep{stojanovic_bayesian_2019}, and the second example is the monthly counts of burglaries in Chicago \citep{clark_class_2021}.

\subsection{Rota Virus Data}
The rotavirus is one of the most frequent causes of diarrhoeal diseases among children, according to the \textcite{rki_rota_2010}. The viruses are easily transmitted through smear infections, contaminated food or water. After an incubation period of around 1 to 3 days, the symptoms typically last for 2 to 6 days. Infected people are usually infectious for up to 8 days, but individual cases may remain infectious much longer. Vaccination for infants under the age of 6 months is recommended in Germany since July 2013.

\subsubsection{Data Description}
The dataset we use contains the weekly number of Rota virus infections reported by the RKI in $p = 412$ urban and rural districts in Germany. The observation period ranges from 2001 to 2018 ($T = 903$ weeks). The data were obtained from the RKI website (\url{https://survstat.rki.de/}) and is available in pre-processed format at \url{https://github.com/ostojanovic/BSTIM}. Demographic information is available in the form of population size in the regions, which we assume to be constant for the observed time period. A descriptive analysis of the data is provided by \textcite{stojanovic_bayesian_2019}.

The administrative subdivision of Germany into the urban and rural districts gives rise to an irregular grid on a natural way. We consider two regions to be neighbours if they share a common border. The number of neighbours of a region varies from $1$ to $12$, with an average of $5.248$. In total, there are 27 regions that only have one neighbour in the German territory. These are regions completely included in another region or located on the border of Germany.

\begin{figure}[t]
    \centering
    \begin{subfigure}[t]{0.4\textwidth}
         \includegraphics[keepaspectratio, width = 1\textwidth]{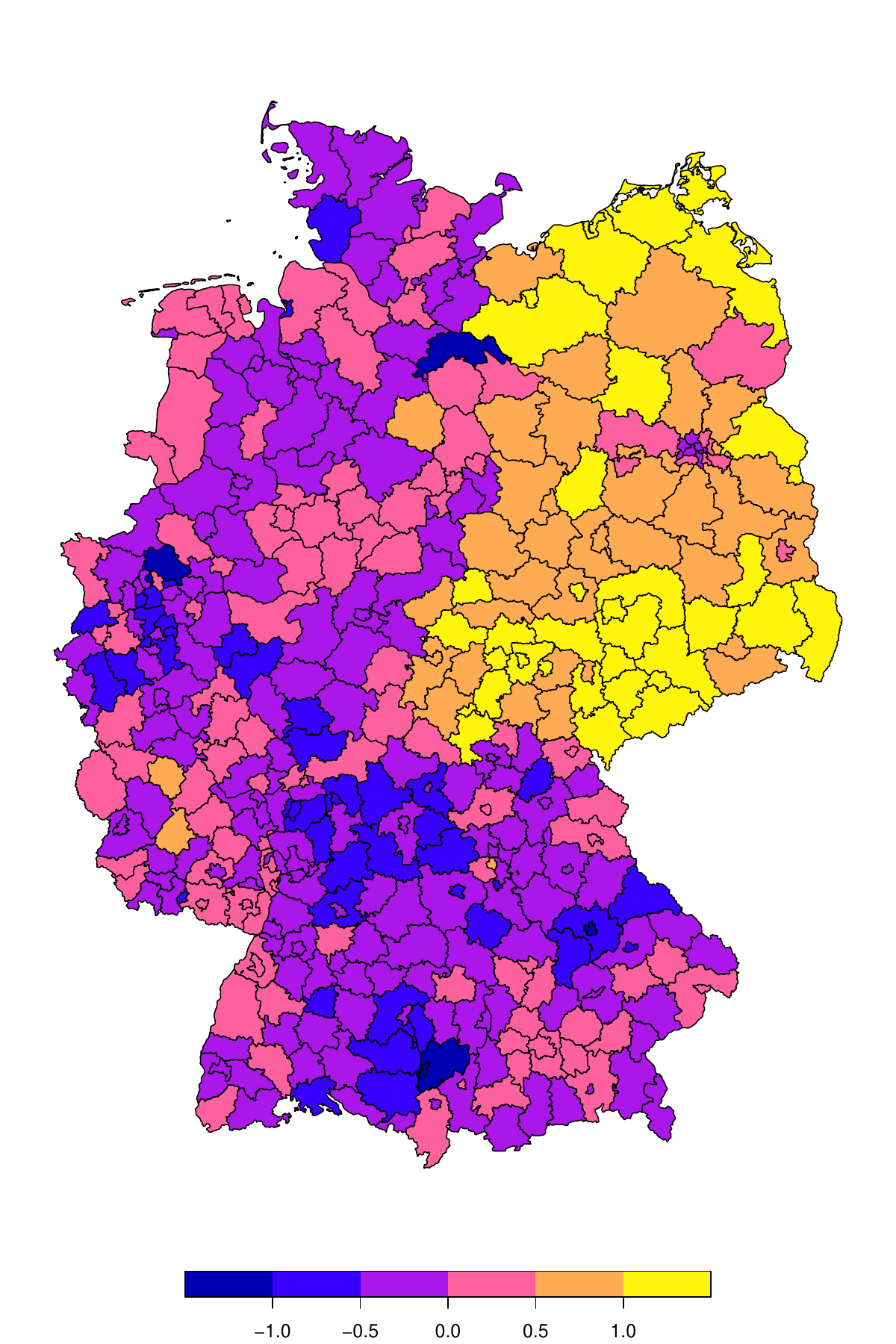}
        \caption{Logarithmic mean incidence of Rota cases in 412 german districts.}
        \label{fig:incidence_rota}
    \end{subfigure}
    \hfill
    \begin{subfigure}[t]{0.5\textwidth}
        \includegraphics[keepaspectratio, width = 1\textwidth]{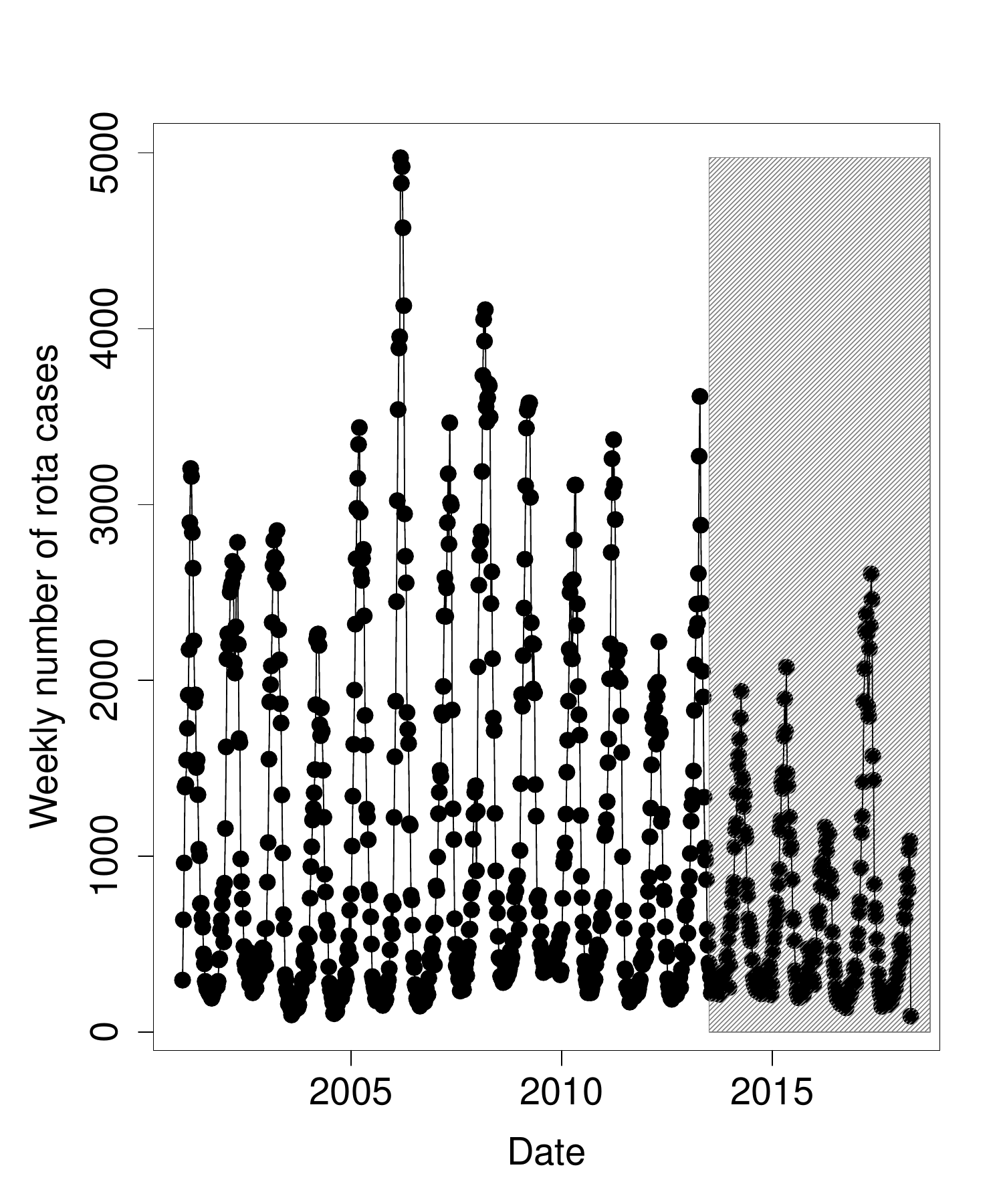}
        \caption{Weekly total number of rota virus cases in Germany from 2001 to 2018.}
        \label{fig:rota_cases}
    \end{subfigure}
    \caption{Spatial and temporal variation of rota virus cases in Germany. Information about the shape of counties within Germany is publicly provided by the German federal agency for cartography and geodesy (Bundesamt für Kartographie und Geodäsie) (GeoBasis-DE / BKG 2018) under the dl-de/by-2-0 license.}
    \label{fig:rota}
\end{figure}

The reported infection numbers within the individual regions varies from $0$ to $201$. On average, $2.4137$ cases are registered per region and week. Figure~\ref{fig:incidence_rota} shows the spatial distribution of the local incidences on a logarithmic scale, i.e., the number of cases per 100,000 inhabitants, averaged over all time points. It reveals high values for regions belonging to the former German Democratic Republic (GDR) in the northeast of Germany. These regions were administered in a fundamentally different way politically and structurally than the other regions for a long period of time.

Figure~\ref{fig:rota_cases} shows the aggregated weekly infection cases for Germany in total. It illustrates seasonal effects with peaks during spring and dips towards the end of summer. The time period highlighted in grey shows the time since RKI recommended vaccination. We observe that there exists a decrease in the number of infected people after July 2013, but we still observe periodicity in the data.

Our analysis of the data aims at several aspects. First, we aim to identify time-lagged spatial dependencies in the data. This is likely because it is an easily transmittable disease and people may still be infectious after the symptoms have subsided. New infections occur with a time delay, for example during doctor's appointments, in schools/nurseries or at work. Infections may also be spread to neighouring regions. Our analysis further aims to investigate and evaluate the effects of the vaccination recommendation, and to quantify the seasonal effects and as well as the influence of population size on a local level.

\subsubsection{Model Fitting and Interpretation}
We confine ourselves to homogeneous log-linear PSTARMAX processes for the analysis. We use the logarithmic population size of the regions as a covariate, a binary spatially constant variable that indicates the period after the start of the vaccination recommendation (July 2013), two spatially constant sinusoidal covariates $x_{t} = \sin(2\pi t / 52 )$ and $\tilde{x}_t = \cos(2\pi t / 52 )$ to capture the yearly seasonal effects and a time constant variable indicating whether the region is located in the former territory of the GDR (1) or was part of West Germany (0). For the regions in the capital Berlin, as in \textcite{stojanovic_bayesian_2019}, we use the value 0.5, as the city was divided until Germany was reunited in 1990.

We use two neighbourhood matrices to capture spatial dependencies. Their weights are set as in equation~\eqref{eq:weights}, i.e., all neighbours are weighted equally. As previously mentioned, we define a region's first-order neighbours as all spatial contiguous regions. The second-order neighbourhood includes all regions that can be reached by crossing one other region. If $\mathcal{N}_i^{(1)}$ is the set of all regions spatially contiguous to region $i$, then $\mathcal{N}_i^{(2)}$ is given by
\begin{align}
    \mathcal{N}_i^{(2)} = \bigcup_{j \in \mathcal{N}_i^{(1)}} \mathcal{N}_j^{(1)} \setminus \left(\mathcal{N}_i^{(1)} \cup \lbrace i \rbrace \right).
\end{align}

Table~\ref{tab:rota_qic_train} shows the QIC, the computation times for fitting in seconds, and the mean square prediction error (MSPE) for the entire data set for different temporal model orders. The mean square prediction error is calculated via $\text{MSPE} = (p(T - r)^{-1} \sum_{t = r + 1}^T \sum_{i = 1}^p (y_{i, t} - \lambda_{i, t}(\hat{\bm{\theta}}))^2$, where $\bm{\hat{\theta}}$ is the QMLE of $\bm{\theta}$, see Section \ref{sec:estimation}. For models including a feedback process ($q = 1$), we use the spatial order 1, i.e., $a_1 = 1$. We set $b_i = 2, i = 1, \ldots, r$ for the observation process. All covariates are included with spatial order 0.

A reduction of the QIC is evident for models with and without a feedback term by increasing the maximum order $r$ for the observation process. Adding the feedback process ($q = 1$) improves the QIC for $r < 4$. However, the smallest MSPE is achieved on the data set for $q = 1$ and $r = 8$, indicating overfitting. This is reflected in the parameter estimates. For $r \geq 4$, the two parameters $\alpha_{0,1}$ and $\alpha_{1, 1}$ are estimated to be slightly different from 0 and are not significant. As a result, the remaining parameter estimates of these models differ just slightly from those without a feedback term. 

\begin{table}[htb]
    \centering
    
\begin{tabular}{|c|ccc|ccc|}
	\hline
	\multirow{2}{*}{\textbf{Model}} & \multicolumn{3}{c|}{$q = 0$} & \multicolumn{3}{c|}{$q = 1$}  \\ \cline{2-7}
    & QIC & MSPE & Time & QIC & MSPE & Time \\ 
	\hline
   \textbf{$r = 1$} & 1,335,035 & 10.392 & \textbf{4.590} & 1,307,423 & 10.046 & 20.472 \\ 
   \textbf{$r = 2$} & 1,310,159 & 10.027 & 5.101 & 1,305,397 & 10.029 & 30.177 \\ 
   \textbf{$r = 3$} & 1,302,990 & 9.998 & 6.522 & 1,302,826 & 9.997 & 40.323 \\ 
   \textbf{$r = 4$} & 1,299,847 & 10.016 & 8.505 & 1,299,963 & 9.957 & 45.385 \\ 
   \textbf{$r = 5$} & 1,297,988 & 10.031 & 10.707 & 1,298,102 & 9.942 & 58.810 \\ 
   \textbf{$r = 6$} & 1,296,207 & 10.046 & 13.853 & 1,296,327 & 9.925 & 68.686 \\ 
   \textbf{$r = 7$} & 1,294,439 & 10.070 & 14.647 & 1,294,582 & 9.901 & 82.287 \\ 
   \textbf{$r = 8$} & \textbf{1,292,571} & 10.104 & 31.291 & 1,292,718 & \textbf{9.876} & 98.291 \\
	\hline
\end{tabular}
    \caption{QIC, MSPE and computation time of log-linear PSTARMAX processes with different temporal orders for regression on past observations ($r$) and feedback process ($q$).}
    \label{tab:rota_qic_train}
\end{table}

Table~\ref{tab:params_rota} contains the parameter estimations for a model without a feedback term ($q = 0$ and $r = 4$) and a model with a feedback term ($q = 1$ and $r = 3$). For reasons of simplicity, we restrict attention to these model orders, as the additional parameters of higher order models are estimated close to~0.

Both models are in agreement and show a significant and decreasing autoregressive dependence structure in the data. Observations less distant, both spatially and temporally, are more important for predicting the next time point than those further away. While observations within the same region exhibit a longer temporal dependence over several weeks, the direct neighbors have an impact of only one week. More distant regions had no significant impact at the 5\% level.

\begin{table}[tb]

\centering
\begin{tabular}{|l|ccc|ccc|}
  \hline
  \multirow{2}{*}{Coefficients} & \multicolumn{3}{c|}{$q = 0, r = 4$} & \multicolumn{3}{c|}{$q = 1, r = 3$} \\ \cline{2-7}
  & Estimate & Std. Error & $p$-value & Estimate & Std. Error & $p$-value \\ 
  \hline
  $\delta_0$ & -0.686 & 0.015 & $<2\cdot10^{-16}$ & -0.587 & 0.037 & $<2\cdot10^{-16}$ \\ 
  $\alpha_{0, 1}$ & - & - & - & 0.160 & 0.042 & $1.3\cdot10^{-4}$ \\
  $\alpha_{1, 1}$ & - & - & - & -0.011 & 0.030 & 0.722 \\
  $\beta_{0, 1}$ & 0.422 & 0.006 & $<2\cdot10^{-16}$ &  0.426 & 0.006 & $<2\cdot10^{-16}$ \\
  $\beta_{1, 1}$ & 0.126 & 0.010 & $<2\cdot10^{-16}$ &  0.114 & 0.011 & $<2\cdot10^{-16}$ \\
  $\beta_{2, 1}$ & 0.040 & 0.020 & 0.050 &  0.032 & 0.021 & 0.118 \\
  $\beta_{0, 2}$ & 0.203 & 0.005 & $<2\cdot10^{-16}$ &  0.140 & 0.020 & $8.57\cdot10^{-13}$ \\
  $\beta_{1, 2}$ & 0.002 & 0.010 & 0.853 &  0.000 & 0.020 & 0.982 \\
  $\beta_{2, 2}$ & 0.000 & 0.019 & 0.990 &  -0.001 & 0.020 & 0.977 \\
  $\beta_{0, 3}$ & 0.110 & 0.005 & $<2\cdot10^{-16}$ &  0.092 & 0.012 & $1.22\cdot10^{-14}$ \\
  $\beta_{1, 3}$ & -0.001 & 0.011 & 0.927 &  -0.005 & 0.014 & 0.749 \\
  $\beta_{2, 3}$ & -0.002 & 0.019 & 0.900 &  -0.012 & 0.019 & 0.533 \\
  $\beta_{0, 4}$ & 0.053 & 0.005 & $<2\cdot10^{-16}$ &  - & - & - \\
  $\beta_{1, 4}$ & -0.020 & 0.011 & 0.061 &  - & - & - \\
  $\beta_{2, 4}$ & -0.011 & 0.019 & 0.549 &  - & - & - \\
  GDR            & 0.333 & 0.011 & $<2\cdot10^{-16}$ & 0.280 & 0.017 & $<2\cdot10^{-16}$ \\ 
  Population     & 0.350 & 0.006 & $<2\cdot10^{-16}$ & 0.297 & 0.017 & $<2\cdot10^{-16}$ \\ 
  $\cos(2\pi t / 52)$ & 0.109 & 0.015 & $<2\cdot10^{-16}$ & 0.105 & 0.012 & $<2\cdot10^{-16}$ \\ 
  $\sin(2\pi t / 52)$ & 0.487 & 0.013 & $<2\cdot10^{-16}$ & 0.410 & 0.027 & $<2\cdot10^{-16}$ \\ 
  Vaccine & -0.098 & 0.022 & $4.78\cdot10^{-6}$ & -0.082 & 0.019 & $1.53\cdot10^{-5}$\\ 
   \hline
\end{tabular}
\caption{Parameter estimates of log-linear PSTARMAX processes on the rota virus data.}
\label{tab:params_rota}
\end{table}

In both models, all covariates are significant at the 5\% level. Using the model without feedback term, we expect about $(\exp(0.333) - 1)\cdot 100 \approx 39.5\%$ more infections per week in the former GDR regions than in the other regions, while keeping the other variables constant. With all other factors held constant, regions with twice the population size are expected to have $(2^{0.333} - 1)\cdot 100 \approx 27.5\%$ more infections per week. After the start of the vaccination recommendation, around $9.34\%$ less infections are expected.

In summary, the analysis by the PSTARMAX processes provides a deeper insight into the data and confirms a significant reduction of weekly infections after the start of the vaccination recommendation. Infections primarily cause new infections in their local region and to a lesser extent in spatial contiguous regions. An intuitive explanation for this could be that the Rota virus primarily affects children. Children, especially infants, for whom vaccination is recommended, are in general less mobile than adults, meaning that new infections tend to occur locally.

\subsection{Chicago Crime Data}
Next, we analyse the Chicago Crime Dataset, which was studied by \textcite{clark_class_2021}. The dataset is freely available on GitHub, see \url{https://github.com/nick3703/Chicago-Data}. 
It contains the aggregated monthly number of burglaries in the period from 2010 to 2015 ($T = 72$ months) in $p = 552$ census block groups. Census block groups are small areas used by the United States Census Bureau as a geographic unit. In addition to the data, socio-economic covariates such as population size, the number of young men (between the ages of 15 and 20), income per capita and the unemployment rate are available. Temperature data for Chicago were obtained by the website \url{weather.gov}, for the same time period.

Analysing such data is of interest in order to identify trends in crime, seasonal effects and other driving factors, such as spatio-temporal autocorrelation or socio-economic parameters. This is important e.g. for taking preventive actions or identification of areas with an elevated risk. 

\subsubsection{Data Description}
The 552 census block groups in the dataset are located on the south side of Chicago and are described by \textcite{clark_class_2021} as \enquote{relatively racial and socio-economic homogeneous}. The number of adjacent blocks varies from 1 to 14, with an average of 4,812 neighbours. The population size in the blocks varies from 12 to 3057, and the proportion of young men from 0 to $19.5\%$ whilst the unemployment rate takes values from $0\%$ to $54.3\%$.

The correlation between income and population size is $0.895$, indicating a strong linear relationship between the two variables. Wealthy people tend to live in more populated blocks. The correlation between wealth and the unemployment rate is $-0.026$, i.e., there is almost no linear relationship.

\begin{figure}[t]
    \centering
    \begin{minipage}[t]{0.45\textwidth}
    \centering
     \includegraphics[width=1\textwidth]{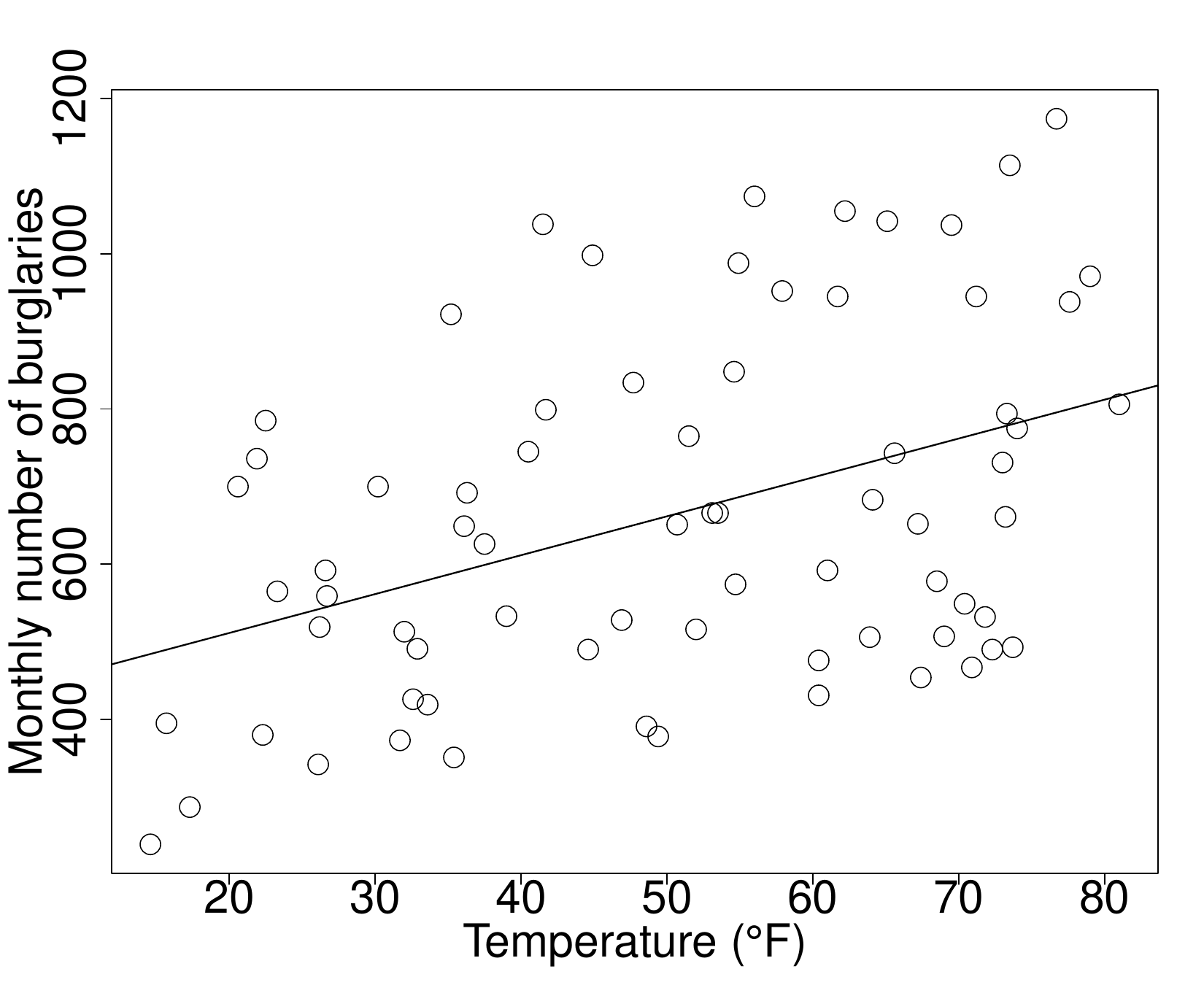} 
        \caption{Monthly accumulated number of burglaries against temperature (°F)}
        \label{fig:crime_temperature}
    \end{minipage}\hfill
    \begin{minipage}[t]{0.45\textwidth}
    \centering
        \includegraphics[width=1\textwidth]{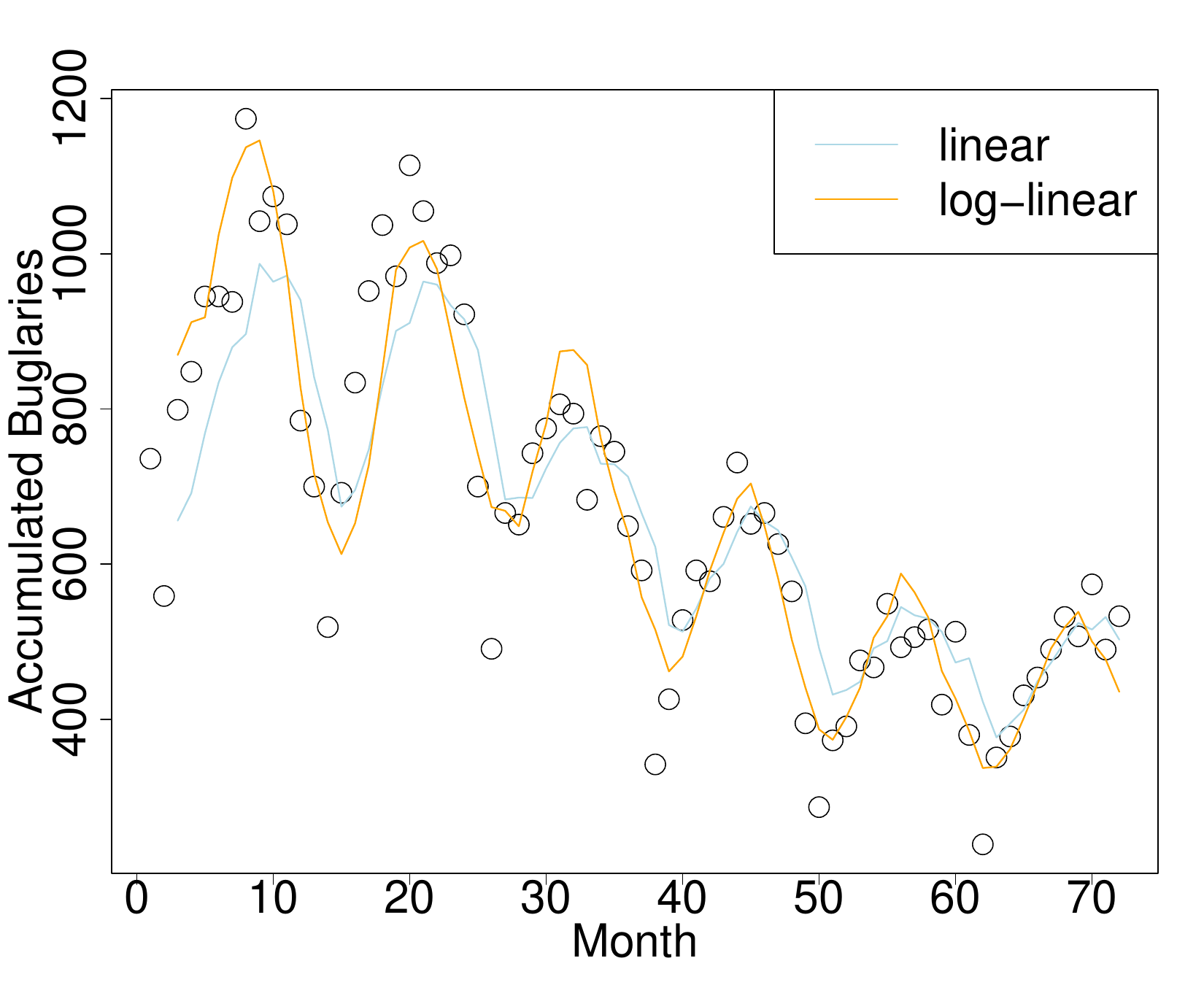} 
        \caption{Monthly accumulated number of burglaries and accumulated predictions of models IV (lightblue), VIII (orange) in Table~\ref{tab:crime_coef}.}
        \label{fig:crime_prediction}
    \end{minipage}
\end{figure}

In the entire observation period, there were a total of 47836 burglaries in the 552 blocks, which is equivalent to an average of 86,659 burglaries per block and 664,389 burglaries per month. On average, there were 1,204 burglaries per month per block in the entire period. Figure~\ref{fig:crime_prediction} shows the cumulative number of burglaries by month across all blocks. Overall, there is a decreasing trend with seasonal effects. The trend might be explained by the temperature, see Figure~\ref{fig:crime_temperature}. In warmer months, there tend to be more burglaries than in colder months.

\subsubsection{Model Fitting and Interpretation}

We consider PSTARMA and PSTARMAX models of different orders. We select the neighbourhood matrices as in the Rota Virus application, i.e., regions with a common boundary are first-order neighbours, second-order neighbours are the regions that can be reached by passing exactly one other region. 

We include socio-economic variables in the linear PSTARMAX models in a different way than in the log-linear ones. For the linear model, we use the number of inhabitants per 1000 as Population variable. We define the variable Unemployed using the unemployment rate multiplied by the number of inhabitants. Young Males is set to the number of young men between the ages of 15 and 20. The income per capita is only available in centred and standardised form. To avoid negative values, we use the transformation $\tilde{x} = \log(\exp(x) + 1)$ in the linear PSTARMAX processes for the variable Wealth, which creates values on a similar scale.

Log-linear models typically perform better on relative scales. Therefore, in the log-linear PSTARMAX processes, we utilise the Population variable as the logarithmised number of inhabitants in the blocks, the variables Unemployed and Young Males denote the relative shares of these groups within the population of the block. The variable Wealth is inserted in its centred and standardised form.

For both processes, we include two spatially constant variables. The first one is Temperature, which indicates the monthly average temperature in °F for each month. The second one is Trend, which is defined by $x_t = 72 - t, t = 1, \ldots, 72$.

\begin{table}[htb]

	\centering
 %\tiny
 \begin{adjustbox}{width=\textwidth}
	\begin{tabular}{lcccccccc}
		\toprule
		                                                   &                                                 \multicolumn{8}{c}{\textbf{Model}}                                                  \\
		                                                   &                    \multicolumn{4}{c}{\textbf{linear}}                    &         \multicolumn{4}{c}{\textbf{log-linear}}         \\
		\cmidrule(lr){2-5} \cmidrule(lr){6-9}
		Estimation &        I         &        II        &       III        &        IV        &        V         &        VI        &   VII   &  VIII   \\ \midrule
		$\boldsymbol{\delta_0}$                            & 0.0447 $(\star)$  & 0.0000 & 0.0486 $(\star)$ & 0.0000 & -0.1699 $(\star)$ & -0.5529 $(\star)$ & -0.2268 $(\star)$ & -0.7466 $(\star)$ \\
		$\boldsymbol{\alpha_{0,1}}$                        & 0.6200 $(\star)$ & 0.5452 $(\star)$ & 0.1403 & 0.0913 & 0.6661  $(\star)$& 0.6065 $(\star)$ & 0.1495 & 0.1249 \\
		$\boldsymbol{\alpha_{1,1}}$                        & 0.0000 & 0.0000 & 0.0000 & 0.0000 & 0.0035 & 0.0034 & 0.0032 & 0.0043 \\
		$\boldsymbol{\alpha_{0,2}}$                        & - & - & 0.3631 $(\star)$ & 0.2888 $(\star)$ & - & - & 0.3975 $(\star)$ & 0.3501 $(\star)$ \\
		$\boldsymbol{\alpha_{1,2}}$                        & - & - & 0.0000 & 0.0000 & - & - & 0.0031 & 0.0034 \\
        $\boldsymbol{\beta_{0,1}}$                         & 0.1917 $(\star)$ & 0.2011 $(\star)$ & 0.1838 $(\star)$& 0.1919 $(\star)$& 0.3135 $(\star)$& 0.3028 $(\star)$ & 0.3137 $(\star)$ & 0.2894 $(\star)$ \\
		$\boldsymbol{\beta_{1, 1}}$                        & 0.0748 $(\star)$ & 0.0798 $(\star)$ & 0.0900 $(\star)$& 0.0964 $(\star)$& 0.0104 & 0.0813 $(\star)$ & 0.0067 & 0.0715 $(\star)$ \\
        $\boldsymbol{\beta_{2, 1}}$                        & 0.0685 $(\star)$ & 0.0451 $(\star)$ & 0.0879 $(\star)$& 0.0593 $(\star)$& 0.0036 & 0.0035 & 0.0032 & 0.0033 \\
		$\boldsymbol{\beta_{0,2}}$                         & - & - & 0.0847 $(\star)$ & 0.1024  $(\star)$& - & - & 0.1096 $(\star)$ & 0.1383 $(\star)$ \\
		$\boldsymbol{\beta_{1, 2}}$                        & - & - & 0.0000 & 0.0000 & - & - & 0.0037 & 0.0062 \\
        $\boldsymbol{\beta_{2, 2}}$                        & - & - & 0.0000 & 0.0000 & - & - & 0.0024 & 0.0026 \\

		Population                                         & - & 0.0066 & - & 0.0000 & - & 0.1170 $(\star)$ & - & 0.1555 $(\star)$ \\
		Unemployed                                         & - & 0.0013 & - & 0.0002 & - & 0.0656 $(\star)$ & - & 0.0896 $(\star)$ \\
		Wealth                                             & - & 0.0360 $(\star)$ & - & 0.0486 $(\star)$ & - & -0.0005 & -  & -0.0023 \\
		Young Males                                        & - & 0.0390 $(\star)$ & - & 0.0493 $(\star)$ & - & 0.2394 $(\star)$ & - & 0.3103 $(\star)$ \\
		Temperature                                        & - & 0.0011 & - & 0.0017 & - & 0.0047 $(\star)$ & - & 0.0065 $(\star)$ \\
		Trend                                              & - & 0.0009 $(\star)$ & - & 0.0011 & - & 0.0022 $(\star)$ & - & 0.0029 $(\star)$ \\ \midrule
		\textbf{QIC ($\boldsymbol{\times 10^{-3}}$)}       & 112.29 & 111.73 & 110.61 & 110.05 & 113.88 & 111.35 & 112.09 & 109.59 \\ 
        \textbf{Time (in s)}                               & 1.211 & 4.735 & 1.489 & 4.864 & 1.111 & 2.126 & 3.860 & 5.958 \\
        \textbf{MSPE}                                                & 1.7493 & 1.7295 & 1.7180 & 1.7006 & 1.8091 & 1.7155 & 1.7851 & 1.6980 \\
        \bottomrule
	\end{tabular}
 \end{adjustbox}
\caption{Estimated parameters and QIC values of linear PSTARMA($1_1, 1_2$) (I), PSTARMAX($1_{1}, 1_2$) (II), PSTARMA($2_{1, 1}, 2_{2, 2}$) (III), PSTARMAX($2_{1, 1}, 2_{2, 2}$) (IV) and log-linear PSTARMA($1_1, 1_2$) (V), PSTARMAX($1_{1}, 1_2$) (VI), PSTARMA($2_{1, 1}, 2_{2, 2}$) (VII), PSTARMAX($2_{1, 1}, 2_{2, 2}$) (VIII) models}
\label{tab:crime_coef}
\end{table}

Table~\ref{tab:crime_coef} shows the estimated coefficients for PSTARMA($1_1, 1_2$), PSTARMA($2_{1, 1}, 2_{2, 2}$) and PSTARMAX processes of the same order. Significant parameters ($p < 0.05$) are marked with a star. Due to the limited number of observation times ($T = 72$), we restrict our attention to models with a maximal time order of $2$.

The models considered appear to fit the data similarly well. While the linear models has advantages due to increased time information, the log-linear models improves by adding covariates. This is evident in the QIC of the models, but is also reflected in the MSPE on the training dataset.

All models show significant autoregressive effects. For the feedback process this is limited to the spatial order $0$. For the observation process, the models identify a significant spatial correlation at time lag 1. In case of the linear processes, both parameters $\beta_{1, 1}$ and $\beta_{1, 2}$ are significant, in the log-linear case this applies only to $\beta_{1, 1}$.

The results show considerable differences regarding the significance of the covariates. The effect of the number/proportion of young men is significant in all models, but the results for other covariate effects are different in the linear and the log-linear models.  In the log-linear models the population size is significant, while the linear models show a significant effect for the variable Wealth. This can be explained by the high correlation of these two variables. Removing one of them results in the significance of the other one in all cases, with no relevant impact on the model performance.

Overall, the conclusions that can be drawn from this dataset are limited. The fitted models agree that there is a time-delayed dependence on neighbouring blocks. This might be explained by criminal hotspots. Due to the small size of the blocks, it is likely that an adjacent block also has a high rate of criminal activity. Moreover, a large number of young men affects the number of burglaries significantly.

However, no conclusions can be drawn on the effects of wealth and population size, as both variables are strongly correlated. %This means that significant effects cannot be attributed to one of the variables.  

A substantial part of the autoregressive dependencies is captured by the feedback process. This becomes visible in the parameter estimates for the feedback process, which are large compared to those of the observation process. As a result, the model is highly variable, indicating that there might be further explanatory variables, not included in the model, or that the model specification is too inflexible in general. The negative trend identified in the descriptive analysis may imply, for example, time-dependent parameters for the variables which would require more complex models.

\subsection{Alternative Models}
We consider two models, common in the literature for spatio-temporal data, for comparison with the PSTARMAX processes on the two datasets above. In addition, we consider univariate INGARCH and log-linear time series models for comparison.

We utilise the \textit{hhh4} model framework from \textcite{meyer_hhh4_2016}. This framework distinguishes between endemic and epidemic components. Similar to the PSTARMA processes, past observations are included in the epidemic components. By using a neighbourhood matrix, spatial dependencies lagged in time can be included as well. The endemic component does not contain any information on past observations. All components can be enriched with covariates. The observations $Y_{i, t}$ are assumed to be Poisson distributed with expectation $\lambda_{i, t}$, defined by
\begin{align}
    \lambda_{i, t} = e_{i, t} \nu_{i, t} + \varphi_{i, t} Y_{i, t - 1} + \rho_{i, t} \sum_{i \neq j} w_{ij} Y_{j, t - 1}.
\end{align}
In this model $e_{i, t}$ denotes a potential offset and it holds $\log(\nu_{i, t}) = \alpha^{(\nu)} + \bm{\beta}^{(\nu)\mathrm{T}}\bm{x}_{i, t}^{(\nu)}$, $\log(\varphi_{i, t}) = \alpha^{(\varphi)} + \bm{\beta}^{(\varphi)\mathrm{T}}\bm{x}_{i, t}^{(\varphi)}$ and $\log(\rho_{i, t}) = \alpha^{(\rho)} + \bm{\beta}^{(\rho)\mathrm{T}}\bm{x}_{i, t}^{(\rho)}$. The unknown parameters are given by $\alpha^{(\nu)}, \alpha^{(\rho)}, \alpha^{(\varphi)}, \bm{\beta}^{(\nu)}, \bm{\beta}^{(\rho)}, \bm{\beta}^{(\varphi)}$. This framework only allows a single neighbourhood matrix to be used. We use $\bm{W}^{(1)}$ here as for the PSTARMA processes. The estimation is done using the \textit{surveillance} \citep{meyer_surveillance_2017} package in R. In all components, we use the same covariates as in the applications above.

The second model for comparison is given in \textcite[4.5]{cressie_spatiotemporal_2019}. In this model, the expected value of the observations is expressed by spatiotemporal covariates and a tensor product of spatial and temporal basis functions. Assuming a Poisson distribution for the count $Y_{i, t}$ with expectation $\lambda_{i, t}$ the model is defined by 
\begin{align}
    \log(\lambda_{i, t}) = \delta_0 + \bm{x}_{i, t}'\bm{\gamma} + \sum_{m = 1}^{r_1}\sum_{n = 1}^{r_2} \alpha_{mn}\phi_{1m}(i)\phi_{2n}(t).
\end{align}
Here, $\bm{x}_{i, t}$ denotes a vector of covariates at location $i$ and observation time $t$. The spatial basis functions are denoted by $\phi_{1m}(i), m = 1, \ldots, r_1$, and the temporal basis functions by $\phi_{2n}(t), n = 1, \ldots, r_2$. Both data sets analysed are observed on discrete grids. Thus, we use a Markov Random Field (MRF) smoothing function for the spatial basis functions. The temporal basis functions use the cubic regression spline basis. The remaining parameters $\delta_0, \bm{\gamma}$ and $\alpha_{mn}, m = 1, \ldots, r_1, n = 1, \ldots, r_2$ have to be estimated. For estimation we use the function \texttt{gam} from the R package \textit{mgcv} \citep{wood_mgcv_2017}.

Besides the spatio-temporal models, univariate time series models can be applied to the data at each location. We use the univariate variants of \eqref{method:linear_with_covariates} and \eqref{method:log_with_covariates}. The linear model, see \textcite{ferland_integer-valued_2006}, defines the conditional expectation $\lambda_{i, t}$ at location $i$ at time $t$ via 
\begin{align}
 \lambda_{i, t} = \delta_i + \sum_{k = 1}^q \alpha_{i, k} \lambda_{i, t - k} + \sum_{l = 1}^r \beta_{i, l} Y_{i, t - l} + \bm{x}_{i, t}'\bm{\gamma}_{i},
\end{align}
with $\bm{x}_{i, t}$ an $m$-dimensional vector of location and time-dependent covariates, as well as parameters to be estimated $\delta_i$, $\alpha_{i, k}, k = 1, \ldots, q$, $\beta_{i, l}, l = 1, \ldots, r$ and $\bm{\gamma}_{i}$ for each location. As in model~\eqref{method:linear_with_covariates}, the unknown parameters and the covariates are restricted to positive values. With some abuse of notation, the log-linear model, see \textcite{fokianos_log-linear_2011}, is defined via
\begin{align}
 \nu_{i, t} = \delta_i + \sum_{k = 1}^q \alpha_{i, k} \nu_{i, t - k} + \sum_{l = 1}^r \beta_{i, l} \log(Y_{i, t - l} + 1) + \bm{x}_{i, t}'\bm{\gamma}_{i},
\end{align}
with $\nu_{i, t} = \log(\lambda_{i, t})$. The parameter estimation of both models is done with the package \textit{tscount} \citep{liboschik_tscount_2017}.

We compare the models according to several criteria. First, we consider the computation time on the complete data sets analysed in the previous sections. We also calculate the mean square error and the proportion of explained deviance, see \textcite{guisan_deviance_2000}, on these data sets. The latter is calculated via
\begin{align}
    \text{\%Deviance} = 1 - \frac{\log(P(Y_{i, t} = y_{i, t} | \lambda_{i, t} = \hat{\lambda}_{i, t})) - \log(P(Y_{i, t} = y_{i, t} | \lambda_{i, t} = y_{i, t}))}{\log(P(Y_{i, t} = y_{i, t} | \lambda_{i, t} = \Bar{y})) - \log(P(Y_{i, t} = y_{i, t} | \lambda_{i, t} = y_{i, t}))}.
\end{align}
Here, $P(Y_{i, t} = y | \lambda_{i, t} = \lambda)$ is the density of a Poisson distribution with parameter $\lambda$ and $\Bar{y} = (pT)^{-1}\sum_{i, t} y_{i, t}$ is the mean of all observations in the data set. The explained deviance can be considered as a proportion of explained uncertainty in the data. 

In order to analyse the models with regard to overfitting, we split both data sets into training and test sets. Using the models estimated on the training data, we make running one-step predictions for the test dataset and then determine the MSPE and the explained deviance on both datasets. For the Rota virus data, we use the years 2001 to 2016 ($T = 835$ weeks) for model estimation and predict the remaining $68$ weeks. For the Crime dataset, we use the first 5 years ($T = 60$ months) for model estimation and predict the last year (12 months) for prediction.

For the rota data, we use the log-linear PSTARMAX model with time order $q = 0$ and $r = 4$, as no further significant parameters were found for higher model orders. For the crime data, we use the linear and log-linear PSTARMAX($2_{1, 1}, 2_{2, 2}$) models, as both showed similar performance. For the other models, we use the same covariates as in the PSTARMAX processes. We include them in all components (endemic and epidemic) of the hhh4 model. In the mgcv models, we tried different combinations of numbers of basis functions and used the ones with the best overall performance. For the rota data this is $r_1 = 5, r_2 = 15$, and for the crime data $r_1 = 20$ and $r_2 = 5$. 
Because an individual intercept is estimated for each location in the univariate models, we only use covariates that vary over time in these models for reasons of identifiability. In the Rota data, these are the covariates for seasonality and the indicator for the vaccination recommendation. We also restrict our analysis to univariate log-linear models with the same temporal orders as the PSTARMAX processes, i.e. $q = 0$ and $r = 4$.
In the crime dataset, we use temperature and the trend variable for the univariate models. We consider linear and log-linear models with the orders $q = r = 1$.

Table~\ref{tab:compare} contains the results. The PSTARMAX processes perform well on both data sets consistently. The univariate time series models, applied to each location, overfit strongly and also have a high computation time due to the high number of parameters. Model fitting by the hhh4 model takes about twice as long when compared to the PSTARMAX processes. In addition the fitting of the models with the mgcv package required more computation time. However, we must keep in mind that the computation time of the PSTARMAX processes depends on the model orders and the number of covariates, see Tables~\ref{tab:rota_qic_train} and~\ref{tab:crime_coef}, i.e., the computation time increases quadratically with the number of parameters. Processes with a feedback term have higher computation times than processes without, as in the model fitting the design matrix changes in each optimization step and the update has linear computation costs. Especially in high dimensional settings with many observation times this might be crucial.

Overall, hhh4 models perform inferior to the PSTARMAX processes, although they have a similar structure. This is because higher order models can be fitted using PSTARMAX processes.
The models estimated with the mgcv package show the worst performance among the models considered here. They give large errors during the fitting and also for prediction. This might be caused by the model structure, which does not include autoregressive dependencies and therefore does not allow straightforward predictions. In addition, the basis functions require estimation of many parameters, which results in high computation times and highly volatile models.

Similar remarks hold for univariate time series models. A reasonably well goodness of fit is achieved, which is indicated by low MSPE and a high proportion of explained deviance in the training. Yet, these models do not have a good generalization ability which is reflected in the higher errors on the test data. 

\begin{table}[tb]
    \centering
    
    \begin{adjustbox}{width=\textwidth}
    \begin{tabular}{|c|c|c|cc|cc|cc|} \hline
    \multirow{2}{*}{Data}& \multirow{2}{*}{Model}& \multirow{2}{*}{Time} & \multicolumn{2}{c}{Full} & \multicolumn{2}{c}{Training} & \multicolumn{2}{c|}{Test} \\ \cline{4-9}
    & & & MSPE & \%Deviance & MSPE & \%Deviance & MSPE & \%Deviance \\ \hline
       \multirow{4}{*}{Rota} & PSTARMAX & \textbf{8.505}  & 10.016 & 0.666 & 9.974 & 0.672 & \textbf{9.569} & \textbf{0.583} \\
        & hhh4 & 17.961 & 10.380 & 0.651 & 10.388 & 0.656 & 10.162 & 0.575 \\
        & univariate log & 282.904 & \textbf{9.189} & \textbf{0.682} & \textbf{9.197} & \textbf{0.687} & 9.838 & 0.578 \\
        & mgcv & 415.892 & 15.682 & 0.546 & 15.201 & 0.557 & 15.682 & 0.379 \\ \hline
        \multirow{6}{*}{Crime} & linear PSTARMAX & \textbf{4.864} & 1.701 & 0.207 & 1.814 & 0.203 &  \textbf{1.141} & \textbf{0.135} \\
        & log-linear PSTARMAX & 5.958 & 1.6979 & 0.212 & 1.810 & 0.210 &  1.150 & 0.133 \\
        & hhh4 & 11.813 & 1.755 & 0.189 & 1.874 & 0.186 & 1.174 & 0.113 \\
        & mgcv & 55.328 & 1.824 & 0.169 & 1.951 & 0.161 & 1.237 & 0.073 \\ 
        & univariate log & 386.455 & \textbf{1.433} & \textbf{0.328} & \textbf{1.485} & \textbf{0.341} & 1.753 & -0.155 \\
         & univariate linear & 402.212 & 1.510 & 0.298 & 1.591 & 0.301 & 1.224 & -0.011 \\
        \hline
    \end{tabular}
    \end{adjustbox}
    \caption{MSPE and explained deviance of different models on the Rota and Crime datasets. The lowest value for Time and MSPE, and the highest for \%Deviance are marked in bold.}
    \label{tab:compare}
\end{table}

	\section{Summary}
In this paper, we have developed models for spatio-temporal count data, derived from the multivariate linear and log-linear models proposed by \textcite{fokianos_multivariate_2020}. We replace high-dimensional parameter matrices with sparsely parameterized linear combinations of given neighborhood matrices, reminiscent of the approach used in the STARMA models introduced by \textcite{pfeifer_three-stage_1980}.

The model classes studied in our contribution are more flexible than those studied by \textcite{armillotta_count_2023, armillotta_nonlinear_2023} or \textcite{jahn_approximately_2023}, because of the integration of a feedback term (regression on past conditional expectations), the capability to fit spatially anisotropic models, and the inclusion of time-dependent covariates. We point out that the results obtained by \textcite{armillotta_count_2023, armillotta_nonlinear_2023} are under a purely autoregressive model and assume that $p$ and $T \to \infty$ in a suitable way.

To facilitate these advancements, we have formulated less restrictive assumptions, aiming to establish consistent and asymptotically normally distributed parameter estimation in the models. Our simulation studies verify these properties in the considered settings.

The application of these models to two real datasets illustrates their utility. The first dataset, which is the weekly number of reported Rota virus infections in Germany, revealed spatial dependencies on neighboring locations, significant effects in the former GDR territory, as well as seasonal and demographic influences (\textcite{stojanovic_bayesian_2019}). Further, our analysis reveals a reduction of the infection numbers after vaccination recommendation in Germany.

For the second dataset,  we also identified spatial dependencies, and a positive relationship between the number of young men in a region and the number of burglaries. Linear and log-linear models were able to model the data set to a similar level of accuracy.

A comparison with existing models in the literature demonstrates the usefulness of the PSTARMAX processes. The option of fitting higher-order models allows to capture spatio-temporal structures better. The regression on the latent feedback term permits fitting parsimonious models without a relevant decrease in performance.

We currently develop an R package to provide convenient implementations of the models presented in this paper, as well as extensions to related models employing other commonly encountered distributions.

	\newpage
	\appendix
    \numberwithin{equation}{section}
    \renewcommand{\theequation}{A-\arabic{equation}}
	\printbibliography[heading=bibintoc, title={Literature}]

    \newpage
    \renewcommand{\theequation}{S-\arabic{equation}}
\renewcommand{\thetable}{S-\arabic{table}}
\renewcommand{\thefigure}{S-\arabic{figure}}

\appendix
\renewcommand{\thesection}{S}
\section{Supplementary Material}
In this supplementary material, we present further theoretical properties of PSTARMA processes and simulation results, which complement Section \ref{sec:simulation}.

\subsection{Definition STARMA processes}
\label{definition:starma}
Let $\lbrace \bm{Z}_{t} = (Z_{1, t}, \ldots, Z_{p, t})'\rbrace$ denote a $p$-dimensional stochastic process and let $\lbrace \bm{\varepsilon}_t = (\varepsilon_{1, t}, \ldots, \varepsilon_{p, t})' \rbrace$ denote a $p$-dimensional white noise process, i.e. $\mathbb{E}(\bm{\varepsilon}_t) = \bm{0}_p$ and 
\begin{align}
    \mathbb{E}(\bm{\varepsilon}_t \bm{\varepsilon}_{t + s}') = \begin{cases}
        \bm{\sigma} \in \mathbb{R}^{p \times p}, & \text{if } s = 0, \\
        \bm{0}_{p \times p}, & \text{otherwise}
    \end{cases}
\end{align}

We refer to $\lbrace \bm{Z}_{t} \rbrace$ as STARMA($v_{u_1, \ldots, u_v}, q_{a_1, \ldots a_q}$) process with autoregressive order $v$ and moving-average order $q$ with associated spatial orders $u_i$ and $a_j$, see \textcite{pfeifer_three-stage_1980}, if it holds
\begin{align}
    \bm{Z}_{t} = \sum_{i = 1}^{v} \sum_{\ell = 0}^{u_i} \phi_{i\ell} \bm{W}^{(\ell)} \bm{Z}_{t - i} - \sum_{j = 1}^{q} \sum_{\ell = 0}^{a_j} \theta_{j\ell} \bm{W}^{(\ell)} \bm{\varepsilon}_{t - j} + \bm{\varepsilon}_t.
\end{align}

\subsection{Stationarity and Ergodicity}
\label{sec:stationarity}
\textcite{fokianos_multivariate_2020} provide conditions for stationarity and ergodicity in models with regression on a single previous time point, i.e., generalised versions of~\eqref{method:linear_extended} and~\eqref{method:log_without_covariates} with $q = 1$ and $p = 1$. These conditions are thus also valid for~\eqref{method:linear_extended} and \eqref{method:log_without_covariates} but can be simplified further in our framework. Stability conditions for models with a temporal order larger than 1 have been investigated by \textcite{debaly_note_2021}. These are summarised in the following lemma.
\paragraph{Lemma}
Define matrices $\bm{A}_i = \sum_{\ell = 0}^{a_i} \alpha_{i\ell} \bm{W}^{(\ell)}, i = 1, \ldots, q$ and $\bm{B}_j = \sum_{\ell = 0}^{b_i} \beta_{j\ell} \bm{W}^{(\ell)}, j = 1, \ldots, r$ and let $|\bm{A}_i|_{\text{elem}}$ be the element-wise application of the absolute value function to the matrices $\bm{A}_i, i = 1, \ldots, q$; $|\bm{B}_i|_{\text{elem}}, i = 1, \ldots, r$ are defined analogously. Then it is sufficient for the stationarity, ergodicity and existence of all moments of the (log-)linear PSTARMA process \eqref{method:linear_extended} (respectively\eqref{method:log_without_covariates}) that
\begin{align}
\label{eq:log_stationarity}
    \left\Vert \sum_{i = 1}^{\max\lbrace q, r \rbrace}  (|\bm{A}_i|_{\text{elem}} + |\bm{B}_i|_{\text{elem}}) \right\Vert_2 < 1,
\end{align}
where, $\Vert \cdot \Vert_2$, denotes the spectral norm. Due to the restriction of the parameters to non-negative values in the linear model, the use of the absolute value function is not relevant in this case. In the master's thesis by \textcite{maletz_spatio-temporal_2021}, simplified conditions for first-order models were derived from the work of \textcite{fokianos_multivariate_2020}. An extension to the above lemma is easily possible, which leads us to the following proposition.

\paragraph{Proposition}
Let a linear or log-linear PSTARMA$(q_{a_1, \ldots, a_q}, r_{b_1, \ldots, b_r})$ process satisfy
\begin{align}
\label{eq:linear_log_stationarity_extendet}
    \sum_{i = 1}^q \sum_{\ell = 0}^{a_i} |\alpha_{i\ell}| + \sum_{j = 1}^r \sum_{\ell = 0}^{b_j} |\beta_{j\ell}| < \frac{1}{\sqrt{\tau}}
\end{align}
with $\tau = \max_{\ell} \Vert \bm{W}^{(\ell)} \Vert_1$, the column sum norm. Then the joint process $\lbrace (\bm{Y}_t, \bm{\lambda}_t) \rbrace$ or $\lbrace (\bm{Y}_t, \bm{\nu}_t) \rbrace$ is stationary, ergodic and all moments exist. If $\bm{W}^{(\ell)}$ is symmetric for all $\ell$, we have $\tau = 1$, while we have $\tau \in [1, p - 1]$ in general. 

Conditions less conservative than \eqref{eq:linear_log_stationarity_extendet} can be derived for special cases of the general model classes considered here. For PNAR processes, the sum in \eqref{eq:linear_log_stationarity_extendet} only needs to be less than 1 \citep{armillotta_count_2023}, and this condition guarantees second order stationarity of a linear PSTARMA process since it fulfills a STARMA model, see \citet{pfeifer_stationarity_1980}. However, restricting the sum in \eqref{eq:linear_log_stationarity_extendet} to be less than 1 does not necessarily imply \eqref{eq:log_stationarity}.

\sloppy The aforementioned conditions only apply to models without covariates and models with time-constant covariates. PSTARMAX processes with time-dependent deterministic covariates are not stationary in general. Stationarity of PSTARMAX processes with stochastic covariate processes requires stationarity of the covariate  process $\lbrace \bm{\Tilde{X}}_t \rbrace$ with $\bm{\Tilde{X}}_t = \sum\limits_{k = 1}^m \sum\limits_{\ell = 0}^s \gamma_{k,\ell} \bm{W}^{(\ell)} \bm{X}_{k, t}$. This applies for example if all covariate processes are stationary. Conditions for stationarity and ergodicity of PSTARMAX processes may be derived from the works of \textcite{doukhan_stationarity_2023} and \textcite{agosto_modeling_2016}.

\subsection{Second Order Properties}
As discussed in Section~\ref{subsec:starma}, the linear PSTARMA$(q_{a_1, \ldots, a_q}, r_{b_1, \ldots, b_r})$ process conforms to the model equation of a STARMA$(v_{u_1, \ldots, u_v}, q_{a_1, \ldots, a_q})$ process, where $v = \max \lbrace q, r \rbrace$ and $u_i = \max \lbrace a_i, b_i \rbrace$ for $i = 1, \ldots, v$, with $a_i = -1$ for $i > q$ and $b_i = -1$ for $i > r$. When considered stationary, i.e. fulfilling \eqref{eq:log_stationarity}, various properties such as the unconditional expected values $\mathbb{E}(\bm{Y}_t)$ and $\mathbb{E}(\bm{\lambda}_t)$, or the autocorrelation function, can be derived from the STARMA processes.
To be more precise, for the linear PSTARMA process \eqref{method:linear_extended} it holds
\begin{align}
\label{eq:expectation_linear}
    \mathbb{E}(\bm{Y}_t) = \mathbb{E}(\bm{\lambda}_t) = \left(\bm{I}_p - \sum_{i = 1}^q \bm{A}_i - \sum_{j = 1}^r \bm{B}_j \right)^{-1} \bm{\delta}.
\end{align}
In the case of a homogeneous PSTARMA process, \eqref{eq:expectation_linear} reduces to
\begin{align}
    \mathbb{E}(\bm{Y}_t) = \mathbb{E}(\bm{\lambda}_t) = \delta_0 \cdot \left(1 - \sum_{i = 1}^q \sum_{\ell = 0}^{a_i} \alpha_{i\ell} - \sum_{j = 1}^r \sum_{\ell = 0}^{b_j} b_{j\ell} \right)^{-1} \bm{1}_p.
\end{align}
If the covariates in the linear PSTARMAX process \eqref{method:linear_with_covariates} are stationary or constant in time, so that $\bm{\gamma} := \mathbb{E}\left( \sum_{k = 1}^m \sum_{\ell = 0}^{s_k} \gamma_{k, \ell} \bm{W}^{(\ell)} \bm{X}_{k, t} \right)$ exists and is independent of $t$, then it holds
\begin{align}
    \mathbb{E}(\bm{Y}_t) = \mathbb{E}(\bm{\lambda}_t) = \left(\bm{I}_p - \sum_{i = 1}^q \bm{A}_i - \sum_{j = 1}^r \bm{B}_j \right)^{-1} (\bm{\delta} + \bm{\gamma}).
\end{align}

As STARMA processes are derived from modified VARMA processes, they exhibit identical autocovariance functions and expected values. In general, there's no straightforward expression for the autocovariance. However, for low-temporal-order temporal models, such as the linear PSTARMA($1_{a_1}, 1_{b_1}$) process, we can do exact calculations.

Define $\bm{\Gamma}_0 \in \mathbb{R}^{p \times p}$ via 
$$\vect(\bm{\Gamma}_0) := \vect(\bm{\Sigma}) + \left( \bm{I}_p \otimes \bm{I}_p - (\bm{A}_1 + \bm{B}_1) \otimes (\bm{A}_1 + \bm{B}_1) \right)^{-1} \cdot \vect(\bm{B}_1 \bm{\Sigma} \bm{B}_1'),$$
with $\vect$ as the vec-operator, and $\otimes$ the Kronecker product.
Here, $\bm{\Sigma} := \mathbb{E}\left[\bm{\Sigma}_t \right]$ denotes the expected value of the covariance matrix conditioned on the past $\bm{\Sigma}_t = \var(\bm{Y}_t \vert \mathcal{F}_{t - 1}^{\bm{Y}, \bm{\lambda}, \bm{X}})$.
Consequently, the autocovariance function of the linear process in \eqref{method:linear_without_covariates} can be expressed as follows:
\begin{align}
	\label{eq:autocovariance}
	\bm{\Gamma}(h) = \begin{cases}
		\bm{\Gamma}_0, & h = 0 \\
		(\bm{A}_1 + \bm{B}_1) \bm{\Gamma}_0 - \bm{A}_1\bm{\Sigma}, & h = 1 \\
		(\bm{A}_1 + \bm{B}_1)^{h - 1} \bm{\Gamma}(1), & h \geq 2
	\end{cases}.
\end{align}

When the components of the conditional distribution $\bm{Y}_t \vert \mathcal{F}_{t - 1}^{\bm{Y}, \bm{\lambda}, \bm{X}}$ are uncorrelated, the covariance matrix $\bm{\Sigma}_t$ becomes a diagonal matrix with the diagonal elements determined by $\bm{\lambda}_t$. As a result, the covariance matrix $\bm{\Sigma}$ is also diagonal, with the elements of \eqref{eq:expectation_linear} on its main diagonal.

The expected values and autocovariance function can be explicitly specified solely for the linear PSTARMA process. For instance, considering the expected value of $(\bm{Y}_t, \bm{\nu}_t)$ in a stationary log-linear PSTARMA($1_{a_1}, 1_{b_1}$) process as $(\bm{\mu}, \bm{\nu})$, we have
\begin{align}
\bm{\nu} := \mathbb{E}\left[ \bm{\nu}_t \right] = \left(\bm{I}_p - \bm{A}\right)^{-1} \left[ \bm{d} + \bm{B}\cdot \mathbb{E}\left[ \log\left( \bm{Y}_t + \bm{1}_p \right) \right] \right] \preceq \log(\bm{\mu}).
\end{align}
However, the expected value $\mathbb{E}\left[ \log\left( \bm{Y}_t + \bm{1}_p \right) \right]$ of $\log\left( \bm{Y}_t + \bm{1}_p \right)$ is unknown. Consequently, we cannot rely solely on the process's parameters to determine its expected value. Nevertheless, if the elements of $\bm{\mu}$ are large (i.e., $\mu_i \gg 0$), then $\mathbb{E}\left[ \log\left( \bm{Y}_t + \bm{1}_p \right) \right]$ is approximately equal to $\bm{\nu}$, which can be further approximated by $\log(\bm{\mu})$.  Then $\bm{\nu}$ ist approximately equal to~\eqref{eq:expectation_linear}. Moreover, the autocovariance function~\eqref{eq:autocovariance} is almost equal to that of $\log\left( \bm{Y}_t + \bm{1}_p \right)$ with using $\bm{\Sigma}_t = \var(\log\left( \bm{Y}_t + \bm{1}_p \right) | \mathcal{F}_{t - 1}^{\bm{Y}, \bm{\lambda}})$.

\subsection{Asymptotic Properties of the QMLE}
\label{subsec:asymptotic}
In the case of PSTARMA processes, the same asymptotic properties apply to parameter estimation as in the work by \textcite{fokianos_multivariate_2020}. If the true parameter vector $\bm{\theta}_{\text{true}}$ satisfies the stability condition \eqref{eq:log_stationarity}, then, as stated in the main part, the quasi-maximum likelihood estimator defined by $\bm{\hat{\theta}} := {\arg\max}_{\bm{\theta}}\ell(\bm{\theta})$ is strongly consistent and asymptotically normally distributed as $T \to \infty$,
\begin{align}
\label{eq:asymptotics_suppl}
    \sqrt{T} \left(\bm{\hat{\theta}} - \bm{\theta}_{\mathrm{true}} \right) \; \overset{D}{\to}\; \mathcal{N}(\bm{0}, \bm{H}^{-1} \bm{G} \bm{H}^{-1}).
\end{align}

The condition that  $\bm{\theta}_{\mathrm{true}}$ is in the interior of the parameter space means that the linear PSTARMA-processes are only allowed to have positive parameter values $\bm{\theta}_{\mathrm{true}} \succ \bm{0}$. The matrices $\bm{H}$ and $\bm{G}$ in~\eqref{eq:asymptotics_suppl} for the linear PSTARMA-process~\eqref{method:linear_extended} are given by 
\begin{align}
	\label{eq:var_linear_appendix}
	\bm{G}(\bm{\theta}) = \mathbb{E} \left[ \frac{\partial \bm{\lambda}_t'}{\partial \bm{\theta}} \bm{D}_t^{-1}(\bm{\theta}) \bm{\Sigma}_t(\bm{\theta}) \bm{D}_t^{-1}(\bm{\theta}) \frac{\partial \bm{\lambda}_t}{\partial \bm{\theta}} \right] \text{ and } \bm{H}(\bm{\theta}) = \mathbb{E} \left[ \frac{\partial \bm{\lambda}_t'}{\partial \bm{\theta}} \bm{D}_t^{-1}(\bm{\theta}) \frac{\partial \bm{\lambda}_t}{\partial \bm{\theta}} \right],
\end{align}
and for the log-linear PSTARMA-process~\eqref{method:log_without_covariates} by
\begin{align}
	\label{eq:var_log_appendix}
	\bm{G}(\bm{\theta}) = \mathbb{E} \left[ \frac{\partial \bm{\nu}_t'}{\partial \bm{\theta}}  \bm{\Sigma}_t(\bm{\theta})  \frac{\partial \bm{\nu}_t}{\partial \bm{\theta}} \right] \text{ and } \bm{H}(\bm{\theta}) = \mathbb{E} \left[ \frac{\partial \bm{\nu}_t'}{\partial \bm{\theta}} \bm{D}_t(\bm{\theta}) \frac{\partial \bm{\nu}_t}{\partial \bm{\theta}} \right].
\end{align}

For the calculation of standard errors, the matrices $\bm{H}$ and $\bm{G}$ can be estimated by their empirical counterparts by substituting $\bm{\hat{\theta}}$.

If covariates are included in PSTARMAX processes, they must fulfil certain conditions to ensure consistent parameter estimation. First, the model has to be identifiable, see subsection~\ref{sec:identifiability}. In addition, the PSTARMAX processes are related to multivariate GLMs, so that it can be assumed that the conditions that ensure the consistency and asymptotic normality of maximum and quasi-maximum likelihood estimates, as elaborated in the work of \textcite{fahrmeir_consistency_1985, gao_asymptotic_2012, chen_strong_1999, zhang_asymptotic_2011, yin_asymptotic_2006}, are also sufficient here. In total we suppose that the QMLE $\bm{\hat{\theta}}$ in the PSTARMAX-processes for the parameter vector $\bm{\theta}_{\mathrm{true}}$  is consistent and asymptotically normal~\eqref{eq:asymptotics_suppl}, if the following conditions hold:
\begin{enumerate}
    \item The parameters of the true underlying process fulfil  condition~\eqref{eq:log_stationarity}.
    \item All covariate processes are bounded and have finite variance, i.e.,
    \begin{itemize}
        \item for deterministic covariate processes $\lbrace \bm{X}_{k, t} \rbrace, k = 1, \ldots, m$, there is a $c > 0$ such that for all $t, k: \, \Vert\bm{X}_{k, t}\Vert_2^2 < c$ .
        \item for random covariate processes $\lbrace \bm{X}_{k, t} \rbrace, k = 1, \ldots, m$, there is a $c > 0$ such that $\forall t, \forall k: \, \mathbb{E} \left[\Vert\bm{X}_{k, t}\Vert_2^2 \right] < c.$
    \end{itemize}
    \item The processes $$\lbrace \bm{W}^{(\ell)}\bm{X}_{k, t} \rbrace,\; k = 1, \ldots, m, \, \ell = 0, \ldots, c_k, $$ are linearly independent.
    \item There is a $\varepsilon > 1$ such that $\frac{\overline{\sigma_T}^{0.5} \log(\overline{\sigma_T})^{0.5\varepsilon}}{\underline{\sigma_T}} \to 0$ as $T \to \infty$, where $\overline{\sigma_T}( \underline{\sigma_T})$ is the largest (smallest) eigenvalue of the matrix $\bm{F}_T = \sum_{t = 1}^T \bm{\Tilde{X}}_t' \bm{\Tilde{X}}_t$ with $$\bm{\Tilde{X}}_t = \left(\bm{1}_p, \bm{W}^{(0)}\bm{X}_{1, t}, \ldots, \bm{W}^{(s_1)}\bm{X}_{1, t}, \ldots, \bm{W}^{(s_m)}\bm{X}_{m, t} \right).$$
\end{enumerate}
\paragraph{Remark to conditions:} Condition 1 ensures the stability of the autoregressive parts of the model, while condition 2 ensures the stability of the covariates. Condition 3 implies that the covariate parameters are identifiable. Condition 4 implies that the \textit{information} from the covariate processes increases fast enough as the number of observation points increases and thus the consistency of the parameter estimates, see \textcite{zhang_asymptotic_2011}. In the case of an inhomogeneous intercept structure the vector $\bm{1}_p$ is replaced by the matrix $\bm{I}_p$ to verify condition 4.

In practice, conditions 2-4 are satisfied by numerous examples of covariate processes. For instance, consider a time-constant covariate $\bm{X}_{k, t} = \bm{\tilde{X}} \quad \forall t$, where $\bm{\tilde{X}} \in \mathbb{R}^p$ and $\bm{\tilde{X}} \neq c \cdot \bm{1}_p, c \in \mathbb{R}$. Such a covariate, for example the population size of a region, with maximum spatial order $s_k = 0$ satisfies the conditions. Furthermore, two spatially constant covariates $\bm{X}_{1, t} = \sin(2\pi t /n) \bm{1}_p$ and $\bm{X}_{2, t} = \cos(2\pi t / n) \bm{1}_p$ also fulfill these conditions. In time series analysis, these covariates are employed to model a seasonality of length $n \in \mathbb{N}$, $n > 1$. For $T = kn, k \in \mathbb{N}$, the matrix $\bm{F}_T$ from condition 4 becomes a diagonal matrix with $\text{diag}(\bm{F}_T) = (Tp, 0.5Tp, 0.5Tp)'$, thereby satisfying condition 4.

\subsection{Further Simulations}

We provide further simulation results and extend simulations from \textcite{maletz_spatio-temporal_2021}, by adjusting underlying parameters and increasing the number of repetitions to 1,000. A summary of these simulations is presented in Table~\ref{tab:settings_master}. Except for the \textit{Copula} simulation, all simulations are using the same copula in the data-generating process as in the simulations in Section~\ref{sec:simulation}.

	\begin{table}[p]
		\centering
		\resizebox{0.9\textwidth}{!}{%
			\begin{tabular}{|C{0.09\textwidth}|p{0.3\textwidth}|p{0.35\textwidth}|@{}p{0.45\textwidth}@{}|}
				\hline
				\multicolumn{1}{|c|}{\textbf{Study}} & \multicolumn{1}{|c|}{\textbf{Description}} & \multicolumn{1}{|c|}{\textbf{Model Orders \& Data}} &\multicolumn{1}{|c|}{\textbf{True Parameters}} \\ \hline
			 \parbox[t]{2mm}{\multirow{2}{*}{\rotatebox[origin=c]{90}{Normal Approximation}}}
			 & 
			 \multirow[t]{2}{*}{\parbox{\linewidth}{%
			 \par\vspace{-3\baselineskip}
			 Verification of \eqref{eq:wald} and \eqref{eq:asymptotics_suppl}}}
			 & 
			 \multirow[t]{2}{*}{\parbox{\linewidth}{%
			 	\par\vspace{8\baselineskip}
			 \begin{itemize}[left=0.2em, topsep=0pt]
			 	\itemsep0.5pt
			 	\item $T \in \lbrace 5, 10, 20 \rbrace \cup $ \newline $\lbrace50, 100, 250, 500, 750 \rbrace$
			 	\item Grid sizes: \newline $n\times n$ $(p = n^2)$ \newline $n \in \lbrace 5, 7, 9, 10, 20, 30, 40, 50 \rbrace$
			 	\item Maximum time lags: \newline$(q, r) \in \lbrace (1, 1), (0, 1) \rbrace$
			 	\item Number of covariates: \newline$m \in \lbrace 0, 1 \rbrace$
			 	\item Neighborhood matrices: \newline$\bm{W}^{(0)} = \bm{I}_p$, \newline
			 	$\bm{W}^{(1)} = \bm{W}_{\text{4NN}}$
			 \end{itemize}}}
		 	 & 
			 \begin{tabular}{@{}C{0.2\linewidth}@{}|@{}C{0.175\linewidth}@{}C{0.175\linewidth}@{}|@{}C{0.225\linewidth}@{}C{0.225\linewidth}@{}}
			 	& \multicolumn{2}{c}{linear} & \multicolumn{2}{c}{log-linear} \\
			 	\hline
			 	$\delta_0$ & \multicolumn{2}{c|}{5} & \multicolumn{2}{c}{0.6} \\
			 	$\alpha_{0,1}$ & $[0, 0.5]$ & - & $[-0.25, 0.25]$ & {-} \\
			 	$\alpha_{1,1}$ & 0.1 & {-} & 0.1 & {-} \\
			 	$\beta_{0,1}$ & \multicolumn{2}{c|}{$[0, 0.5]$}  & \multicolumn{2}{c}{$[-0.25, 0.25]$} \\
			 	$\beta_{1,1}$ & 0.1  & {0.25} & 0.1 & {0.25} \\
			 	$\gamma_{0,1}$ & \multicolumn{2}{c|}{$[0, 0.5]$} & \multicolumn{2}{c}{$[-0.25, 0.25]$}   \\ \hline
			 	%\hline& & 
			 \end{tabular} \\
			 & & & 
			 \multicolumn{1}{c|}{
			 \parbox{0.37\linewidth}{
			 	\vspace{8pt}
			 \textit{Remark}: At any time, only one parameter is varied in the above intervals. The other two parameters are set to a fixed value ($\alpha_{0,1} = 0.2$, $\beta_{0,1} = 0.2$ and $\gamma_{0,1} = 2$ for linear and $\gamma_{0,1} = 0.9$ for log-linear models). The parameter sets without the feedback process are only used for $p \geq 100$. \vspace{1.25em} }}  \\
            \hline
				\parbox[t]{2mm}{\multirow{2}{*}{\rotatebox[origin=c]{90}{Copula}}}
				& 
				\vspace{-3\baselineskip}Influence of different copulas on parameter estimation and comparison with contemporary independent data. Copulas used with parameters:
						\begin{itemize}[left=0.2em, topsep=0pt]
							\item Frank: $\lbrace 0.5, 1, \ldots, 3 \rbrace$
							\item Clayton: $\lbrace 0.5, 1, \ldots, 3 \rbrace$
							\item Joe: $\lbrace 1.5, 2, 2.5, 3 \rbrace$
							\item Independent
				\end{itemize}
				& 
				\vspace{-3\baselineskip}
						\begin{itemize}[left=0.2em, topsep=0pt]
							\itemsep0.5pt
							\item $T \in \lbrace 50, 100, 250, 500 \rbrace$
							\item Grid size: $9 \times 9$ $(p = 81)$
							\item Maximum time lags: \newline $(q, r) = (1, 1)$
							\item Number of covariates: \newline $m = 0$
							\item Neighborhood matrices: \newline$\bm{W}^{(0)} = \bm{I}_p$, \newline$\bm{W}^{(1)} = \bm{W}_{\text{4NN}}$
				\end{itemize}
				&
				\begin{tabular}{@{}C{0.2\linewidth}@{}|@{}C{0.4\linewidth}@{}C{0.4\linewidth}@{}}
					& linear & log-linear \\
					\hline
					$\delta_0$ & 5 & 0.6 \\
					$\alpha_{0,1}$ & 0.2 & 0.2 \\
					$\alpha_{1,1}$ & 0.1 & 0.1 \\
					$\beta_{0,1}$ & 0.2 & 0.2 \\
					$\beta_{1,1}$ & 0.1 & 0.1 \\\hline
				\end{tabular} \\
				\hline
				\parbox[t]{2mm}{\multirow{2}{*}{\rotatebox[origin=c]{90}{Intercept}}}
				& 
				\multirow[t]{2}{*}{\parbox{\linewidth}{%
						\par\vspace{-2.5\baselineskip}
						Estimation of models with wrong intercept structure.}}
				& 
				\multirow[t]{2}{*}{\parbox{\linewidth}{%
						\par\vspace{5\baselineskip}
						\begin{itemize}[left=0.2em]
							\itemsep0.5pt
							\item $T \in \lbrace 50, 100, 250, 500 \rbrace$
							\item Grid size: $9 \times 9$ $(p = 81)$
							\item Maximum time lags: \newline $(q, r) = (1, 1)$
							\item Number of covariates: \newline $m = 0$
							\item Neighborhood matrices: \newline$\bm{W}^{(0)} = \bm{I}_p$, \newline$\bm{W}^{(1)} = \bm{W}_{\text{4NN}}$
				\end{itemize}}}
				& 
				\begin{tabular}{@{}C{0.2\linewidth}@{}|@{}C{0.35\linewidth}@{}C{0.45\linewidth}@{}}
					& linear & log-linear \\
					\hline
					$\bm{\delta}$ & $\bm{\delta}_{\text{linear}}$ & $\bm{\delta}_{\text{log}}$ \\
					$\alpha_{0,1}$ & 0.2 & 0.2 \\
					$\alpha_{1,1}$ & 0.1 & 0.1 \\
					$\beta_{0,1}$ & 0.2 & 0.2 \\
					$\beta_{1,1}$ & 0.1 & 0.1 \\ \hline
					%\hline& & 
				\end{tabular} \\
				& & & 
				\multicolumn{1}{c|}{
					\parbox{0.37\linewidth}{
						\vspace{8pt}
						\textit{Remark}: We define the intercept vectors  elementwise with $\bm{\delta}_{\text{linear}} = (\delta_i)$ and $\bm{\delta}_{\text{log}} = (\tilde{\delta}_i)$ for $i = 1, \ldots, 81$ with $\delta_i = 2 + 3 \cdot \frac{i}{81}$ and $\tilde{\delta}_i = 0.6 \cdot \frac{i}{81}$\vspace{1em}}
      } \\ \hline
				% \parbox[c]{5mm}{\rotatebox[origin=c]{90}{\multirow{3}{*}{\shortstack[l]{Misspecification of \\ Spatial Dependencies}}}}
				\parbox{7mm}{\multirow{2}{*}{\rotatebox[origin=c]{90}{Link}}}
				& 
				\vspace{-3\baselineskip}%
				Estimation of wrong model type (linear/log-linear)	 
				&
				\vspace{-3\baselineskip}
						\begin{itemize}[left=0.2em, topsep=0pt]
							\itemsep0.5pt
							\item $T \in \lbrace 50, 100, 250, 500 \rbrace$
							\item Grid size: $9 \times 9$ $(p = 81)$
							\item Maximum time lags: \newline $(q, r) = (1, 1)$
							\item Number of covariates: \newline $m = 0$
							\item Neighborhood matrices: \newline$\bm{W}^{(0)} = \bm{I}_p$, \newline$\bm{W}^{(1)} = \bm{W}_{\text{4NN}}$
				\end{itemize}
				& 
				\begin{tabular}{@{}C{0.2\linewidth}@{}|@{}C{0.3\linewidth}@{}|@{}C{0.25\linewidth}@{}@{}C{0.25\linewidth}@{}}
					& linear & \multicolumn{2}{c}{log-linear} \\
					\hline
					$\delta_0$ & 5 & 0.6 & 0.6\\
					$\alpha_{0,1}$ & 0.2 & 0.2 & -0.2 \\
					$\alpha_{1,1}$ & 0.1 & 0.1 & 0.1 \\
					$\beta_{0,1}$ & 0.2 & 0.2 & -0.2 \\
					$\beta_{1,1}$ & 0.1 & 0.1 & 0.1 \\ \hline
					%\hline& & 
				\end{tabular}
				\\ \hline
		\end{tabular}}
		\caption{Summary of additional simulations. Contemporary correlations between the counts are generated by the data-generating process proposed by \textcite{fokianos_multivariate_2020}, using the Clayton copula with parameter 2, for the \textit{Normal Approximation}, \textit{Intercept} and the \textit{Link} Simulations.}
		\label{tab:settings_master}
	\end{table}

\subsubsection{Normal Approximation}
\label{sim:asymptotics}
\begin{figure}[htbp]
    \centering
    \includegraphics[keepaspectratio, width=0.85\textwidth]{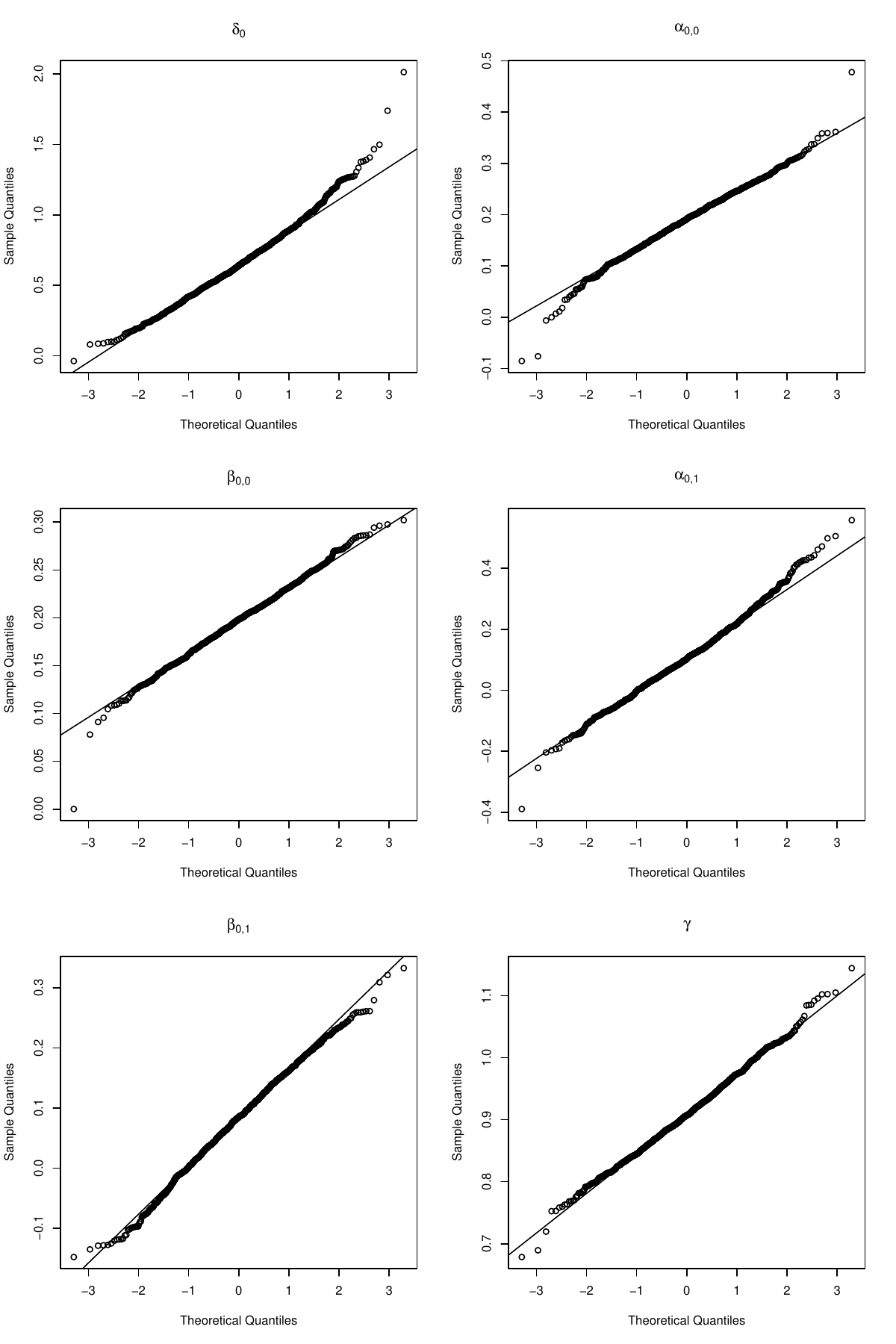}
    \caption{QQ-plots for the QMLE in case of the log-linear PSTARMAX process on a $5 \times 5$ grid and $T = 100$ observation times in the \enquote{Normal Approximation} simulation from Table~\ref{tab:settings_master}.}
    \label{fig:qqplot_log_100}
\end{figure}

\begin{figure}[htbp]
    \centering
    \includegraphics[keepaspectratio, width=0.85\textwidth]{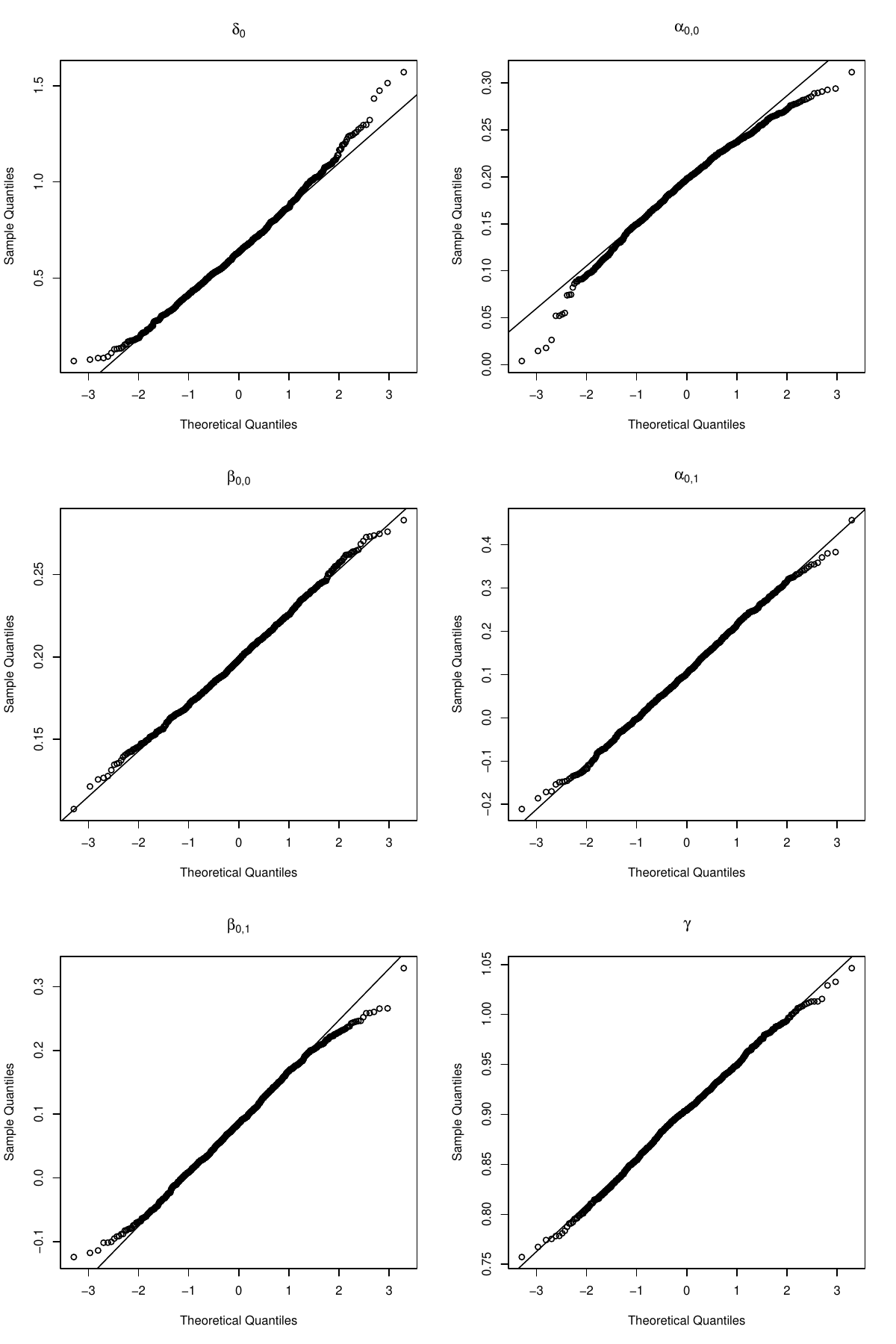}
    \caption{QQ plots of QMLE in case of the log-linear PSTARMAX process on a $9 \times 9$ grid for $T = 100$ observation times for the \enquote{Normal Approximation} simulation from Table~\ref{tab:settings_master}.}
    \label{fig:qqplot_log_100_9x9}
\end{figure}

\begin{figure}[htbp]
    \centering
    \includegraphics[keepaspectratio, width=0.85\textwidth]{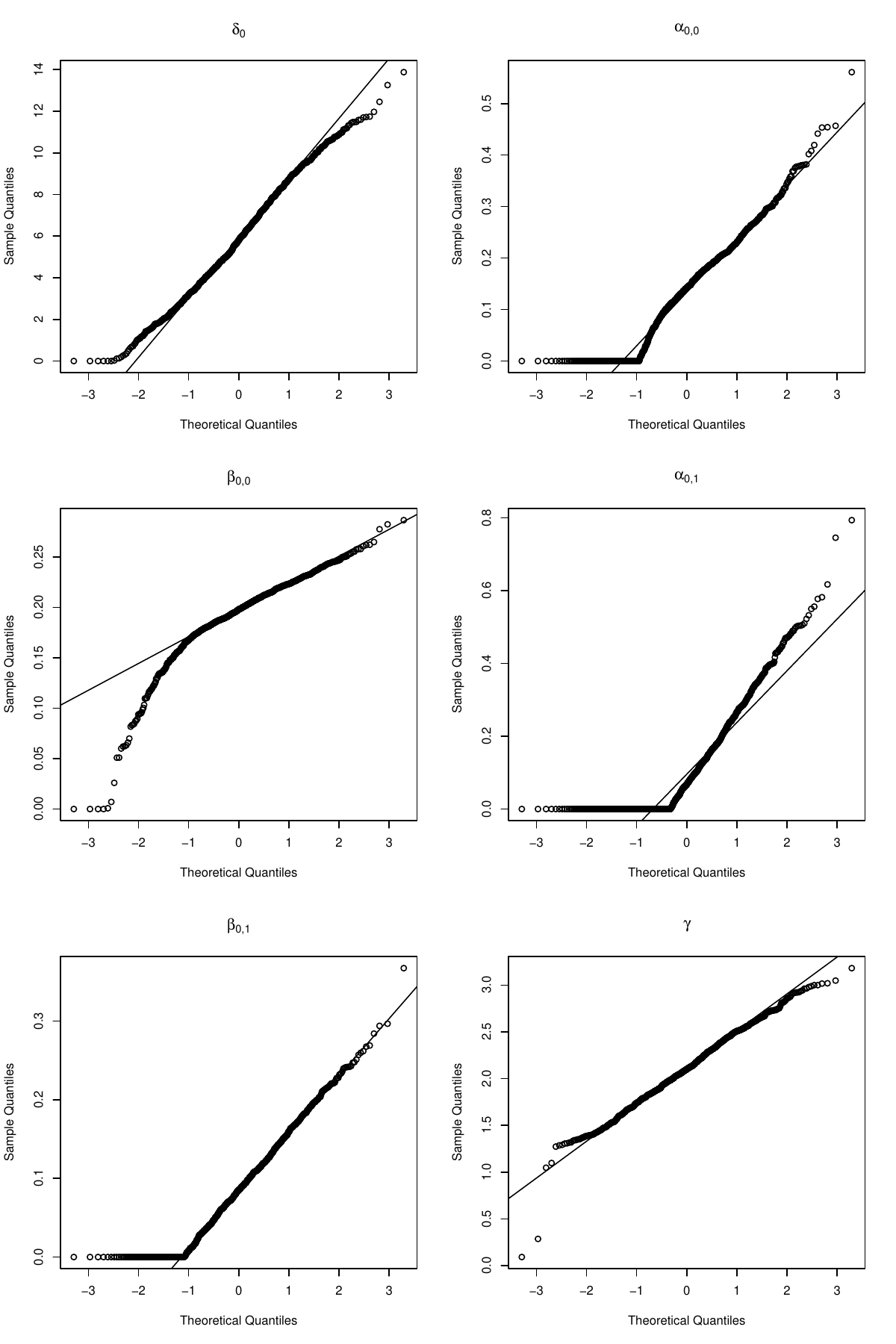}
    \caption{QQ plots of QMLE in case of the linear PSTARMAX process on a $9 \times 9$ grid for $T = 100$ observation times for the \enquote{Normal Approximation} simulation from Table~\ref{tab:settings_master}.}
    \label{fig:qqplot_linear_100}
\end{figure}

\begin{figure}[htbp]
    \centering
    \includegraphics[keepaspectratio, width=0.85\textwidth]{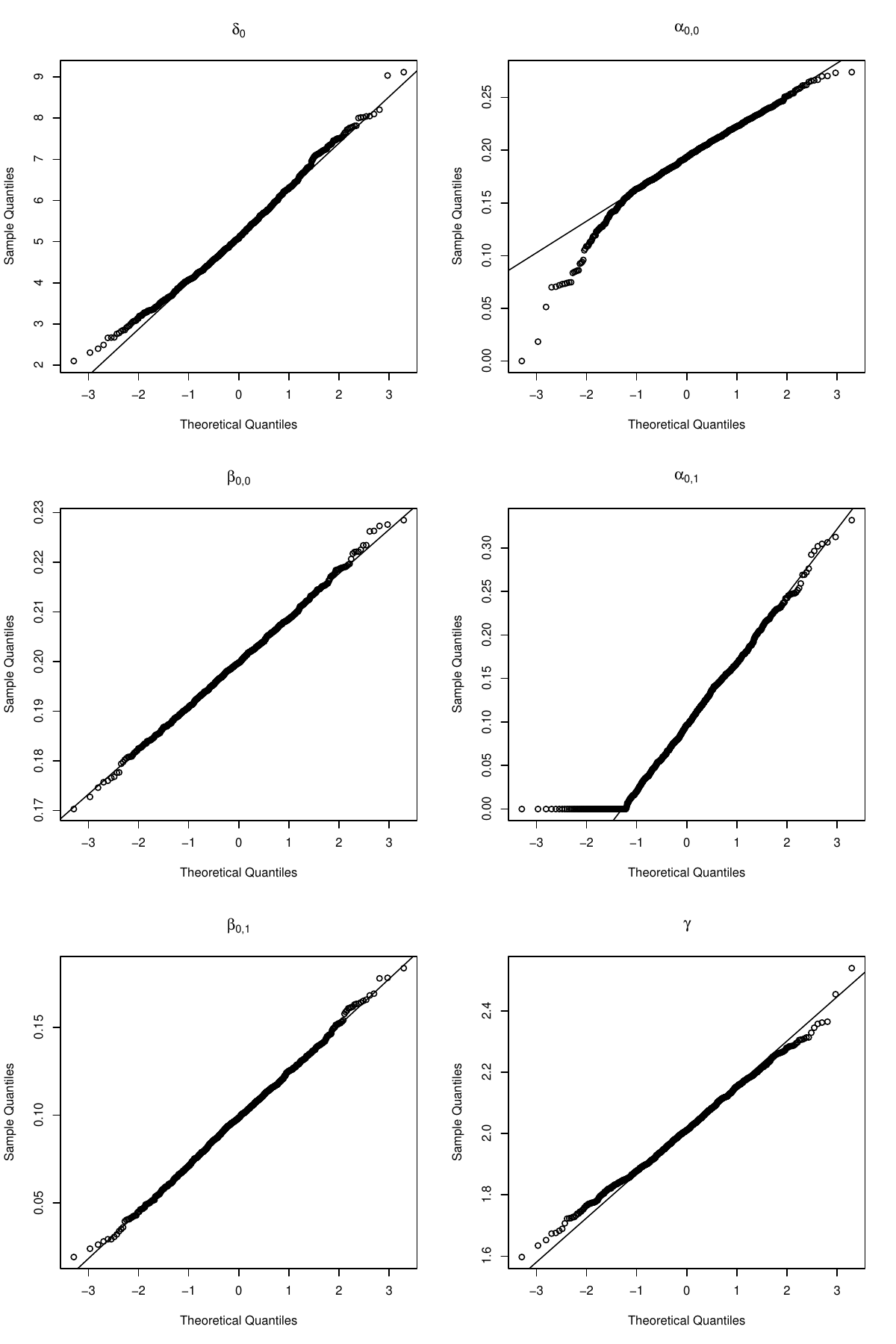}
    \caption{QQ plots of QMLE in case of the linear PSTARMAX process on a $9 \times 9$ grid for $T = 750$ observation times in the \enquote{Normal Approximation} simulation from Table~\ref{tab:settings}.}
    \label{fig:qqplot_linear_750}
\end{figure}

Figure~\ref{fig:qqplot_log_100} shows QQ-plots of the parameter estimates in the case of $T = 100$ observations on a $5 \times 5$ grid in the log-linear model. This plot shows adequate approximation to the normal distribution. This can also be observed for the other dimensions of the processes considered here, see Figures~\ref{fig:qqplot_log_100_9x9}, \ref{fig:qqplot_linear_100} and~\ref{fig:qqplot_linear_750}.

For the linear model, we observe stronger deviations from the asserted asymptotic normal distribution, which is due to the restrictions imposed on the parameters. For $T = 250$, the distributions of all parameter estimators, with the exception of $\alpha_{0, 0}$ and $\alpha_{0, 1}$, appear to be well approximated by a normal distribution. For $T = 750$ in the $5 \times 5$ grid, approx. 12\% of the values for $\alpha_{0, 1}$ are still estimated in the range $[0, 10^{-5})$, i.e. almost 0, although the true value is 0.2.

Figure~\ref{fig:qqplot_log_100_9x9} illustrates QQ plots of the simulated asymptotic distribution of parameter estimation for the log-linear PSTARMAX model across a grid of dimensions $9 \times 9$. This is similar to the QQ plots presented in Figure~\ref{fig:qqplot_log_100}, which depict parameter estimates on a grid of dimensions $5 \times 5$. Only slight changes can be seen. Figures~\ref{fig:qqplot_linear_100} and \ref{fig:qqplot_linear_750} showcase outcomes derived from linear PSTARMAX models applied to a $9 \times 9$ grid, yet with variations in the number of observation times. While substantial deviations from normal distribution are evident for $T = 100$, which is due to to parameter constraints, the approximation for $T = 750$ is adequate except $\alpha_{0, 0}$ and $\alpha_{0, 1}$.

\subsubsection{Wald Test}

We proceed with investigating the properties of the asymptotic  Wald test. Table~\ref{tab:empirical_size} illustrates its Type I errors at the nominal 5\% significance level across different dimensions and observation times. For each data generating process, one of the parameters $\alpha_{0, 1}$, $\beta_{0, 1}$, or $\gamma_{0,1}$ is set to 0, while the remaining parameters are kept constant; recall  Table~\ref{tab:settings_master}. For a sample size of $T = 50$, the results for most parameter settings deviate from the target value 0.05. However, as $T$ increases the nominal significance level is approached. Notably, the test for the parameter $\alpha_{0, 1}$ in the linear model seems to be somewhat conservative as the empirical size is about 0.04. This is probably due to the slower convergence to the asymptotic normal distribution for this specific parameter, which has been discussed above.

{
Fast convergence to the nominal significance level is obtained for models without a feedback term, see Table~\ref{tab:empirical_size_high_auto}. For $T = 5$, we observe values close to the target value of 0.05 for the autoregressive parameter $\beta_{0, 1}$ in the linear model, which indicates that the test is less conservative than for a processes with the feedback term. The test for detecting covariate effects requires a larger $T$. In the linear model, this yields adequate convergence at $T = 50$. A slower convergence to the value 0.05.

A straight comparison of the linear and log-linear models shows that the linear model reaches the nominal 5\% significance level quicker, i.e. less observation times are required. This also applies to high-dimensional settings with only a few observation times, see Table~\ref{tab:empirical_size_high}. It appears that the number of locations only has an impact on the inference of the feedback term. In case of few observation times ($T = 5$), larger Type I errors arise for larger grids, e.g. in the linear model $0.345$ on the $50 \times 50$ grid compared to $0.279$ on the $10 \times 10$ grid.
}

\begin{table}[tb]
    \centering
    \begin{tabular}{|c|ccccccc|}
    	\hline
      \multirow{2}{*}{Parameter} & \multirow{2}{*}{Model}& \multirow{2}{*}{Gridsize} & \multicolumn{5}{c|}{$T$} \\ \cline{4-8}
    %	           Parameter             &            Model             & Gridsize &  
    & & & 50   &  100  &  250  &  500  &  750  \\ \hline
    	\multirow{6}{*}{$\gamma_{0, 1}$} & \multirow{3}{*}{log-linear} &  $5 \times 5$  & {0.076} & {0.061} & {0.052} & 0.052 & 0.052 \\
    	                                 &                             &  $7 \times 7$  & {0.057} & {0.045} & 0.051 & 0.053 & 0.042 \\
    	                                 &                             &  $9 \times 9$  & {0.066} & {0.058} & 0.064 & 0.052 & 0.063 \\
    	       \cline{2-8}        &   \multirow{3}{*}{linear}   &  $5 \times 5$  & 0.067 & 0.054 & 0.056 & 0.042 & 0.047 \\ 
    	                                 &                             &  $7 \times 7$   & 0.062 & 0.054 & 0.053 & 0.057 & 0.059 \\
    	                                 &                             &  $9 \times 9$  & 0.060 & 0.066 & 0.050 & 0.046 & 0.046 \\ \hline
    	\multirow{6}{*}{$\alpha_{0, 1}$} & \multirow{3}{*}{log-linear} &  $5 \times 5$ & {0.151} & {0.108} & {0.075} & {0.064} & {0.062} \\
    	                                 &                             &  $7 \times 7$  & {0.149} & {0.096} & {0.069} & {0.052} & {0.050} \\ 
    	                                 &                             &  $9 \times 9$  & {0.196} & {0.116} & 0.066 & 0.065 & 0.051 \\
    	       \cline{2-8}        &   \multirow{3}{*}{linear}   &  $5 \times 5$  & {0.047} & {0.035} & 0.038 & {0.040} & {0.035} \\
    	                                 &                             &  $7 \times 7$  & {0.054} & {0.035} & {0.037} & {0.034} & {0.041} \\ 
    	                                 &                             &  $9 \times 9$  & {0.046} & 0.043 & 0.048 & 0.037 & 0.042 \\ \hline
    	\multirow{6}{*}{$\beta_{0, 1}$}  & \multirow{3}{*}{log-linear} &  $5 \times 5$  & {0.128} & {0.093} & {0.081} & {0.056} & {0.051} \\
    	                                 &                             &  $7 \times 7$  & {0.131} & {0.101} & {0.085} & {0.049} & {0.057} \\ 
    	                                 &                             &  $9 \times 9$  & {0.145} & 0.115 & 0.074 & 0.058 & 0.046 \\
    	       \cline{2-8}        &   \multirow{3}{*}{linear}   &  $5 \times 5$  & {0.043} & {0.043} & {0.053} & {0.057} & {0.055} \\
    	                                 &                             &  $7 \times 7$  & {0.044} & {0.044} & {0.0046} & {0.070} & {0.057} \\ 
    	                                 &                             &  $9 \times 9$  & 0.046 & 0.042 & 0.037 & 0.058 & 0.045 \\ \hline
    \end{tabular}
    \caption{Empirical size of the asymptotic Wald test for different parameters and different numbers of locations.}
    \label{tab:empirical_size}
\end{table}

\begin{table}[tbp]
    \centering
    
    \begin{tabular}{|c|ccccccc|}
    	\hline
      \multirow{2}{*}{Parameter} & \multirow{2}{*}{Model}& \multirow{2}{*}{Gridsize} & \multicolumn{5}{c|}{$T$} \\ \cline{4-8}
    %	           Parameter             &            Model             & Gridsize &  
    & & & 5   &  10  &  20  &  50  &  100  \\ \hline
    	\multirow{10}{*}{$\gamma_{0, 1}$} & \multirow{5}{*}{log-linear} &  $10 \times 10$ & 0.211 & 0.111 & 0.091 & 0.063 & 0.057 \\ 
                                                                        && $20 \times 20$ & 0.238 & 0.141 & 0.106 & 0.061 & 0.051 \\ 
                                                                        && $30 \times 30$ & 0.190 & 0.153 & 0.094 & 0.069 & 0.067 \\ 
                                                                        && $40 \times 40$ & 0.193 & 0.175 & 0.109 & 0.073 & 0.051 \\ 
                                                                        && $50 \times 50$ & 0.196 & 0.134 & 0.122 & 0.067 & 0.064 \\ 
    	       \cline{2-8}                 &   \multirow{5}{*}{linear}   &  $10 \times 10$ & 0.122 & 0.107 & 0.079 & 0.070 & 0.057 \\ 
                                                                        && $20 \times 20$ & 0.149 & 0.109 & 0.072 & 0.057 & 0.054 \\ 
                                                                        && $30 \times 30$ & 0.108 & 0.129 & 0.070 & 0.069 & 0.061 \\ 
                                                                        && $40 \times 40$ & 0.111 & 0.108 & 0.082 & 0.086 & 0.080 \\ 
                                                                        && $50 \times 50$ & 0.114 & 0.129 & 0.087 & 0.075 & 0.059 \\ \hline

    	\multirow{10}{*}{$\alpha_{0, 1}$} & \multirow{5}{*}{log-linear} &  $10 \times 10$ & 0.232 & 0.319 & 0.296 & 0.156 & 0.110 \\ 
                                                                        && $20 \times 20$ & 0.289 & 0.350 & 0.285 & 0.177 & 0.110 \\ 
                                                                        && $30 \times 30$ & 0.282 & 0.337 & 0.301 & 0.202 & 0.135 \\ 
                                                                        && $40 \times 40$ & 0.317 & 0.354 & 0.302 & 0.205 & 0.119 \\ 
                                                                        && $50 \times 50$ & 0.313 & 0.347 & 0.294 & 0.204 & 0.134 \\ 
    	       \cline{2-8}                 &   \multirow{5}{*}{linear}   &  $10 \times 10$ & 0.279 & 0.222 & 0.134 & 0.071 & 0.048 \\
                                                                        && $20 \times 20$ & 0.284 & 0.240 & 0.149 & 0.089 & 0.061 \\ 
                                                                        && $30 \times 30$ & 0.344 & 0.254 & 0.158 & 0.087 & 0.050 \\ 
                                                                        && $40 \times 40$ & 0.357 & 0.265 & 0.167 & 0.082 & 0.067 \\ 
                                                                        && $50 \times 50$ & 0.345 & 0.273 & 0.162 & 0.066 & 0.051 \\ \hline
    	\multirow{10}{*}{$\beta_{0, 1}$}  & \multirow{5}{*}{log-linear} &  $10 \times 10$ & 0.216 & 0.205 & 0.179 & 0.138 & 0.103 \\ 
                                                                        && $20 \times 20$ & 0.223 & 0.237 & 0.209 & 0.176 & 0.104 \\ 
                                                                        && $30 \times 30$ & 0.193 & 0.210 & 0.219 & 0.165 & 0.128 \\ 
                                                                        && $40 \times 40$ & 0.200 & 0.211 & 0.194 & 0.165 & 0.115 \\ 
                                                                        && $50 \times 50$ & 0.208 & 0.238 & 0.209 & 0.166 & 0.123 \\
    	       \cline{2-8}                 &   \multirow{5}{*}{linear}   &  $10 \times 10$ & 0.006 & 0.032 & 0.038 & 0.043 & 0.057 \\ 
                                                                        && $20 \times 20$ & 0.006 & 0.026 & 0.037 & 0.055 & 0.059 \\ 
                                                                        && $30 \times 30$ & 0.007 & 0.028 & 0.037 & 0.045 & 0.067 \\ 
                                                                        && $40 \times 40$ & 0.010 & 0.023 & 0.028 & 0.053 & 0.071 \\ 
                                                                        && $50 \times 50$ & 0.004 & 0.022 & 0.027 & 0.056 & 0.059 \\ \hline
    \end{tabular}
    \caption{Empirical size of the asymptotic Wald test for different parameters and different numbers of locations in high-dimensonal settings with few observation times.}
    \label{tab:empirical_size_high}
\end{table}

\begin{table}[tbp]
    \centering
    
    \begin{tabular}{|c|ccccccc|}
    	\hline
      \multirow{2}{*}{Parameter} & \multirow{2}{*}{Model}& \multirow{2}{*}{Gridsize} & \multicolumn{5}{c|}{$T$} \\ \cline{4-8}
    %	           Parameter             &            Model             & Gridsize &  
    & & & 5   &  10  &  20  &  50  &  100  \\ \hline
    	\multirow{10}{*}{$\gamma_{0, 1}$} & \multirow{5}{*}{log-linear} &  $10 \times 10$ & 0.222 & 0.126 & 0.096 & 0.072 & 0.060 \\ 
                                                                        && $20 \times 20$ & 0.250 & 0.122 & 0.093 & 0.070 & 0.053 \\ 
                                                                        && $30 \times 30$ & 0.210 & 0.143 & 0.096 & 0.056 & 0.051 \\ 
                                                                        && $40 \times 40$ & 0.222 & 0.156 & 0.100 & 0.073 & 0.066 \\ 
                                                                        && $50 \times 50$ & 0.227 & 0.149 & 0.087 & 0.059 & 0.048 \\ 
    	       \cline{2-8}                 &   \multirow{5}{*}{linear}   &  $10 \times 10$ & 0.133 & 0.094 & 0.086 & 0.063 & 0.055 \\ 
                                                                        && $20 \times 20$ & 0.145 & 0.105 & 0.085 & 0.060 & 0.054 \\ 
                                                                        && $30 \times 30$ & 0.153 & 0.133 & 0.078 & 0.065 & 0.051 \\ 
                                                                        && $40 \times 40$ & 0.119 & 0.118 & 0.084 & 0.062 & 0.063 \\ 
                                                                        && $50 \times 50$ & 0.129 & 0.112 & 0.072 & 0.051 & 0.047 \\ \hline
    	\multirow{10}{*}{$\beta_{0, 1}$}  & \multirow{5}{*}{log-linear} &  $10 \times 10$ & 0.299 & 0.176 & 0.117 & 0.073 & 0.064 \\ 
                                                                        && $20 \times 20$ & 0.329 & 0.192 & 0.111 & 0.074 & 0.069 \\ 
                                                                        && $30 \times 30$ & 0.310 & 0.185 & 0.123 & 0.063 & 0.061 \\ 
                                                                        && $40 \times 40$ & 0.320 & 0.165 & 0.119 & 0.083 & 0.068 \\ 
                                                                        && $50 \times 50$ & 0.333 & 0.210 & 0.125 & 0.086 & 0.070 \\
    	       \cline{2-8}                 &   \multirow{5}{*}{linear}   &  $10 \times 10$ & 0.060 & 0.044 & 0.061 & 0.049 & 0.045 \\ 
                                                                        && $20 \times 20$ & 0.060 & 0.042 & 0.030 & 0.044 & 0.049 \\ 
                                                                        && $30 \times 30$ & 0.050 & 0.039 & 0.032 & 0.054 & 0.054 \\ 
                                                                        && $40 \times 40$ & 0.055 & 0.047 & 0.047 & 0.050 & 0.053 \\ 
                                                                        && $50 \times 50$ & 0.050 & 0.039 & 0.037 & 0.043 & 0.061 \\ \hline
    \end{tabular}
    \caption{Empirical size of the asymptotic Wald test for different parameters and different numbers of locations in high-dimensonal settings with few observation times for processes without feedbackterm.}
    \label{tab:empirical_size_high_auto}
\end{table}

\begin{figure}[tb]
    \centering
    \begin{subfigure}{0.45\textwidth}
        \includegraphics[keepaspectratio, width = 1\textwidth]{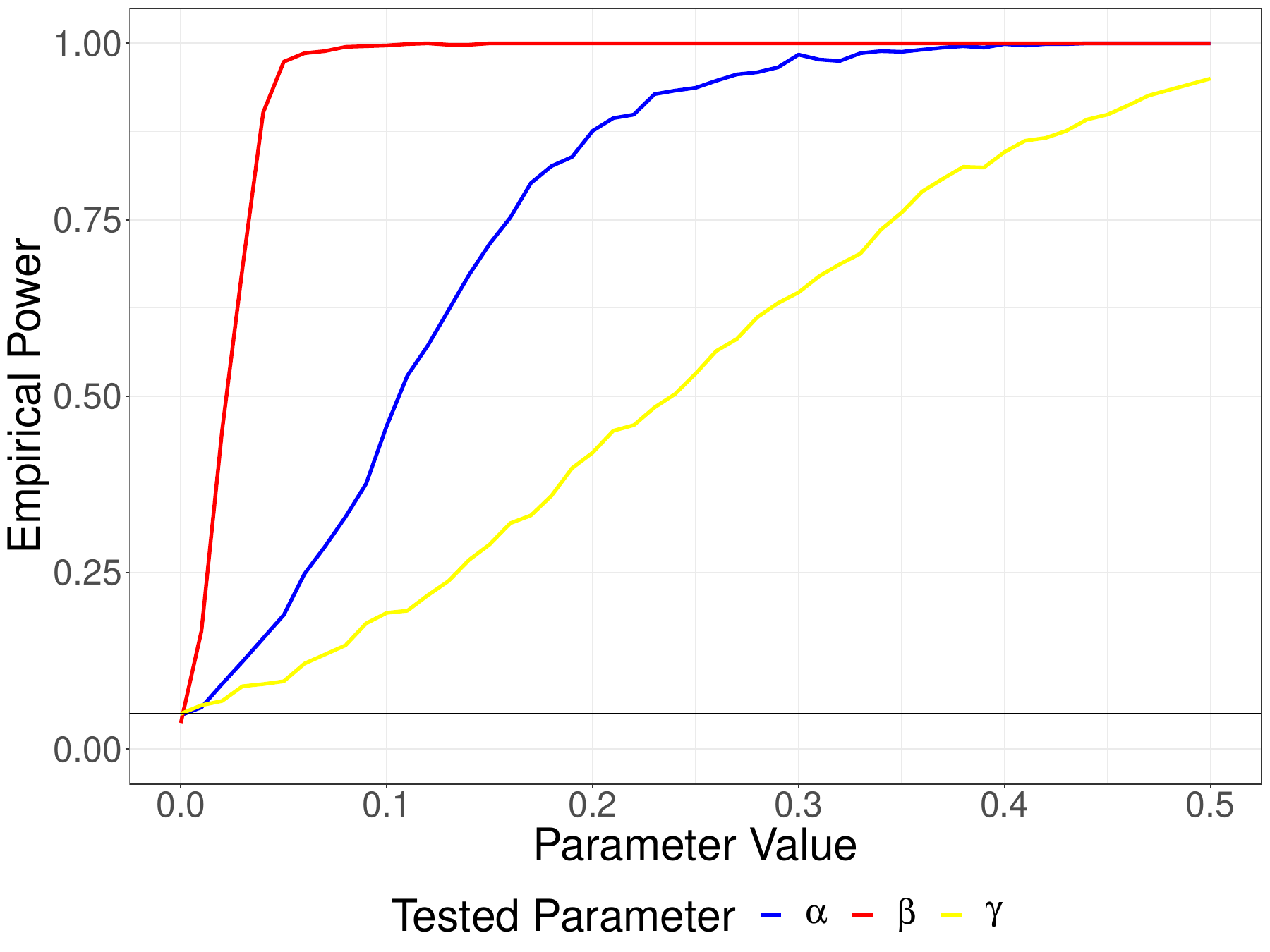}
        \caption{linear}
        \label{fig:power_linear}
    \end{subfigure}
    \hfill
    \begin{subfigure}{0.45\textwidth}
         \includegraphics[keepaspectratio, width = 1\textwidth]{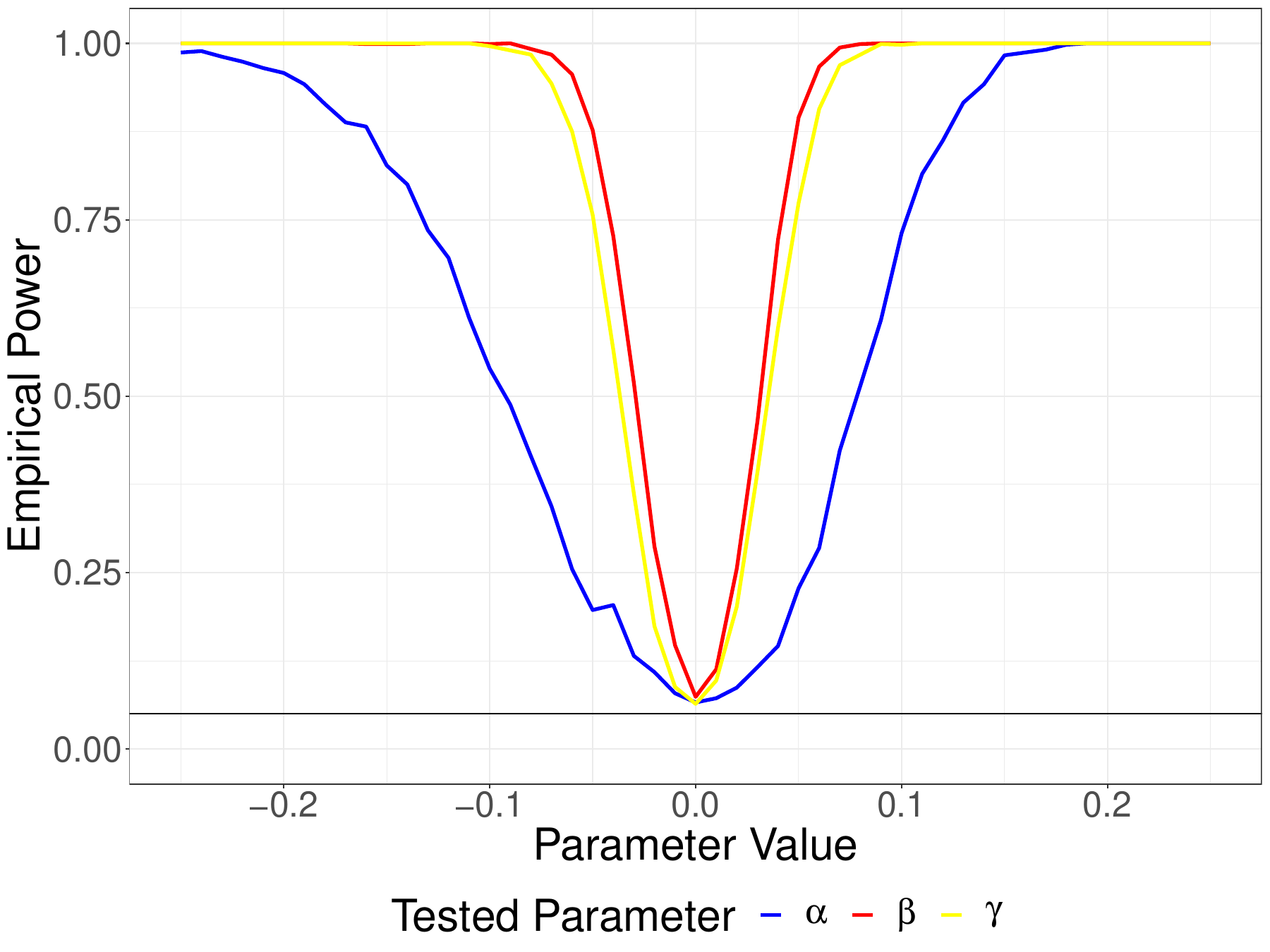}
        \caption{log-linear}
        \label{fig:power_log}
    \end{subfigure}
    \caption{Power curves of PSTARMAX process on a $9 \times 9$ grid with 250 observation times.}
    \label{fig:power_curve}
\end{figure}

{
Summarizing, asymptotic approximations are adequate for models including the feedback process, provided that many observations are available, i.e. $T$ increases. In addition detecting significance for the feedback process is a more complex task than the detection of a significant lagged value of $\bm{Y}_t$. This is also evident from power analysis.
}
Figures~\ref{fig:power_linear} and \ref{fig:power_log} show the results on the power in case of $T = 250$ for the linear and log-linear models on the $9 \times 9$ grid. For the log-linear model, deviations of $\beta_{0, 1}$ and $\gamma_{0, 1}$ from 0 are detected equally well and better than a deviation of $\alpha_{0, 1}$. In case of the  linear model, a deviation of $\alpha_{0, 1}$ from 0 is detected less efficiently than a deviation of $\beta_{0, 1}$, similar to the log-linear model. However, the power of the test for the covariate parameter is lower than in the log-linear model.  In the linear model, covariates affect the expected value linearly, while in the log-linear model, the effect is multiplicative. Hence, the observed differences between the power functions for covariate effects are consistent with the underlying model structure.

\begin{figure}[tb]
    \centering
    \begin{subfigure}{0.45\textwidth}
        \includegraphics[keepaspectratio, width = 1\textwidth]{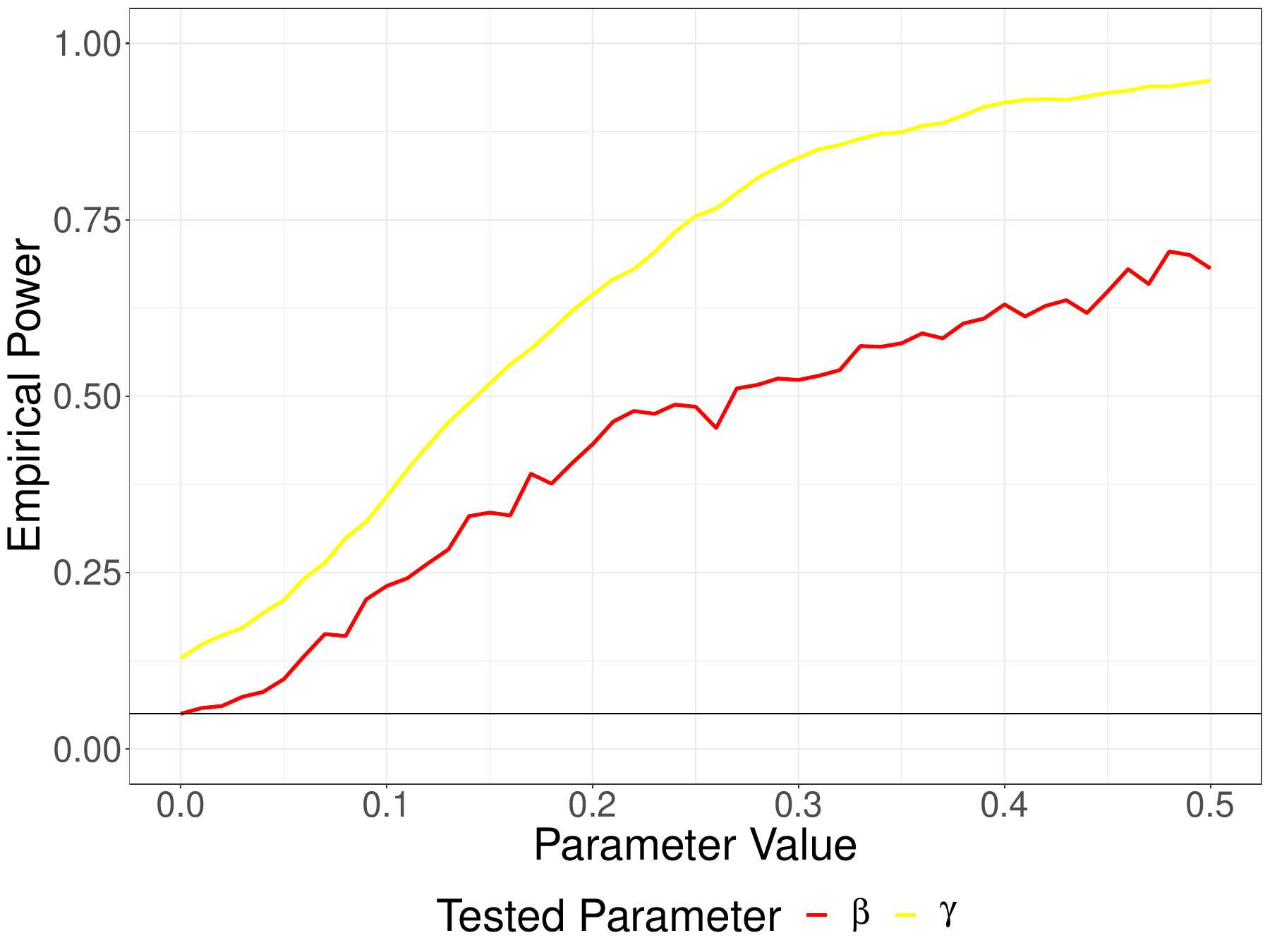}
        \caption{linear}
        \label{fig:power_linear_auto}
    \end{subfigure}
    \hfill
    \begin{subfigure}{0.45\textwidth}
         \includegraphics[keepaspectratio, width = 1\textwidth]{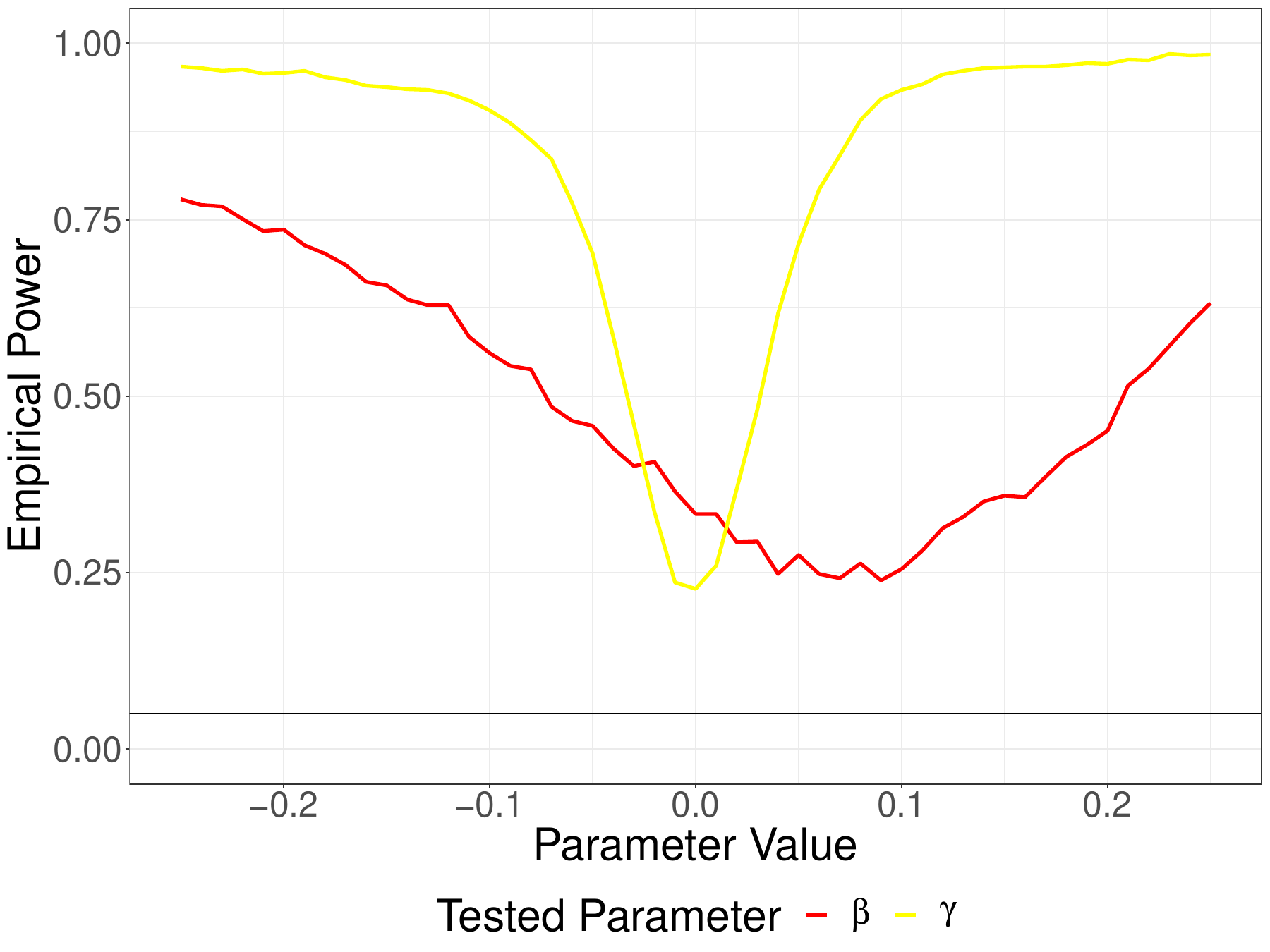}
        \caption{log-linear}
        \label{fig:power_log_auto}
    \end{subfigure}
    \caption{Power curves of PSTARMAX process without regression on the feedback process on a $50 \times 50$ grid with 5 observation times.}
    \label{fig:power_curve_auto}
\end{figure}

If $T$ is small and the observations are take in a large grid, then there should be a strong autoregressive effect in order to detect it reliably. Figures~\ref{fig:power_linear_auto} and \ref{fig:power_log_auto} show the power curves for the parameters $\beta_{0, 1}$ and $\gamma_{0, 1}$ in linear and log-linear PSTARMAX processes without feedback term on a $50 \times 50$ grid for $T = 5$. It is easy to see that the power for both parameters is lower than in the lower dimensional case with feedback term in Figures~\ref{fig:power_linear} and \ref{fig:power_log}.

{
Table~\ref{tab:fitting_time_small} shows the mean calculation times for the models of Table~\ref{tab:empirical_size}. As discussed in Section~\ref{sim:implementation} of the main text, the calculation time is short and increases with the number of observation times and grid size.
}

\begin{table}[tb]
		\centering
		
		\begin{tabular}{|c|cccccc|}
			\hline
			\multirow{2}{*}{Model} & \multirow{2}{*}{Gridsize} & \multicolumn{5}{c|}{$T$} \\ \cline{3-7}
			%	           Parameter             &            Model             & Gridsize &  
			& & 50 & 100   &  250  &  500  &  750  \\ \hline
			\multirow{3}{*}{linear}     & $5 \times 5$ & 0.012 & 0.024 & 0.063 & 0.135 & 0.209 \\
			                            & $7 \times 7$ & 0.022 & 0.045 & 0.121 & 0.262 & 0.409 \\
			                            & $9 \times 9$ & 0.035 & 0.074 & 0.206 & 0.466 & 0.736 \\ \hline
			\multirow{3}{*}{log-linear} & $5 \times 5$ & 0.020 & 0.039 & 0.093 & 0.194 & 0.306 \\
			                            & $7 \times 7$ & 0.036 & 0.071 & 0.179 & 0.388 & 0.593 \\
			                            & $9 \times 9$ & 0.058 & 0.111 & 0.307 & 0.703 & 1.015 \\ \hline
		\end{tabular}
	\caption{Mean computation time (in seconds) of PSTARMAX processes for different grid sizes and observation times including regression on the feedback process}
	\label{tab:fitting_time_small}
\end{table}

\subsubsection{Copula} \label{sim:copula}
In the simulation section~\ref{sec:simulation} in the main text, we employed the Clayton copula with a parameter value of 2, following the data generation process proposed by \textcite{fokianos_multivariate_2020}. This study broadens the scope to explore the impact of varying copulas and their parameters on parameter estimation accuracy. To assess the outcomes, we calculated the MSE of parameter estimation, bias (i.e., $\bm{\theta} - \bm{\theta}_{\text{true}}$), and the Mean Absolute Error (MAE) for the data. This is defined as
\begin{align}
    \text{MAE} = \frac{1}{pT} \sum_{t = 1}^T \sum_{i = 1}^p |y_{i, t} - \lambda_{i, t}(\bm{\hat{\theta}})|,
\end{align}
with $p$ the dimension of the time series, i.e. the number of locations.

\begin{figure}[tb]
    \centering
    \begin{subfigure}[t]{\textwidth}
        \centering
         \includegraphics[width=0.85\textwidth]{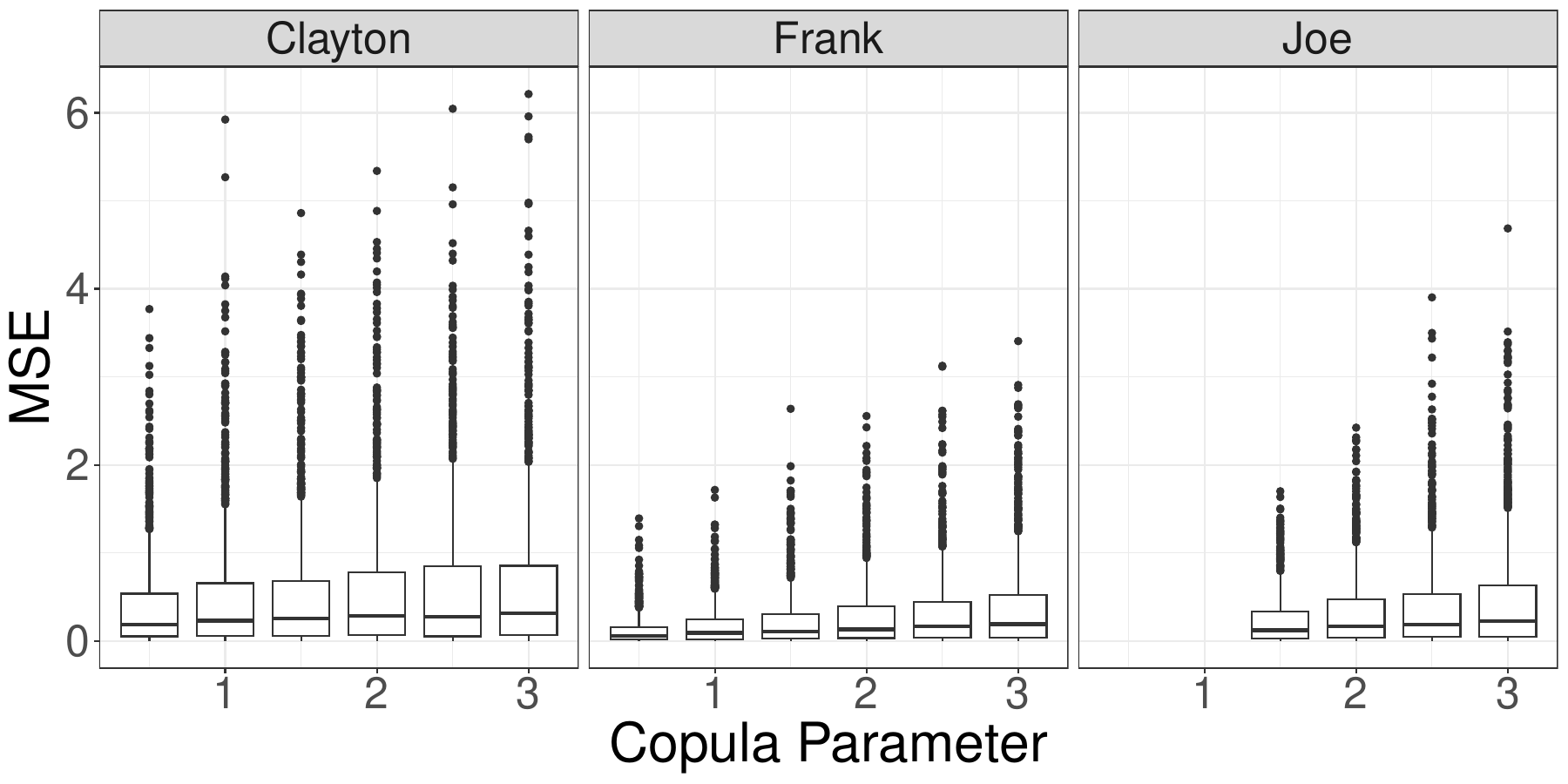} 
    \caption{linear}
    \label{fig:copula_linear}
    \end{subfigure}
    \hfill
    \begin{subfigure}[t]{\textwidth}
    \centering
        \includegraphics[width=0.85\textwidth]{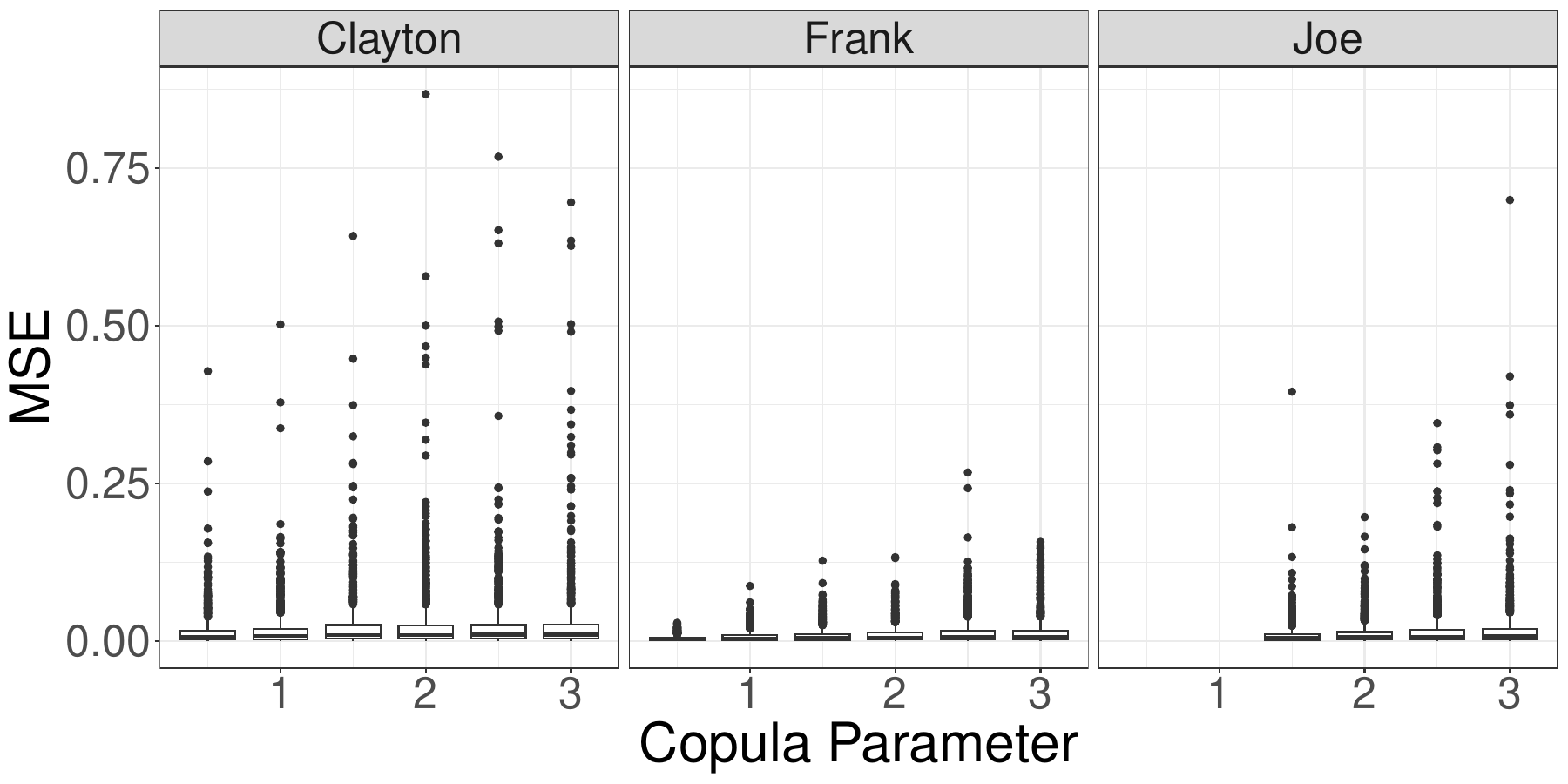} 
    \caption{log-linear}
    \label{fig:copula_log}
    \end{subfigure}
    \caption{MSE of parameter estimation in different copula settings for $T = 250$ observation times.}
    \label{fig:copula_mse}
\end{figure}

It can be seen that all copulas result in higher MSE values with increasing copula parameters and that these also have a higher variance. Table~\ref{tab:copula_bias} shows mean bias values for the parameters in the linear setting for $T = 250$ in the case of the Clayton copula. It can generally be seen that the bias increases slightly as the copula parameter increases. However, the mean squared bias is very small compared to the mean MSE, so that the increasing MSEs are primarily caused by a higher variance of the parameter estimation and the data. For reference, the values in the case of contemporaneous independence are also listed in the table. They are considerably smaller.

\begin{table}[ht]
\centering
\begin{tabular}{|c|ccccc|ccc|}
  \hline
  Copula & $\delta_0$ & $\alpha_{0, 1}$ & $\alpha_{1, 1}$ & $\beta_{0, 1}$ & $\beta_{1, 1}$ & $\text{Bias}^2$ & MSE & Variance  \\ 
  \hline
 0.5 & 0.3556 & -0.0323 & 0.0050 & 0.0008 & -0.0025 & 0.0255 & 0.3959 & 0.3704  \\ 
   1.0 & 0.3855 & -0.0378 & 0.0115 & -0.0001 & -0.0041 & 0.0300 & 0.5210 & 0.4910  \\ 
   1.5 & 0.4317 & -0.0404 & 0.0084 & -0.0004 & -0.0031 & 0.0376 & 0.5500 & 0.5124  \\ 
   2.0 & 0.4919 & -0.0415 & 0.0092 & -0.0011 & -0.0055 & 0.0488 & 0.6057 & 0.5570  \\ 
   2.5 & 0.4848 & -0.0436 & 0.0125 & -0.0010 & -0.0062 & 0.0474 & 0.6326 & 0.5852  \\ 
   3.0 & 0.4455 & -0.0459 & 0.0132 & -0.0003 & -0.0037 & 0.0402 & 0.6614 & 0.6213  \\ \hline
   Indep. & 0.1218 & -0.0176 & 0.0070 & 0.0007 & 0.0001 & 0.0030 & 0.0652 & 0.0622  \\
   \hline
\end{tabular}
\caption{Bias, MSE and Variance of the QMLE in the linear setting in the case of the Clayton copula for different copula parameters at $T = 250$.}
\label{tab:copula_bias}
\end{table}

\begin{figure}[tb]
    \centering
    \begin{subfigure}[t]{0.85\textwidth}
    \centering
         \includegraphics[width=1\textwidth]{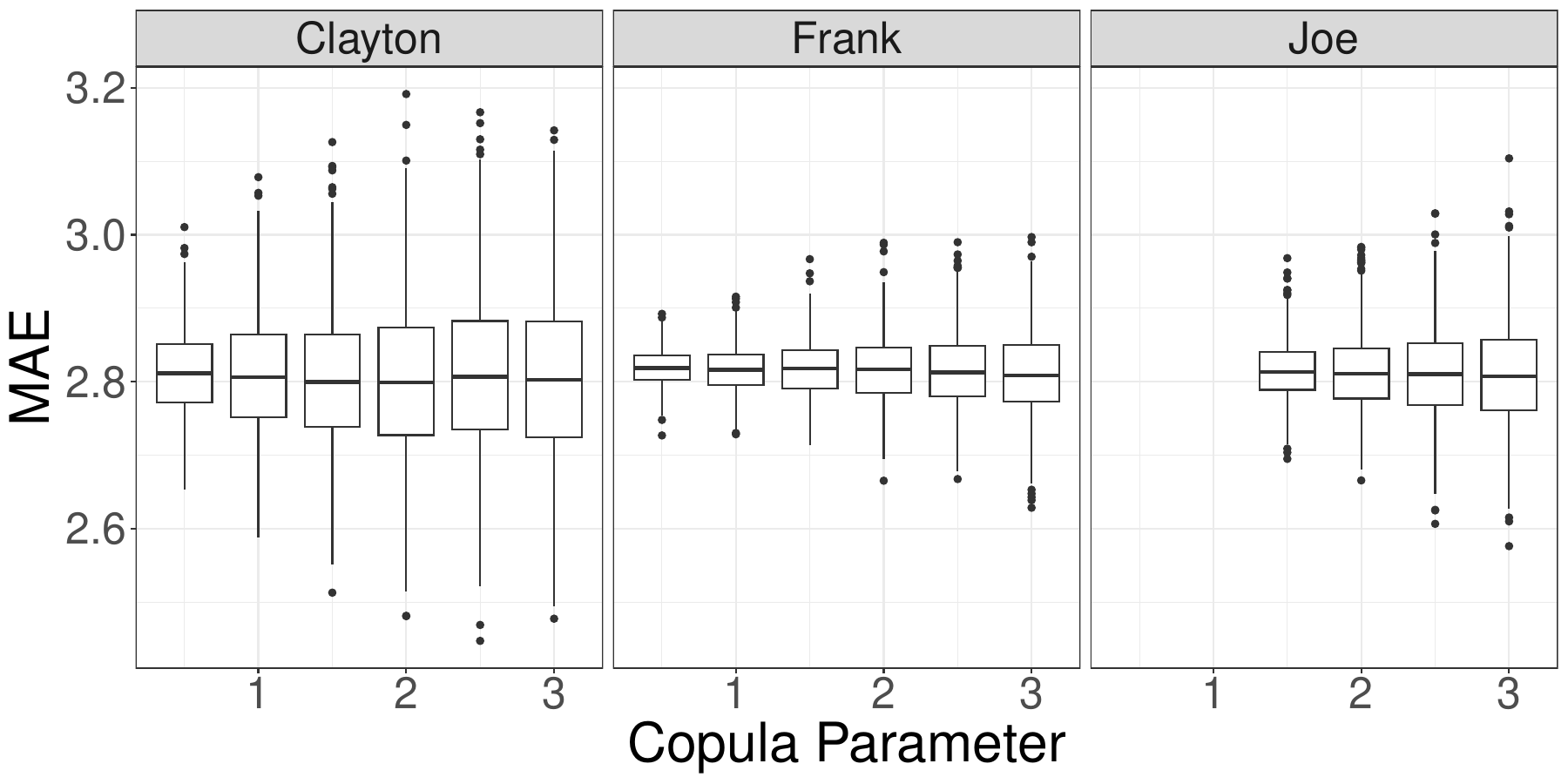} 
    \caption{linear}
    \label{fig:copula_linear_mae}
    \end{subfigure}
    \hfill
    \begin{subfigure}[t]{0.85\textwidth}
    \centering
        \includegraphics[width=1\textwidth]{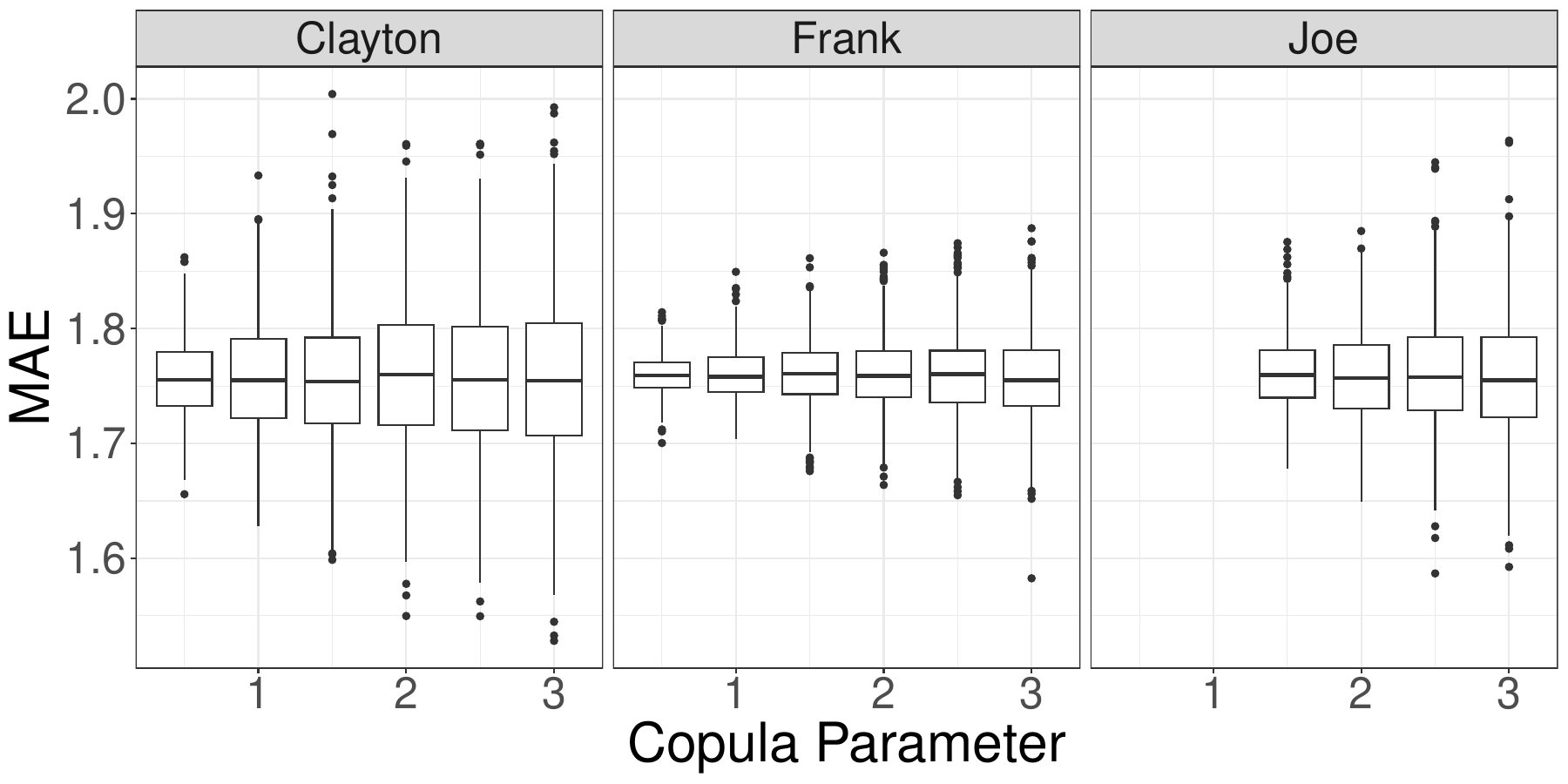} 
    \caption{log-linear}
    \label{fig:copula_log_mae}
    \end{subfigure}
    \caption{MAE in data for different copula settings for $T = 250$ observation times.}
    \label{fig:copula_mae}
\end{figure}

The results are also evident in the MAE in the data. Figure~\ref{fig:copula_mae} shows boxplots of the MAEs from the simulations. These are at roughly the same level, but also show a higher variance for higher copula parameters. Overall, it can be seen that stronger contemporaneous dependencies between the components due to different copulas do not lead to a strongly biased estimate, but that there is more variability as a result.

\subsubsection{Intercept}
For the simulations in Section~\ref{sec:simulation}, only a homogeneous intercept structure was used. This has the benefit of fast computational time, as only a few parameters need to be estimated and it also allows the inclusion of time-constant covariates. In other words demographic effects, such as population size, can also be captured. Such information is not always available. In this study, we investigate the effects of estimating a model with a homogeneous intercept structure even though the data are inhomogeneous.

\begin{table}[ht]
\centering
\begin{tabular}{|l|c|rrrr|rrrr|}
  \hline
  \multirow{2}{*}{Model} & \multirow{2}{*}{$T$} & \multicolumn{4}{c}{inhomogenous} & \multicolumn{4}{|c|}{homogenous} \\
   &  & $\alpha_{0, 1}$ & $\alpha_{1, 1}$ & $\beta_{0, 1}$ & $\beta_{1, 1}$ & $\alpha_{0, 1}$ & $\alpha_{1, 1}$ & $\beta_{0, 1}$ & $\beta_{1, 1}$ \\ 
  \hline
\multirow{4}{*}{linear} & 50 & -0.140 & 0.056 & -0.033 & 0.001 & 0.240 & 0.030 & -0.006 & 0.029 \\ 
    & 100 & -0.108 & 0.048 & -0.013 & 0.001 & 0.306 & -0.001 & -0.012 & 0.022 \\ 
    & 250 & -0.058 & 0.030 & -0.004 & -0.002 & 0.368 & -0.026 & -0.024 & 0.014 \\ 
    & 500 & -0.029 & 0.012 & -0.002 & -0.001 & 0.388 & -0.042 & -0.027 & 0.017 \\ \hline
   \multirow{4}{*}{log-linear} & 50 & -0.174 & -0.075 & -0.043 & -0.005 & 0.297 & 0.000 & -0.012 & 0.049 \\ 
    & 100 & -0.118 & -0.018 & -0.018 & 0.001 & 0.385 & -0.024 & -0.024 & 0.032 \\ 
    & 250 & -0.057 & -0.002 & -0.006 & 0.002 & 0.456 & -0.046 & -0.040 & 0.021 \\ 
    & 500 & -0.028 & 0.004 & -0.003 & 0.000 & 0.493 & -0.061 & -0.051 & 0.014 \\ 
   \hline
\end{tabular}
\caption{Bias for non-intercept parameters of models with inhomogenous and homogenous intercept structure}
\label{tab:bias_intercept}
\end{table}

Table~\ref{tab:bias_intercept} shows the bias of the estimation of the two models for the non-intercept parameters. While the bias for the correctly specified model converges towards 0 as the number of observation points increases, there is a constant bias in the homogeneous model. For the parameters $\beta_{0, 1}$ and $\beta_{1, 1}$ this is relatively low overall, $\beta_{0,1}$ is slightly underestimated on average (-0.0236 for $T = 250$ in the linear setting) and $\beta_{1, 1}$ tends to be overestimated (0.0141 for $T = 250$ in the linear setting). A strong bias can be observed for the parameter $\alpha_{0, 1}$. This results in a bias of 0.3057 for $T = 100$ in the linear setting, which increases to 0.3879 for $T = 500$. The mean bias for the parameter $\alpha_{1, 1}$ is very low compared to this and the parameter tends to be slightly underestimated.

This indicates that the model utilises the feedback process $\lbrace \bm{\lambda}_t \rbrace$ to compensate for the misspecification of the intercept. However, the spatial dependencies still seem to be captured fairly well by the model.

\begin{figure}[tb]
    \centering
    \begin{subfigure}[t]{0.75\textwidth}
         \includegraphics[width=1\textwidth]{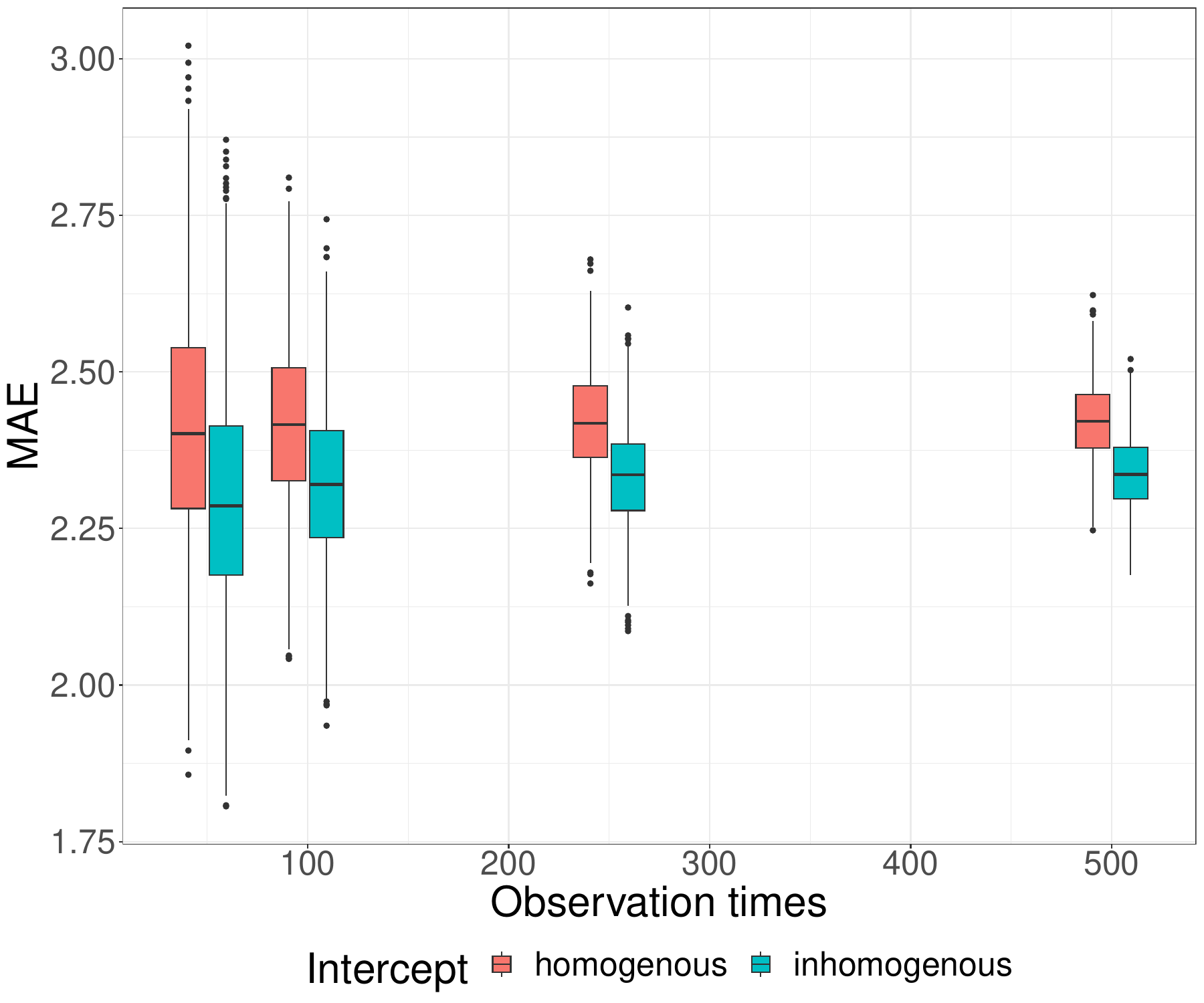} 
    \caption{linear}
    \label{fig:intercept_linear_mae}
    \end{subfigure}
    \hfill
    \begin{subfigure}[t]{0.75\textwidth}
        \includegraphics[width=1\textwidth]{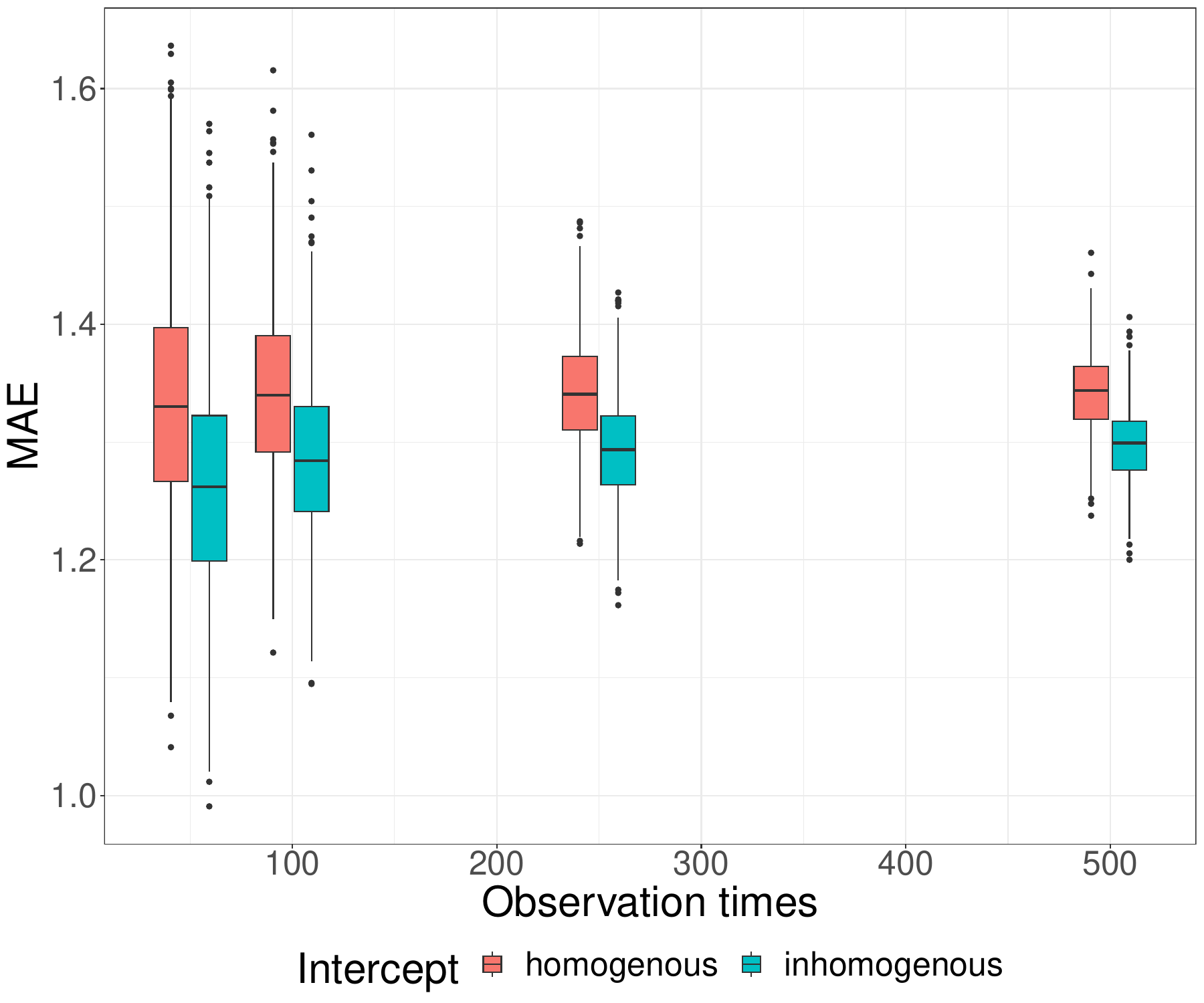} 
    \caption{log-linear}
    \label{fig:intercept_log_mae}
    \end{subfigure}
    \caption{MAE in data for different copula settings for $T = 250$ observation times.}
    \label{fig:intercept_mae}
\end{figure}

This overall bias is also reflected in the mean absolute deviation in the data. These are shown in Figures~\ref{fig:intercept_linear_mae} and \ref{fig:intercept_log_mae}. For a small number of observations, there are still large overlaps between the boxplots of the homogeneous and inhomogeneous models, meaning that the high number of parameters to be estimated in the inhomogeneous model outweigh the bias to a large extent.

\subsubsection{Link}
In the simulations conducted so far, it was assumed that the true model type, i.e. linear or log-linear model, was chosen. In this simulation, we consider the impact if an incorrect model class was selected, i.e. a linear model if a log-linear model is the true model and vice versa.

\begin{table}[tb]
    \centering
    \begin{tabular}{llrrrr}
  \hline
  \multirow{2}{*}{True Model} & \multirow{2}{*}{Parameters} & \multicolumn{4}{c}{$T$} \\
  & & 50 & 100 & 250 & 500 \\ 
  \hline
\multirow{2}{*}{log-linear} & all positive & 0.707 & 0.660 & 0.736 & 0.780 \\ 
 & $\alpha_{0, 1} < 0$ & 0.743 & 0.733 & 0.798 & 0.884 \\ 
linear & all positive & 0.368 & 0.513 & 0.590 & 0.701 \\ 
   \hline
\end{tabular}
    \caption{Relative frequencies favoring a log-linear model over a linear model based on QIC comparison}
    \label{tab:intercept_qic}
\end{table}

Table~\ref{tab:intercept_qic} shows the relative proportions where, based on the QIC, the correct model was chosen. 
It can be seen that the correct model is identified well in most cases based on the QIC. For the log-linear setting with a negative parameter, there is a clear jump in this proportion, i.e. the log-linear model is identified much more frequently than the correct one. This is explained by the fact that the linear model is not flexible enough to capture the data due to the parameter restrictions. This can also be seen in Table~\ref{tab:estimate_link}, which contains the mean estimated parameters of the models. We observe that the linear model often estimates the actually negative parameter $\alpha_{0, 1}$ to values very close to 0. Overall, the very similar mean QIC values in all situations indicate that even misspecified models are able to capture the data relatively well. In practice, care should be taken to consider log-linear models in particular if parameters of the linear model are estimated close to or equal to 0.

\begin{table}[ht]
\centering
\resizebox{\textwidth}{!}{
\begin{tabular}{|c|c|c|ccccc|ccccc|}
  \hline
   \multirow{2}{*}{True Model}& \multirow{2}{*}{Parameters} & \multirow{2}{*}{$T$} & \multicolumn{5}{c}{log-linear} & \multicolumn{5}{|c|}{linear} \\
   &  &  & $\alpha_{0, 1}$ & $\alpha_{1, 1}$ & $\beta_{0, 1}$ & $\beta_{1, 1}$ & QIC &$\alpha_{0, 1}$ & $\alpha_{1, 1}$ & $\beta_{0, 1}$ & $\beta_{1, 1}$ & QIC \\ 
	\hline
	\multirow{4}{*}{linear} & \multirow{4}{*}{all positive}  & 50 & 0.114 & 0.006 & 0.193 & 0.081 &  21324.10 & 0.099 & 0.127 & 0.179 & 0.084 &    21348.15 \\ 
	&  & 100 & 0.152 & 0.038 & 0.203 & 0.092 &  43031.34 & 0.123 & 0.116 & 0.194 & 0.091 &    43036.43 \\ 
	&  & 250 & 0.182 & 0.083 & 0.206 & 0.094 & 107952.99 & 0.156 & 0.115 & 0.199 & 0.094 &   107924.85 \\ 
	&  & 500 & 0.192 & 0.095 & 0.207 & 0.096 & 216235.15 & 0.177 & 0.106 & 0.200 & 0.097 &   216168.35 \\ \hline
	\multirow{8}{*}{log-linear} & \multirow{4}{*}{$\alpha_{0, 1} = -0.2$} & 50 & -0.115 & -0.045 & 0.185 & 0.082 &  14314.99 & 0.024 & 0.116 & 0.117 & 0.069 &    14374.42 \\ 
	&  & 100 & -0.143 & -0.022 & 0.191 & 0.090 &  28805.66 & 0.008 & 0.095 & 0.130 & 0.072 &    28868.96 \\ 
	&  & 250 & -0.170 & 0.031 & 0.197 & 0.096 &  72318.83 & 0.001 & 0.062 & 0.138 & 0.077 &    72415.19 \\ 
	&  & 500 & -0.182 & 0.070 & 0.199 & 0.098 & 144791.65 & 0.000 & 0.035 & 0.139 & 0.080 &   144945.90 \\ \cline{2-13}
	& \multirow{4}{*}{all positive} & 50 & 0.098 & -0.036 & 0.185 & 0.082 &  17516.16 & 0.093 & 0.116 & 0.152 & 0.085 &    17558.87 \\ 
	&  & 100 & 0.136 & 0.000 & 0.195 & 0.097 &  35289.53 & 0.108 & 0.120 & 0.168 & 0.094 &    35326.11 \\ 
	&  & 250 & 0.177 & 0.072 & 0.199 & 0.099 &  88537.88 & 0.144 & 0.113 & 0.173 & 0.099 &    88605.01 \\ 
	&  & 500 & 0.190 & 0.090 & 0.199 & 0.098 & 177117.87 & 0.168 & 0.108 & 0.173 & 0.099 &   177239.26 \\ 
   \hline
\end{tabular}}
\caption{Mean parameter estimates and mean QIC of linear and log-linear models in the link settings.}
\label{tab:estimate_link}
\end{table}

\subsection{Repetition of simulations with a different copula}
In the simulations in Section~\ref{sec:simulation} and \ref{sim:asymptotics}, the Clayton copula with parameter 2 was used in all cases to generate contemporaneous dependencies. To validate the results, we repeated the simulations described in Table~\ref{tab:settings}, but now used the Frank copula with parameter 1 in the data-generating process. Both copulas generate different, not directly comparable, dependency structures. The Frank copula used here tends to lead to weaker contemporaneous dependencies between the components.

The results from Section \ref{sec:simulation} remain consistent overall, with notable differences observed in the Frank copula settings. Particularly, we notice a quicker convergence to the approximate distribution and higher test power, likely due to weaker contemporaneous dependencies that better suit the quasi-likelihood \eqref{eq:qll}. These observations are consistent with those from Section~\ref{sim:copula}, in which it was found that the overall variability increases with increasing contemporaneous dependencies.

\begin{figure}[tb]
    \centering
    \begin{subfigure}[t]{0.6\textwidth}
    \centering
         \includegraphics[width=1\textwidth]{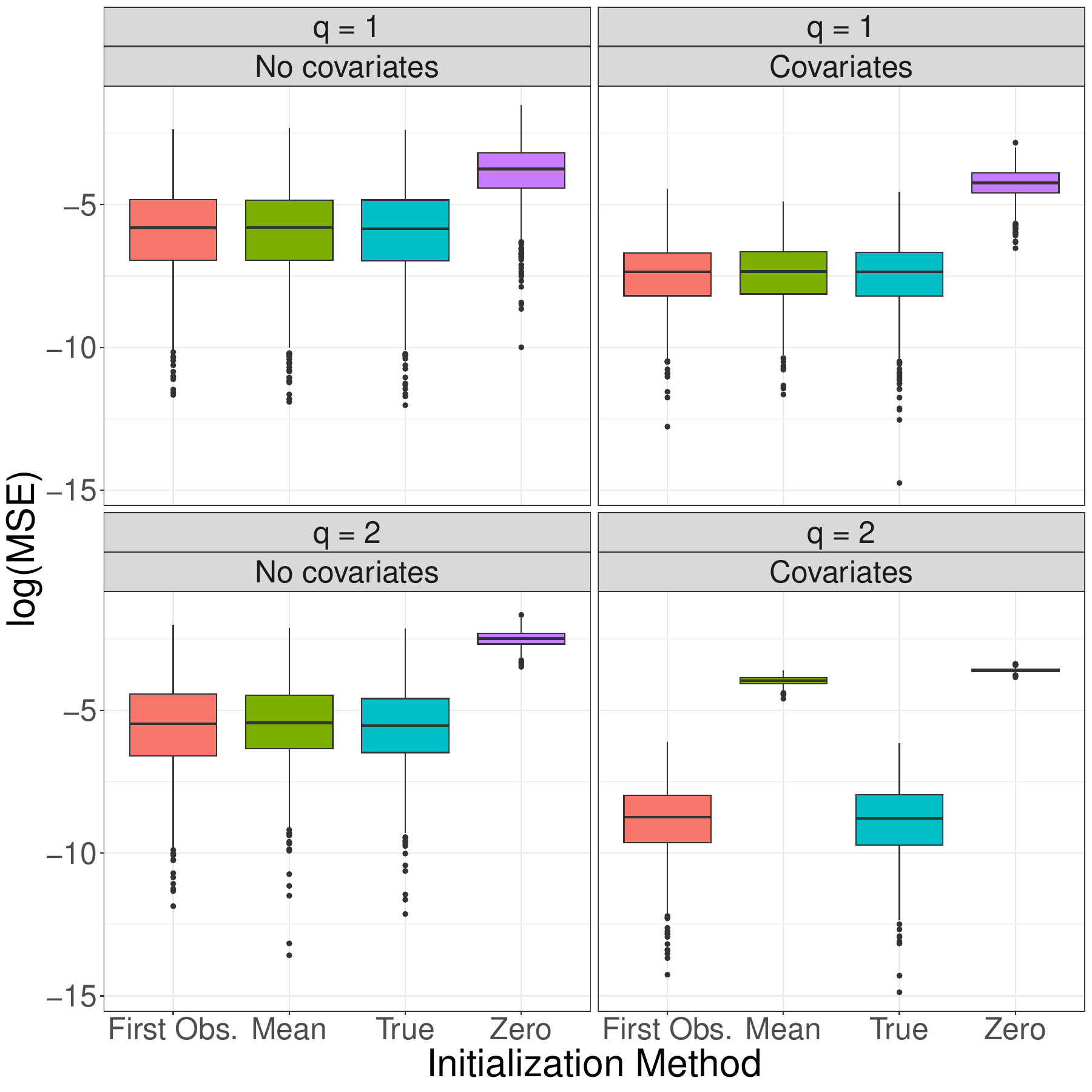} 
    \caption{log-linear}
    \label{fig:init_log_frank}
    \end{subfigure}
    \hfill
    \begin{subfigure}[t]{0.6\textwidth}
    \centering
        \includegraphics[width=1\textwidth]{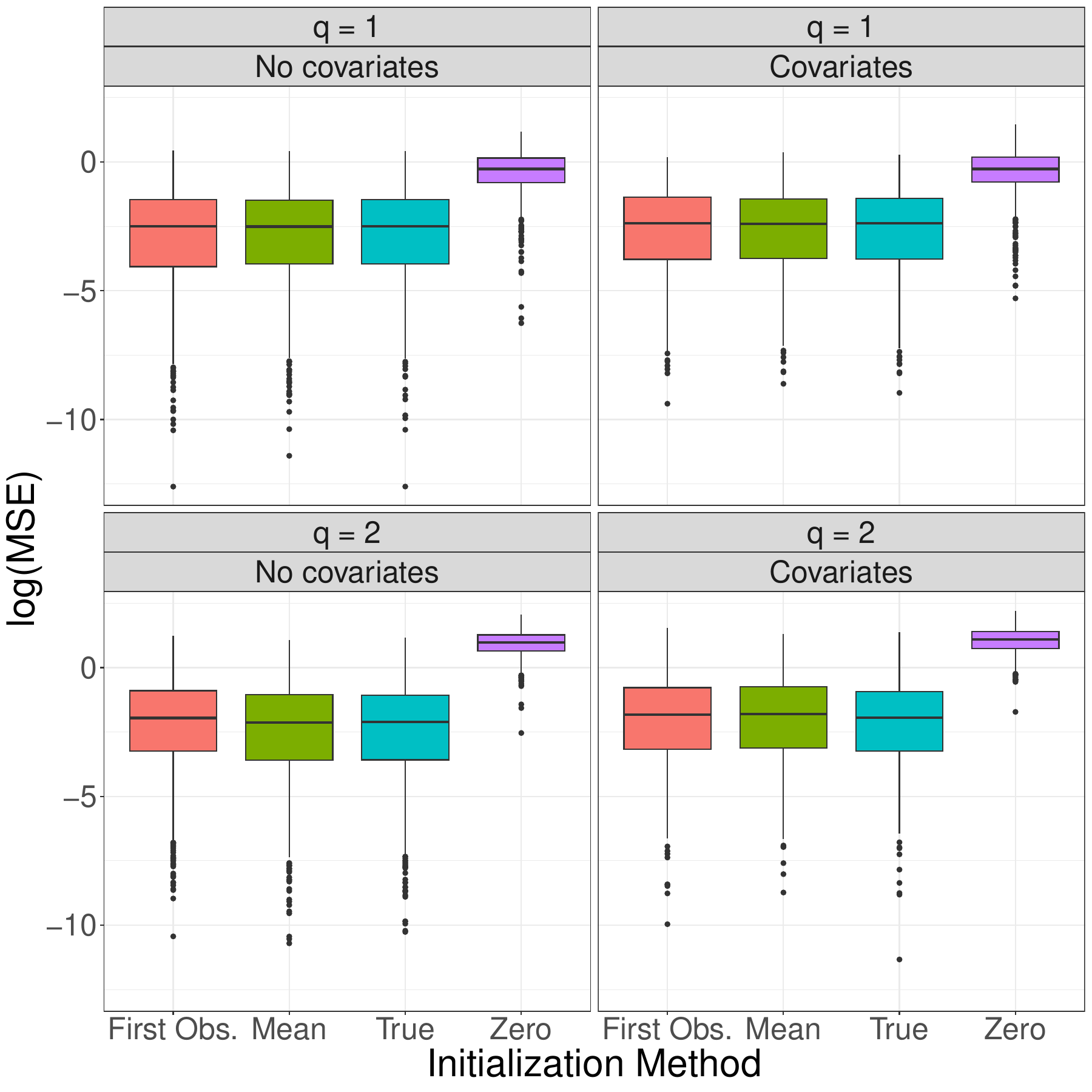} 
    \caption{linear}
    \label{fig:init_linear_frank}
    \end{subfigure}
    \caption{(Log) MSE of QMLE of different initialization methods for the feedback process in the Frank copula setting.}
    \label{fig:init_frank}
\end{figure}

Figures \ref{fig:init_log_frank} and \ref{fig:init_linear_frank} illustrate the behavior of the MSEs under different initializations, showing similar trends to Figures \ref{fig:init_log} and \ref{fig:init_linear}, albeit with slightly lower MSEs in parameter estimation compared to using the Clayton copula for data generation.

\begin{figure}[tb]
    \centering
    \begin{subfigure}{0.45\textwidth}
         \includegraphics[keepaspectratio, width = 1\textwidth]{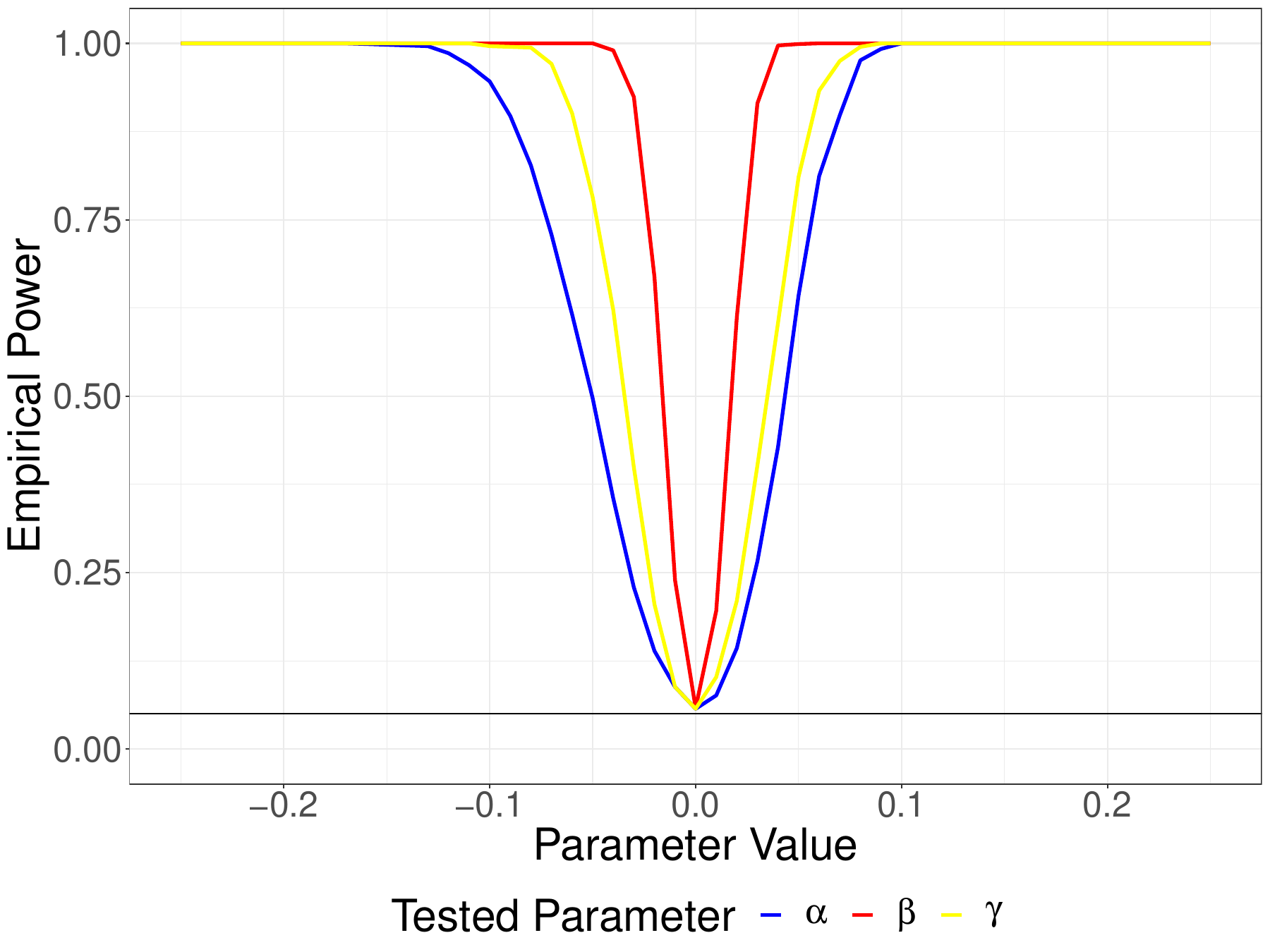}
        \caption{log-linear}
        \label{fig:power_log_frank}
    \end{subfigure}
    \hfill
    \begin{subfigure}{0.45\textwidth}
        \includegraphics[keepaspectratio, width = 1\textwidth]{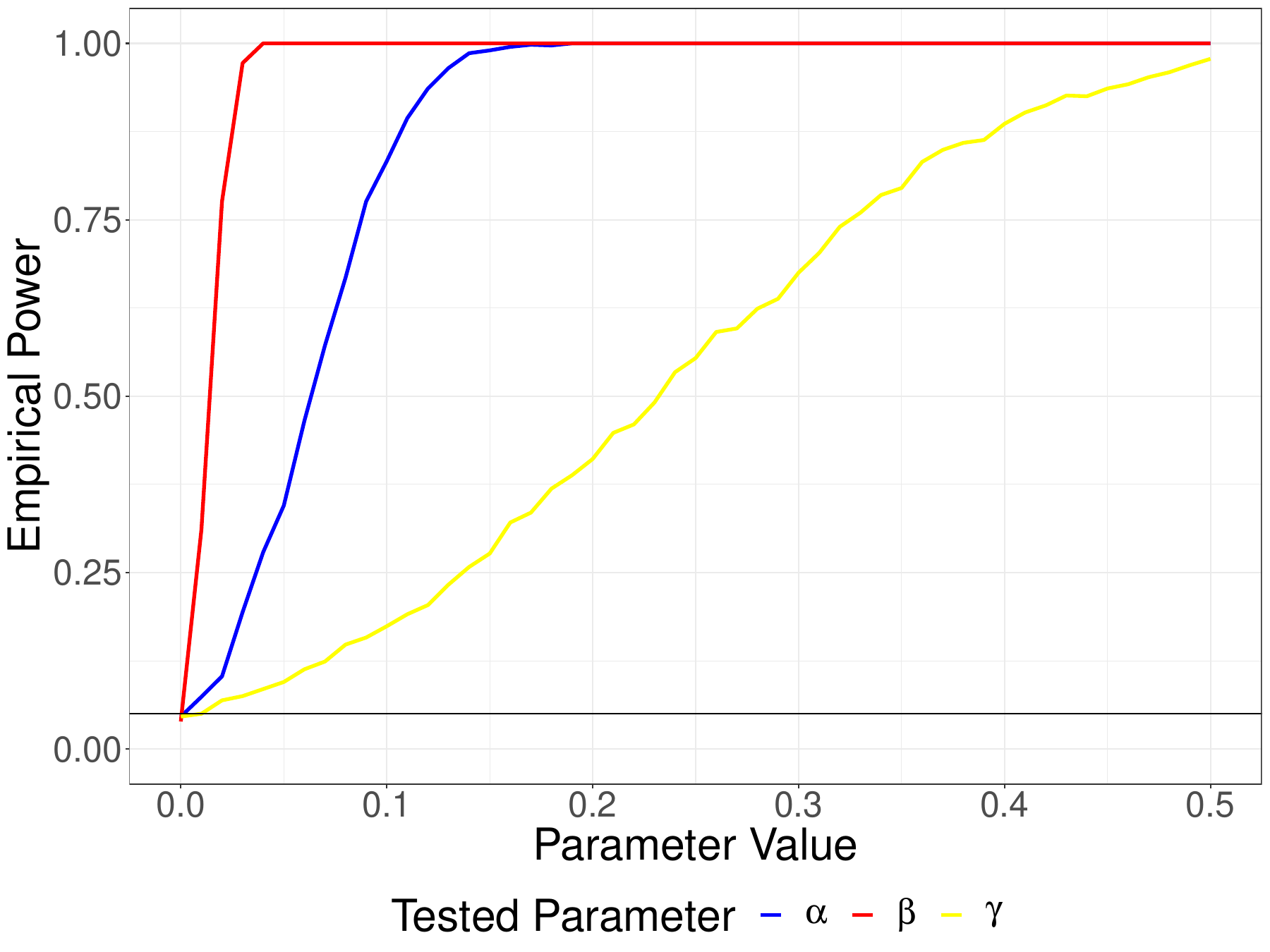}
        \caption{linear}
        \label{fig:power_linear_frank}
    \end{subfigure}
    \caption{Power curves of PSTARMAX process on $9 \times 9$ grid with 250 observation times.}
    \label{fig:power_curve_frank}
\end{figure}

Table~\ref{tab:empirical_size_frank} presents empirical sizes similar to Table~\ref{tab:empirical_size}. Notably, there's a faster convergence to the 5\% significance level, often met at $T = 50$ in most scenarios. Furthermore, power curves in Figures~\ref{fig:power_log_frank} and \ref{fig:power_linear_frank} show consistently higher power, similar to Figures~\ref{fig:power_log} and \ref{fig:power_linear}.
\begin{table}[tb]
    \centering
    \begin{tabular}{|c|ccccccc|}
    	\hline
    	           \multirow{2}{*}{Parameter}             &           \multirow{2}{*}{Model}              & \multirow{2}{*}{Gridsize} &  \multicolumn{5}{c|}{$T$} \\ \cline{4-8}
                &&& 50   &  100  &  250  &  500  &  750  \\ \hline
    	\multirow{6}{*}{$\gamma_{0, 1}$} & \multirow{3}{*}{log-linear} &  $5 \times 5$  & 0.070 & 0.047 & 0.053 & 0.059 & 0.051 \\
    	                                 &                             &  $7 \times 7$  & 0.068 & 0.061 & 0.052 & 0.042 & 0.036 \\
    	                                 &                             &  $9 \times 9$  & 0.061 & 0.060 & 0.057 & 0.039 & 0.052 \\ 
    	       \cline{2-8}        &   \multirow{3}{*}{linear}   &  $5 \times 5$  & 0.067 & 0.054 & 0.056 & 0.042 & 0.047 \\ 
    	                                 &                             &  $7 \times 7$   & 0.062 & 0.054 & 0.053 & 0.057 & 0.059 \\
    	                                 &                             &  $9 \times 9$  & 0.060 & 0.066 & 0.050 & 0.046 & 0.046 \\ \hline
    	\multirow{6}{*}{$\alpha_{0, 1}$} & \multirow{3}{*}{log-linear} &  $5 \times 5$ & 0.095 & 0.072 & 0.054 & 0.043 & 0.035 \\
    	                                 &                             &  $7 \times 7$  & 0.077 & 0.060 & 0.050 & 0.051 & 0.056 \\ 
    	                                 &                             &  $9 \times 9$  & 0.075 & 0.042 & 0.057 & 0.051 & 0.048 \\
    	       \cline{2-8}        &   \multirow{3}{*}{linear}   &  $5 \times 5$  & 0.049 & 0.043 & 0.044 & 0.046 & 0.042 \\
    	                                 &                             &  $7 \times 7$  & 0.059 & 0.066 & 0.052 & 0.044 & 0.056 \\
    	                                 &                             &  $9 \times 9$  & 0.064 & 0.047 & 0.046 & 0.038 & 0.047 \\ \hline
    	\multirow{6}{*}{$\beta_{0, 1}$}  & \multirow{3}{*}{log-linear} &  $5 \times 5$  & 0.047 & 0.054 & 0.044 & 0.072 & 0.068 \\
    	                                 &                             &  $7 \times 7$  & 0.054 & 0.060 & 0.051 & 0.077 & 0.072 \\
    	                                 &                             &  $9 \times 9$  & 0.058 & 0.061 & 0.051 & 0.062 & 0.072 \\
    	       \cline{2-8}        &   \multirow{3}{*}{linear}   &  $5 \times 5$  & 0.049 & 0.049 & 0.054 & 0.038 & 0.054 \\
    	                                 &                             &  $7 \times 7$  & 0.053 & 0.049 & 0.050 & 0.045 & 0.042 \\ 
    	                                 &                             &  $9 \times 9$  & 0.039 & 0.038 & 0.042 & 0.043 & 0.044 \\ \hline
    \end{tabular}
    \caption{Empirical size of the asymptotic Wald test different parameters for different numbers of locations}
    \label{tab:empirical_size_frank}
\end{table}

In situations involving misspecified spatial dependencies, the correct model is readily identified based on the QIC with fewer observation times, as evidenced in Table~\ref{tab:anisotropy_qic_frank}. Consistent with prior tests, this leads to enhanced power in detecting the anisotropic neighborhood structure, as indicated in Table~\ref{tab:anisotropy_test_frank}. Additionally, the isotropic model exhibits a somewhat improved detection of spatial dependency, as demonstrated in Table~\ref{tab:isotropy_test_frank}.

\begin{table}[tb]
    \centering
    \begin{tabular}{lrrrr}
  \hline
  \multirow{2}{*}{Model}& \multicolumn{4}{c}{$T$} \\
  & 50 & 100 & 250 & 500 \\ 
  \hline
log-linear & 0.299 & 0.093 & 0.002 & 0.000 \\ 
linear & 0.271 & 0.082 & 0.000 & 0.000 \\ 
   \hline
\end{tabular}
    \caption{Relative frequencies of cases in which the QIC favors isotropic models over anisotropic models, which are the true ones, in the Frank copula simulations}
    \label{tab:anisotropy_qic_frank}
\end{table}

\begin{table}[tb]
    \centering
    \begin{tabular}{cccccc}
  \hline
  \multirow{2}{*}{Model} & \multirow{2}{*}{Anisotropy-Test}& \multicolumn{4}{c}{$T$} \\
 &  & 50 & 100 & 250 & 500 \\ 
  \hline
\multirow{2}{*}{log-linear}&$\beta$  & 0.410 & 0.672 & 0.965 & 1.000 \\
&$\alpha$  & 0.109 & 0.121 & 0.141 & 0.222 \\ \hline
 \multirow{2}{*}{linear} &$\beta$  & 0.454 & 0.704 & 0.975 & 1.000 \\
  &$\alpha$  & 0.066 & 0.087 & 0.150 & 0.256 \\
   \hline
\end{tabular}
    \caption{Empirical power of Wald tests at 5\% significance level for testing differences between the coefficients of an anisotropic model using separate neighborhood matrices for different directions in the Frank copula simulations;  $\alpha$ refers to the test for the coefficients of the feedback process, while $\beta$ refers to the observation process.}
\label{tab:anisotropy_test_frank}
\end{table}

\begin{table}[tb]
\centering
\begin{tabular}{cccccc}
  \hline
 \multirow{2}{*}{Model} & \multirow{2}{*}{Coefficient}& \multicolumn{4}{c}{$T$} \\
 &  & 50 & 100 & 250 & 500 \\ 
  \hline
\multirow{2}{*}{log-linear}&$\beta_{1, 1}$  & 0.858 & 0.995 & 1.000 & 1.000 \\
&$\alpha_{1, 1}$  & 0.164 & 0.205 & 0.380 & 0.593 \\ \hline
 \multirow{2}{*}{linear} &$\beta_{1, 1}$  & 0.923 & 0.993 & 1.000 & 1.000 \\ 
  &$\alpha_{1, 1}$  & 0.198 & 0.313 & 0.515 & 0.750 \\
   \hline
\end{tabular}
\caption{Relative frequencies of significant coefficients ($\beta_{1, 1}$ and $\alpha_{1, 1}$) ($p < 0.05$) in an isotropic model fitted to data from an anisotropic model, in the Frank copula simulations.}
\label{tab:isotropy_test_frank}
\end{table}

\end{document}